\documentclass[fleqn,usenatbib,useAMS]{mnras}

\usepackage{newtxtext,newtxmath}
\usepackage[T1]{fontenc}

\DeclareRobustCommand{\VAN}[3]{#2}
\let\VANthebibliography\thebibliography
\def\thebibliography{\DeclareRobustCommand{\VAN}[3]{##3}\VANthebibliography}

\usepackage{hyperref}
\usepackage{subcaption}
\usepackage{graphicx}	% Including figure files
\usepackage{amsmath}	% Advanced maths commands
\usepackage{tablefootnote}
\usepackage{gensymb}
\usepackage{orcidlink}
\usepackage{float}

\title[RIDDLER: SNe~Ia explosion scenarios]{Quantitative modelling of type Ia supernovae spectral time series II:\\Exploring the diversity of thermonuclear explosion scenarios }

\author[M. R. Magee]{
M. R. Magee$^{1}$\orcidlink{0000-0002-0629-8931}\thanks{E-mail: mrmagee.astro@gmail.com}
\\
$^{1}$Department of Physics, University of Warwick, Gibbet Hill Road, Coventry CV4 7AL, UK \\ 
}

\date{Accepted 2026 October 23. Received 2026 October 22; in original form 2026 April 24}

\pubyear{2026}

\begin{document}
\label{firstpage}
\pagerange{\pageref{firstpage}--\pageref{lastpage}}
\maketitle

\begin{abstract}
Observations of type Ia supernovae (SNe~Ia) have led to suggestions of multiple progenitor and explosion scenarios. Distinguishing between scenarios and tying specific SNe~Ia to individual scenarios however has so far been challenging. Constraints on the explosion physics are often achieved through empirical modelling of SNe~Ia spectra and qualitative assessments of the level of agreement. While this approach has provided useful insights, it cannot be scaled up to large numbers of SNe~Ia in a robust and systematic way. \textsc{riddler} is a machine learning based framework for automated and quantitative fitting of SNe~Ia designed to overcome these limitations. Neural networks are used as radiative transfer emulators and, in conjunction with nested sampling, emulated spectra are fit to observations of SNe~Ia to determine the best-fitting input parameters and explosion scenario. In this work, we present recent improvements to \textsc{riddler}, including a significantly expanded training dataset covering pure deflagrations, delayed detonations, double detonations, gravitationally confined detonations, and violent mergers. We show that despite the increased complexity and variety of our training data, \textsc{riddler} is able to accurately recover the input parameters and explosion scenario of spectra unseen during training. Using \textsc{riddler}, we fit observations of three SNe~Ia covering different sub-classes: SN~2011fe, SN~2005hk, and SN~2018byg. We detail a number of limitations and assumptions that should be considered when applying similar approaches. Nevertheless, the benefits of this approach and \textsc{riddler} will result in automated fitting playing an increasingly important role in the coming years.
\end{abstract}


%%%%%%%%%%%%%%%%%%%%%%%%%%%%%%%%%%%%%%%%%%%%%%%%%%

%%%%%%%%%%%%%%%%% BODY OF PAPER %%%%%%%%%%%%%%%%%%

\section{Introduction}
\label{sect:intro}
Increasingly large spectroscopic datasets of type Ia supernovae (SNe~Ia) have proved invaluable in furthering our understanding of thermonuclear explosion physics \citep{liu--23b, ruiter--25}. While candidate progenitor scenarios and explosion mechanisms have been identified in a handful of cases (e.g. \citealt{kromer-13, de--19, galbanu--19, dimitriadis--23}), the origins of the majority of SNe~Ia remain unknown. It is widely believed that they are the result of thermonuclear explosions of white dwarfs in binary systems, but multiple pathways to producing such events have been proposed and distinguishing between them is non-trivial. 

\par

To understand the diversity and observable signatures of different explosion mechanisms, highly sophisticated and multi-dimensional explosion simulation and radiative transfer codes have been developed \citep{hoeflich-95, hoeflich--2003a, roepke--hillebrandt--05, kasen-06b, artis, springel--10, hillier--2012, vanrossum--2012, ergon--18}. Beginning from the progenitor white dwarf(s), the explosion can be triggered under the specific desired criteria, the evolution of the explosion followed, and synthetic observables generated at various phases in the SN evolution. These synthetic observables can then be compared directly against observed light curves and spectra, thereby testing whether a given SN is consistent with a given explosion scenario. Significant advances have resulted from this approach (e.g. \citealt{baron--12, dessart--14a, roepke--18, polin--19, shen--21, pakmor--24}), but the high computational expense associated with such modelling means it is not suitable for generating large suites of models or performing detailed characterisation of newly observed SNe~Ia. 

\par

Alternatively, simplified radiative transfer codes have also been developed \citep{mazzali--lucy--93, tardis}. Rather than beginning from the explosion of a white dwarf, modelling with these codes instead follows a more empirical approach and does allow for detailed characterisation. For a given SN, a customised ejecta model is constructed and radiative transfer performed to produce individual snapshot spectra. Input parameters can subsequently be iterated upon to improve agreement and modelling of multiple spectra at different phases enables a more complete picture of the SN ejecta than single epochs. The speed and flexibility of these codes means that they can be used to rapidly converge towards models that reproduce observations and this approach has been highly successful in providing constraints on the explosion physics of individual SNe~Ia and samples \citep{stehle--05, mazzali--07, mazzali--08, tanaka--11, mazzali--14, sasdelli--14, ashall--16, heringer--17, 12bwh, heringer--19, harvey--23, yadavalli--24}. This approach is also limited however as the level of agreement is typically judged based on visual inspection, making it difficult to quantify and reproduce. Likewise, without generating large numbers of spectra across the parameter space it is impossible to determine whether the final model parameters represent the global best-fitting solution.

\par

Overcoming the limitations of both approaches and producing robust comparisons between explosion or empirical models and observations of SNe~Ia has been a challenging prospect, but is required to understand the explosion mechanism(s).
\cite{sasdelli--17} use principal component analysis to characterise SNe~Ia and explosion models, showing that all models have counterparts within the observed SN~Ia sample, but reconstructions of the explosion model spectra show some disagreement. Direct spectral comparisons through quantitative means have also shown promise. \cite{ogawa--23} present a large suite of simulated spectra and define multiple metrics, such as equivalent widths and FWHM, to asses the level of agreement with observations, which are then weighted to broadly reproduce the approach of visual inspection in a semi-automatic way.

\par

Machine learning-based techniques have been increasingly applied to fully automate fitting of stellar \citep{czekala--15, ness--15, ting--19, badenas-agusti--24, badenas-augusti--25} and supernovae spectra \citep{vogl--19, chen--20, kerzendorf--21, obrien--21, chen--24, magee--24, obrien--24}. Typically this involves generative models or emulators producing large numbers of spectra that are compared against observations and physical properties, such as the chemical composition, inferred. \cite{chen--24} use custom neural networks designed to emulate \textsc{tardis} radiative transfer simulations \citep{chen--20} to constrain the $^{56}$Ni abundances of SNe~Ia, while \cite{obrien--24} use the \textsc{tardis} emulator \textsc{dalek} \citep{kerzendorf--21} to investigate differences between the ejecta of normal and 91T-like SNe~Ia \citep{91t--like}. \cite{magee--24} present \textsc{riddler}, a machine learning-based method for automatic and quantitative fitting of SNe~Ia spectra. Unlike previous works, \textsc{riddler} is designed to fit multiple spectra of a given SN~Ia (up to shortly after maximum light) simultaneously and the training data is more directly based on explosion model predictions. \cite{magee--24} show how it is therefore possible to assess the relative likelihood of explosion scenarios based on their agreement with observed spectra. The training data used by \cite{magee--24} however was limited to only two explosion scenarios and did not cover the broad range of scenarios proposed to produce SNe~Ia.

\par

In this paper we describe recent improvements made to \textsc{riddler} and highlight differences with respect to \cite{magee--24}. This includes a significantly expanded training dataset and updated neural network architectures with more robust uncertainty estimates. Our expanded training dataset is presented in Sect.~\ref{sect:training_models} and used to train the new neural networks discussed in Sect.~\ref{sect:nns}. In Sect.~\ref{sect:fitting} we discuss fitting spectra with \textsc{riddler} and the fitting performance. Section~\ref{sect:application} applies this method to observations. Finally, we discuss our results in Sect.~\ref{sect:discussion} and present our conclusions in Sect.~\ref{sect:conclusions}. The updated version of \textsc{riddler} is publicly available\footnote{\href{https://github.com/MarkMageeAstro/riddler}{https://github.com/MarkMageeAstro/riddler}}. A companion paper, \citealt{magee--26b}, explores the use of \textsc{riddler} in fitting samples of cosmological SNe~Ia spectra and implications for SNe~Ia standardisation.

%
%__________________________________________________________________________________________________________________________________________________________________________________________
%__________________________________________________________________________________________________________________________________________________________________________________________
%__________________________________________________________________________________________________________________________________________________________________________________________

\section{Training models}
\label{sect:training_models}

The initial training datasets presented by \cite{magee--24} included spectra generated based on the ejecta structures (density profiles and compositions) predicted by two underlying explosion scenarios: the W7 pure deflagration \citep{nomoto-w7} and the N100 delayed detonation \citep{seitenzahl--13} models. To allow each explosion scenario to fit a variety of SNe~Ia, the kinetic energy ($KE$) for each ejecta model was treated as a free parameter, enabling some differences in structure among the models within each of the respective training datasets.

\par

With this work, we aim to further expand the diversity of objects that can be fit with \textsc{riddler} to include both normal and peculiar SNe~Ia across a wide range of luminosities and spectral features. To that end, we present updated training datasets based on predictions from an increased set of explosion scenarios. This includes pure deflagrations (DEF), delayed detonations (DDT), double detonations (DOD), gravitationally confined detonations (GCD), and violent mergers (VM). In addition, our new pure deflagration and delayed detonation models are no longer based on a single realisation of each explosion scenario. 

\par

The purpose of our training data is to produce models similar to those predicted by explosion simulations, but with varied parameters such that our models may be considered as interpolations between the existing explosion simulations and broadly capturing the diversity predicted by different initial conditions, viewing angles, or modelling systematics. Section~\ref{sect:ejecta} describes the construction of new ejecta structures based on explosion simulations, while Sect.~\ref{sect:simulation_parameters} describes sampling additional simulation parameters used to generate synthetic spectra. In Sect.~\ref{sect:training_data_summary} we summarise the generation of our training data.

\subsection{Ejecta structures}
\label{sect:ejecta}

To generate new ejecta structures based on each underlying explosion scenario, we develop a pipeline to analyse and process the one-dimensional, angle-averaged explosion model predictions available from HESMA\footnote{\href{https://hesma.h-its.org/}{https://hesma.h-its.org/}} \citep{hesma}. These models have been used extensively for direct comparisons with observations and are generally computed with similar methods and/or simulation codes, making them an ideal reference set upon which to base our training data. For the purposes of our pipeline and training data, each explosion model is assumed to contain a `core' and a `shell' (although we note that in most cases the shell mass is 0~$M_{\odot}$), the abundances within which may be updated independently of each other. This process ensures that explosion scenarios are treated in a consistent manner regardless of their complicated underlying structures. The pipeline is designed to be as flexible as possible and allow for new training datasets to be easily generated and added to \textsc{riddler} in future releases. 

\par

For each explosion scenario, rather than the ejecta structure being varied by a single parameter ($KE$) as in \cite{magee--24}, here we introduce an additional ten parameters. The parameters used to calculate the ejecta structure for a specific training sample are therefore defined as:
\begin{itemize}
    \item Explosion strength, $ES$
    \item Total ejecta mass, $M_{\mathrm{ej}}$,
    \item Fraction of the ejecta contained within the core, $f_C$,
    \item Fraction of $f_C$ burned to heavy elements, $f_{C}^{b}$,
    \item Fraction of $f_{C}^{b}$ burned to iron-group elements, $f_{C}^{\mathrm{IGE}}$,
    \item Fraction of $f_{C}^{\mathrm{IGE}}$ burned to $^{56}$Ni, $f_{C}^{\textrm{Ni}}$,
    \item Fraction of the shell, $f_S$, burned to heavy elements, $f_{S}^{b}$,
    \item Fraction of $f_{S}^{b}$ burned to iron-group elements, $f_{S}^{\mathrm{IGE}}$,
    \item Fraction of $f_{S}^{\mathrm{IGE}}$ burned to $^{56}$Ni, $f_{S}^{\textrm{Ni}}$,
    \item Fraction of carbon/oxygen in the unburned shell material, $f_{S}^{\textrm{C/O}}$,
    \item Kinetic energy, $KE$.
\end{itemize}
The definition of explosion strength varies depending on the underlying HESMA models, but represents a parameterised way of controlling how much energy is released by the SN. Here, heavy elements are defined as including iron-group elements (IGE; Ti -- Ni) and intermediate mass elements (IME; Ne -- Ca).

\par 

Each explosion scenario is treated separately, but within each scenario the above parameters are generally highly correlated with the strength of the explosion. Which parameters are correlated and by how much depends on the specific explosion scenario and are discussed within the relevant subsections that follow. For example, total ejecta mass is correlated with explosion strength in the DEF models, but not for the DDT models (which all have the same ejecta mass of 1.4~$M_{\odot}$). For each training dataset (i.e. explosion scenario) typically only one or two ejecta parameters are truly independent -- the parameters that set the strength of the explosion. Each of the remaining parameters is then randomly sampled dependent on the explosion strength, predictions from the respective explosion scenarios, and additional added scatter. This additional scatter is designed to account for intrinsic scatter of the models themselves, variations in abundances that may be expected from, for example, different viewing angles or metallicities, and the uncertainties associated with explosion simulations. The latter is the most difficult to quantify and therefore this additional scatter is somewhat arbitrary. In addition, the level of additional scatter varies depending on the input parameter and explosion scenario. Appendix~\ref{sect:appdx:scatter} provides further discussion of the assumed scatter for each explosion scenario when sampling new input parameters. These parameters are then used to calculate the total masses of $^{56}$Ni ($M_{\mathrm{Ni}}$), iron-group elements ($M_{\mathrm{IGE}}$; excluding $^{56}$Ni), intermediate-mass elements ($M_{\mathrm{IME}}$), and unburned material ($M_{\mathrm{C/O}}$) in the core and the shell. The initial ejecta structure is first based on interpolating the respective underlying explosion models to the relevant explosion strength parameter and then scaled to the new, randomly sampled $KE$ (and $M_{\textrm{ej}}$ for the pure deflagration models; \citealt{hachinger--09, ashall--16, magee--24}). Abundances of each element are then also scaled such that the total masses of the different groups match those from the random sampling. 

\par

\begin{figure}
\centering
\includegraphics[width=\columnwidth]{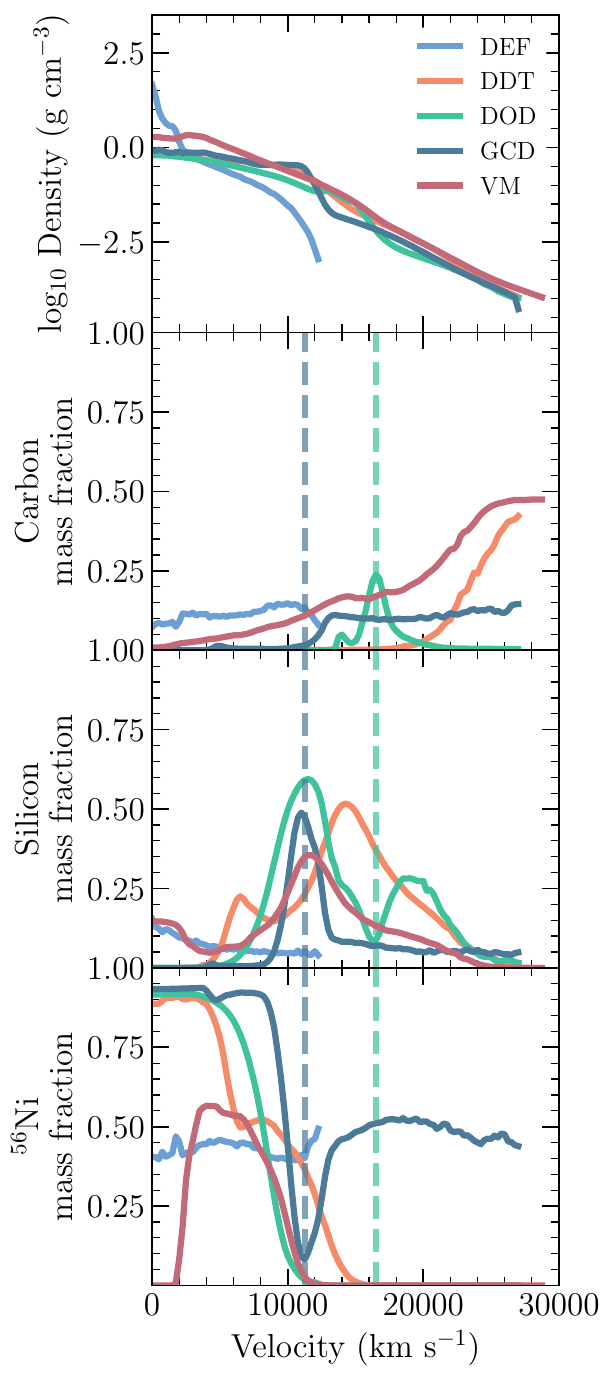}
\caption{Density profiles and mass fractions of carbon, silicon, and $^{56}$Ni for representative models used to construct our training datasets. Models shown are the N5def pure deflagration (DEF; \citealt{fink-2014}), N100 delayed detonation (DDT; \citealt{seitenzahl--13}), M09\_03 double detonation (DOD; \citealt{gronow--21}), r57\_d3.0 gravitationally confined detonation (GCD; \citealt{lach--22b}), and 11\_09 violent merger (VM; \citealt{pakmor-2012}). Densities and mass fractions are shown at 100s after explosion. Vertical dashed lines show the location of the shell for the double detonation and gravitationally confined detonation models. 
}
\label{fig:model_ejecta}
\centering
\end{figure}

The additional parameters used in this work allow for significantly more variation among models within each of the training datasets than those presented by \cite{magee--24}. We note however that each training dataset is still directly tied to predictions from explosion simulations and does not have the freedom to arbitrarily choose any combination of, for example, ejecta mass, $^{56}$Ni mass, etc. In addition, with the exception of $^{56}$Ni, the relative abundances of elements within a specific group (IGEs, IMEs, C/O) are fixed to those predicted by the explosion scenario. For example, while our random sampling and scaling can change the total fraction of IMEs within the ejecta, the relative fraction of silicon to sulphur will remain the same for a given explosion strength. 

\par

In the following subsections, we describe the sampling of parameters and construction of ejecta structures for each explosion scenario considered here. The subscript $x$ refers to values associated with a specific sample of the training data. For example $KE_x$ refers to the kinetic energy of model $x$. Figure~\ref{fig:model_ejecta} shows indicative ejecta structures for each explosion scenario.

\subsubsection{Pure deflagrations (DEF)}
\label{sect:def}
The pure deflagration explosion scenario \citep{nomoto-w7, reinecke--02a, reinecke--02b, gamezo--03, roepke--06, roepke--07} invokes subsonic burning of a near-Chandrasekhar mass white dwarf, known as a deflagration. Within this scenario, the slow speed of burning results in some expansion of the white dwarf fuel and the production of IMEs at lower densities. Rayleigh-Taylor instabilities are formed due to the outward propagation of the deflagration front, accelerating the front and leading to significant mixing of burned and unburned material. Early one-dimensional models of pure deflagrations, such as the W7 model, showed promise in reproducing observations of normal SNe~Ia around maximum light, but were tuned with a parameterised description of the deflagration front and its propagation speed \citep{nomoto-w7}. Unlike \cite{magee--24}, here we do not include the W7 model in our training dataset and instead focus on more realistic multi-dimensional simulations. 

\par

\cite{fink-2014} present a suite of 3D pure deflagration explosions of Chandrasekhar mass white dwarfs. For these models, the explosion is ignited centrally through a number of ignition sparks, $N_k$, ranging from 1 -- 1600. The strength of the explosion is directly correlated with the number of ignition sparks -- more sparks result in stronger deflagrations and more energetic explosions. For weak deflagrations the explosion is insufficient to completely unbind the white dwarf, producing only a small amount of $^{56}$Ni and ejecta, and leaving behind a relatively massive bound remnant. While these models are inconsistent with observations of normal SNe~Ia, they are generally compatible with observations of the peculiar SNe~Ia subclass, SNe~Iax \citep{kromer-13, 15h, camacho-neves--23, maguire--23}. 

\par

The multi-spot ignition approach used by \cite{fink-2014} is a convenient method of controlling the strength of the explosion. \cite{nonaka--12} however argue that such multi-spot ignition is unlikely to occur and instead single-spot ignition is more probable. Nevertheless, \cite{lach--22} explore changes to the initial conditions for single-spot ignition (including the central density, offset radius, and progenitor metallicity) and find similar results to \cite{fink-2014}. For the \cite{fink-2014} models, the ejecta structure parameters described in Sect.~\ref{sect:training_models} are highly correlated with the explosion strength and hence the number of ignition sparks. Given the simplicity of a single parameter controlling the explosion strength, we use the \cite{fink-2014} models as the basis of our pure deflagration training dataset.

\par

For each model $x$ in our pure deflagration training dataset, the explosion strength parameter, $ES$, is defined as the number of ignition sparks $N_k$. We begin by randomly sampling $N_{k,x}$ uniformly between 1 -- 20. While the full set of \cite{fink-2014} models extend from $N_k$ = 1 -- 1600, models with $N_k$ $\lesssim$ 20 are broadly consistent with observations of SNe~Iax and those with $N_k$ $\gtrsim$ 20 show more complicated correlations between ejecta parameters and $N_k$. We therefore do not include models with higher numbers of ignition sparks in our training data. From the \cite{fink-2014} predictions, we sample $M_{\textrm{ej},x}$ based on the randomly selected $N_{k,x}$ for our given model and additional scatter of $-20$ -- $+100$~per cent. The increased upper limit on $M_{\textrm{ej}}$ is somewhat arbitrary and was chosen to ensure that our ejecta masses broadly cover the range observed among SNe Iax as the \cite{fink-2014} generally predict systematically lower masses \citep{fink-2014, obs--05hk--08a, srivastav--22}. The \cite{fink-2014} models do not predict shells within the ejecta, therefore $f_{C,x} = 1$ for all models and all shell parameters are set equal to 0. For $f_{C}^b$, $f_{C}^{\textrm{IGE}}$, and $f_{C}^{\textrm{Ni}}$, we again calculate these as a function of $N_k$ from the \cite{fink-2014} predictions and interpolate to the selected $N_{k,x}$, with scatter of $\pm$20~per~cent. The $KE$ is sampled as a function of ejecta mass with scatter of $\pm$30~per~cent. As mentioned, the density and composition profiles predicted by \cite{fink-2014} are then interpolated to $N_{k,x}$ and the density, velocity, and abundances are scaled to match the sampled values. Appendix~\ref{sect:appdx:def} provides additional figures and details of the correlations between sampled parameters.

\subsubsection{Delayed detonations (DDT)}
\label{sect:dedet}
While simulations invoking pure deflagrations of Chandrasekhar mass white dwarfs are not suitable for normal SNe~Ia, those that eventually undergo a detonation (i.e. the delayed detonation or deflagration-to-detonation transition, DDT, scenario) have shown more promise in reproducing observations \citep{khokhlov--91a, khokhlov--91b, hoeflich-95, hoeflich--2017, blondin--11, blondin--13, sim--13}. In this scenario, the explosion progresses in a similar way to the pure deflagrations described in Sect.~\ref{sect:def} -- subsonic burning of the white dwarf results in some expansion of the fuel and a decrease in density. A subsequent detonation burns through the pre-expanded fuel and the remainder of the white dwarf \citep{khokhlov--89, khokhlov--91a, hoeflich-95, gamezo--04, gamezo--05}. Trigger mechanisms for the detonation are unclear, but it is believed to result from strong turbulent velocity fluctuations \citep{niemeyer--97, woosley--07, woosley--09}. Compared to pure detonations of Chandrasekhar mass white dwarfs, the pre-expanded fuel of the delayed detonation scenario results in the production of less $^{56}$Ni and more IMEs \citep{seitenzahl--17}. 

\par

\cite{magee--24} develop a training dataset for the delayed detonation scenario based purely on the N100 model published by \cite{seitenzahl--13}. Similar to \cite{fink-2014} however, \cite{seitenzahl--13} present a suite of models exploring different explosion strengths, which are again controlled by the number of sparks used to ignite the initial deflagration phase and range from $N_k$ = 1 -- 1600. Contrary to the \cite{fink-2014} pure deflagration models, in the delayed detonation scenario the explosion strength and production of $^{56}$Ni is inversely related to the number of sparks. Ignition with more sparks results in a more vigorous deflagration phase, overall lower densities of the fuel at the moment of detonation, and therefore a less energetic explosion and less $^{56}$Ni being produced. As with the deflagration models, the structure parameters described in Sect.~\ref{sect:ejecta} are highly correlated with the number of ignition sparks. Here we use an expanded set of delayed detonation models from \cite{seitenzahl--13}, covering 10 -- 150 ignition sparks, as the basis of our training dataset. We limit our models to $10 \leq N_k \leq 150$ as initial testing showed that our neural networks struggled to learn features of the spectra outside of this range. This likely results from more complicated relationships between parameters than captured by our sampling. We note however that models outside of this range are generally over/under-luminous relative to normal SNe Ia. Our selected range of $N_k$ = 10 -- 150 should therefore be suitable for the bulk of the SNe~Ia population and models within this range have shown good agreement with observations of normal SNe~Ia \citep{sim--13, bulla--16, harvey--23}. 

\par

As was the case for our pure deflagration models, the explosion strength parameter, ES, is defined as the number of ignition sparks, $N_k$. We uniformly sample $N_{k,x}$ between our chosen limits (10 -- 150). Unlike the pure deflagration scenario, all delayed detonation models presented by \cite{seitenzahl--13} completely unbind the white dwarf and also do not predict a shell. We therefore set $M_{\textrm{ej},x}$ = 1.4~$M_{\odot}$, $f_{C,x} = 1$, and all shell parameters equal to 0 for all models. We follow the same procedure as Sect.~\ref{sect:def} wherein we calculate $f_{C}^b$, $f_{C}^{\textrm{IGE}}$, $f_{C}^{\textrm{Ni}}$, and $KE$ as a function of $N_k$ from the \cite{seitenzahl--13} models, interpolate to $N_{k,x}$, and apply additional random scatter of $\pm$30~per~cent for $KE_x$ and $\pm$20~per~cent for all other parameters. Density and composition profiles predicted by \cite{seitenzahl--13} are interpolated to $N_{k,x}$ and scaled to match the sampled values. Further discussion and figures are provided in Appendix~\ref{sect:appdx:ddt}.

\subsubsection{Double detonations (DOD)}
\label{sect:dodet}
In the double detonation explosion scenario, a sub-Chandrasekhar mass white dwarf accretes material from its companion and accumulates a helium shell on its surface. Under certain conditions, this helium shell may spontaneously ignite and trigger a detonation. \citep{livne--90, livne--91, woosley--94, bildsten--07, fink--10, kromer--10, shen--14b, polin--19, gronow--20}. Once triggered, the detonation front sweeps around the shell and burns it to heavier elements. A secondary detonation of the core that completely disrupts the white dwarf may then be triggered \citep{livne--90, livne--91, moll--13, gronow--20}. Burning in the helium shell can produce a significant amount of IGEs in the outer ejecta that result in strong line blanketing and observables that do not agree with the majority of normal SNe~Ia \citep{kromer--10, polin--19, collins--22}. Some peculiar SNe~Ia however, are consistent with this scenario \citep{jiang--2017, de--19, jacobson-galan--20, dong--22, liu--23, padilla-gonzalez--23}. Models with thin or polluted helium shells generally predict lower amounts of IGEs (or no IGEs) are produced in the outer ejecta and similar observables to normal SNe~Ia \citep{shen--18, polin--19, townsley--19}. 

\par

Multi-dimensional simulations within the double detonation scenario are presented by \cite{gronow--21}. The explosion strength and production of $^{56}$Ni for these models is directly related to the total white dwarf mass (i.e. core + shell), with higher masses resulting in stronger explosions and more $^{56}$Ni. \cite{gronow--21} explore the impact of a range of white dwarf core masses ($M_C$; ranging from 0.8 -- 1.1~$M_{\odot}$) and helium shell masses ($M_S$; ranging from 0.02 -- 0.11~$M_{\odot}$). We use the \cite{gronow--21} models as the basis for our double detonation training dataset. Similar to previous works, models with thick helium shells predict strong line blanketing and observations redder than normal SNe~Ia, but consistent with some peculiar objects \citep{collins--22}.

\par

Here, the explosion strength parameter is defined as the total ejecta mass, $M_{\textrm{ej}}$, which is calculated as the sum of $M_C$ and $M_S$. \cite{gronow--21} do not consider all combinations of $M_C + M_S$, therefore we limit our sampling ranges for $M_{C,x}$ to 0.8 -- 1.0~$M_{\odot}$ and for $M_{S,x}$ to 0.03 -- 0.10~$M_{\odot}$ to avoid extrapolating beyond the models. Both parameters are uniformly sampled within these limits and then used to calculate $M_{\textrm{ej},x}$ and $f_{C,x}$. Within the \cite{gronow--21} models, $f_{C}^b$, $f_{C}^{\textrm{IGE}}$, $f_{C}^{\textrm{Ni}}$, and $KE$ are highly correlated with $M_{\textrm{ej}}$. Using the previously sampled value of $M_{\textrm{ej},x}$, we therefore calculate these parameters allowing for scatter of $\pm$30~per~cent for $KE_x$ and $\pm$20~per~cent for all other parameters. For the composition of the shell we follow a somewhat different approach. As previously mentioned, our spectra are generated using the one-dimensional code \textsc{tardis}, whereas the \cite{gronow--21} models were calculated in 3D and show significant viewing angle variations (in particular, the composition of the helium shell is highly asymmetric). In addition, \cite{shen--14b} and \cite{townsley--19} show how helium shells polluted with $^{14}$N do not produce significant amounts of burned material in the shell. We wish to capture these effects and therefore produce models with lower amounts of burned material than the angle-averaged \cite{gronow--21} models. Hence we calculate $f_{S}^b$ as a function of $M_{\textrm{ej}}$, but allow for scatter of $-80$~per~cent and $+20$~per~cent when determining $f_{S,x}^b$. From the \cite{gronow--21} models, there is no strong correlation between the fraction of the IGEs in the burned shell material and other parameters controlling the explosion strength. Instead, most models range from $f_{S}^{IGE} \simeq$15 -- 60~per~cent, with the lowest mass model predicting only $\sim$0.01~per~cent. We therefore uniformly sample $f_{S,x}^{IGE}$ between 0.01 -- 60~per~cent. The fraction of IGEs burned to $^{56}$Ni however is weakly correlated with the amount of IGEs in the shell, therefore we calculate $f_{S}^{Ni}$ as a function of $f_{S}^{IGE}$ and allow for scatter of $\pm$90~per~cent to cover all models. The fraction of C/O in the unburned material is inversely correlated with the fraction of the shell burned to heavy elements, but again shows some scatter. We calculate $f_{S}^{C/O}$ as a function of $f_{S}^{b}$, interpolate to $f_{S,x}^{b}$, and allow for scatter of $\pm$80~per~cent. Finally, the density and composition profiles from \cite{gronow--20} are interpolated to $M_{C,x}$ and $M_{S,x}$. Mass coordinates within $M \leq M_C$ are scaled to the sampled abundances of the core and those above $M \textgreater M_S$ are scaled to the sampled abundances of the shell. This process introduces a discontinuity between the abundances of the core and the shell, therefore we also apply a Gaussian smoothing function to the composition profiles to avoid abrupt and unrealistic changes in the composition. Appendix~\ref{sect:appdx:dod} shows the relationships between sampled parameters and provides further discussion.

\subsubsection{Gravitationally confined detonations (GCD)}
\label{sect:gcd}
The gravitationally confined detonation scenario is a variation of the delayed detonation scenario \citep{plewa--04}. Similar to the pure deflagration and delayed detonation scenarios, the explosion begins as the subsonic deflagration of a Chandrasekhar mass white dwarf. As the burning products from the deflagration rise towards the white dwarf surface, they wrap around the core and converge on the opposite side. Compression and heating of unburned fuel at this point may trigger a detonation \citep{plewa--07, meakin--09}. Generally these models produce relatively large amounts of $^{56}$Ni and are somewhat over-luminous compared to most SNe~Ia, but also show some similarities with bright SNe~Ia and peculiar 91T-like SNe~Ia \citep{kasen--07b, seitenzahl--16, lach--22b}. 

\par

Following from \cite{lach--22}, \cite{lach--22b} present a suite of gravitationally confined detonation models in which the initial conditions for the deflagration ignition are varied and a subsequent detonation is ignited. The overall strength of the explosion and amount of $^{56}$Ni produced are governed by the amount of nuclear energy released during the deflagration, $E^{\textrm{def}}_{\textrm{nuc}}$, which is itself determined by the ignition conditions and ranges from 1.29 -- $3.15 \times 10^{50}$~erg. As in the delayed detonation scenario, models with more vigorous deflagrations (higher $E^{\textrm{def}}_{\textrm{nuc}}$) result in less $^{56}$Ni being produced. We use the \cite{lach--22b} models as the basis of our training dataset, but only consider models with $E^{\textrm{def}}_{\textrm{nuc}} \leq 2.81 \times 10^{50}$~erg. This excludes the most vigorous deflagration ignition model, which deviates significantly from the other models in terms of the parameters discussed in Sect.~\ref{sect:training_models}. \cite{lach--22b} find good agreement with their models and observations of SN~1991T \citep{91t--disc} within the first few weeks after explosion, but the level of spectroscopic agreement diminishes closer to and after maximum light.

\par

For our GCD models, the explosion strength parameter is defined as $E^{\textrm{def}}_{\textrm{nuc}}$ and uniformly sampled from 1.29 -- 2.81$\times 10^{50}$~erg. As with the delayed detonation scenario, $M_{\textrm{ej},x} = 1.4~M_{\odot}$ for all models. Gravitationally confined detonations result in an ejecta with a shell-like structure in which the material produced by burning in the core (mostly $^{56}$Ni) is surrounded by the deflagration ash. In particular, the angle-averaged $^{56}$Ni distributions show a bimodal structure (see Fig.~\ref{fig:model_ejecta}). For the \cite{lach--22b} models we use the minimum of the $^{56}$Ni distribution to define the point at which the ejecta transitions from the core to the shell. We calculate $f_C$ (which ranges from $\sim$60 -- 90~per~cent) as a function of $E^{\textrm{def}}_{\textrm{nuc}}$ and interpolate to $E^{\textrm{def}}_{\textrm{nuc},x}$, allowing for additional scatter of $\pm$20~per~cent. For most other parameters, including those of the shell, we calculate them as a function of $E^{\textrm{def}}_{\textrm{nuc}}$, interpolate to $E^{\textrm{def}}_{\textrm{nuc},x}$, and randomly sample scatter between $\pm$20~per~cent. Exceptions to this are $f_S^{b}$, for which we allow $-80 -- +20$~per~cent scatter to account for asymmetric distributions, $f_S^{\textrm{Ni}}$, which is calculated as a function of $f_S^{\textrm{IGE}}$, and $f_S^{\textrm{C/O}}$, which is equal to 1 for all models (no helium is present). Similar to the double detonation models, the density and composition profiles from \cite{lach--22b} are interpolated to our selected $E^{\textrm{def}}_{\textrm{nuc},x}$, the abundances in the core and shell scaled to their respective values, and a smoothing function applied to remove discontinuities. The relationships used to sample new parameters for gravitationally confined detonation models are discussed further in Appendix~\ref{sect:appdx:gcd}.

\subsubsection{Violent mergers (VM)}
\label{sect:vm}
The violent merger scenario involves the merging and subsequent detonation of two sub-Chandrasekhar mass white dwarfs \citep{iben--84, webbink--84, pakmor--10}. In this scenario, the (typically lower mass) secondary white dwarf is disrupted and accreted violently onto the primary, leading to the formation of hot spots and a subsequent detonation that completely disrupts the primary \citep{iben--84, webbink--84, nomoto--91}. \cite{pakmor--10} show that mergers of white dwarfs with equal masses can result in detonations of the primary during the merger process and therefore at sub-Chandrasekhar masses. These simulations result in observables that are generally consistent with sub-luminous 91bg-like SNe~Ia. Subsequent simulations covering white dwarfs of similar (but not necessarily equal) masses also showed that this scenario could produce observables more comparable to normal SNe~Ia \citep{pakmor-2012}. 

\par

Multi-dimensional violent merger simulations have been presented by a number of previous works. \cite{pakmor--10} invoke the merger of two 0.9~$M_{\odot}$ white dwarfs, which results in the production of only $\sim$0.1~$M_{\odot}$ of $^{56}$Ni, and show that this scenario broadly reproduces the light curves and spectra of sub-luminous SNe~Ia. \cite{pakmor-2012} simulate the merger of a 0.9~$M_{\odot}$ white dwarf with a higher mass 1.1~$M_{\odot}$ primary. The higher mass (and therefore density) of the primary in this scenario results in 0.61~$M_{\odot}$ of $^{56}$Ni and better agreement with the light curves and spectra of normal SNe~Ia. \cite{kromer--13b} however present the simulation of a 0.9~$M_{\odot}$ primary merging with a lower mass 0.76~$M_{\odot}$ secondary that produces 0.18~$M_{\odot}$ of $^{56}$Ni. \cite{kromer--13b} find that this scenario broadly reproduces the faint and slowly-evolving SN~2010lp \citep{taubenberger--13}. We use these models as the basis of our violent merger training dataset. Models across these works were calculated with different explosion simulation codes, rather than a single suite of models as was the case for those described previously, therefore parameterising the merger models is more uncertain and the models are likely subject to more systematics and biases. In addition, our training dataset is more limited as fewer models are available from HESMA. Nevertheless, the parameters described in Sect.~\ref{sect:training_models} are generally correlated with the total mass of the system (i.e. primary + secondary mass).

\par

The explosion strength parameter is defined as the total ejecta mass for our violent merger models. The limited set of violent merger simulations available on HESMA have mass ratios ranging from 0.8 -- 1.0 and total ejecta masses ranging from 1.7 -- 2.0~$M_{\odot}$. We uniformly sample mass ratios and ejecta masses for our models from within these ranges. Merger models do not predict a shell and therefore we set $f_{C,x} = 1$ and all shell parameters equal to 0 for all models. Most parameters are generally correlated with the ejecta mass. We sample $f_C^b$, $f_C^{\textrm{IGE}}$, and $KE$ as a function of $M_{\textrm{ej}}$, allowing for scatter of $\pm20$~per~cent, $\pm50$~per~cent, and $\pm30$~per~cent. Larger scatter is used for $f_C^{\textrm{IGE}}$ compared to other scenarios to bracket the range of values predicted by the models. It is unclear however whether the scatter in this limited set of models is related to intrinsic properties of the violent mergers or different simulation approaches. The amount of IGE burned to $^{56}$Ni, $f_C^{Ni}$, is not strongly correlated with other parameters and therefore we uniformly sample between 0.75 -- 0.95. Again this brackets the range of values among the existing explosion models. Composition and density profiles are interpolated to the primary and secondary masses, given by $M_{\textrm{ej},x}$ and the mass ratio, and scaled to the appropriate abundances. Appendix~\ref{sect:appdx:vm} presents further discussion of the sampling used for our violent merger models.

\subsection{Simulation parameters}
\label{sect:simulation_parameters}
Having developed models for the SN ejecta, we now require additional simulation parameters with which to generate synthetic spectra that can be used to train our neural networks. As in \cite{magee--24}, all of our model spectra are calculated using \textsc{tardis} \citep{tardis}. Briefly, \textsc{tardis} is a one-dimensional Monte Carlo radiative transfer code designed for modelling spectra of SNe~Ia. Monte Carlo packets are injected at the photosphere assuming a blackbody distribution. Their propagations through and interactions with the model ejecta are followed, and escaping packets are binned to construct the synthetic spectrum. This process is iterated upon until the total luminosity of escaping packets matches the user-requested luminosity. During each iteration, the propagation of packets throughout the ejecta is used to calculate updated temperatures and opacities of the ejecta for the subsequent iteration and the blackbody temperature of the injected packets is updated. Further details are given by \cite{tardis}.

\par

\textsc{tardis} is fast, flexible, and sufficiently physically-motivated to make it ideal for generating large sets of spectra that can be used as training data for neural networks \citep{chen--20, chen--24, kerzendorf--21, obrien--21, obrien--24, magee--24}. Each \textsc{tardis} simulation requires a model of the SN ejecta and a number of additional simulation parameters. New ejecta models are generated following the procedure outlined in Sect.~\ref{sect:ejecta}. Other simulation parameters required by \textsc{tardis} include the time since explosion $t_{\mathrm{exp}}$, the luminosity $L$, and the photosphere (inner boundary) velocity $v_i$. As in \cite{magee--24}, for each model we uniformly sample the time since explosion $t_{\textrm{exp}}$, but with a slightly expanded range of 5 -- 30\,d. The requested luminosity of the model is determined by the rise time, luminosity at maximum light, and time since explosion. Similarly, the inner boundary is determined by the velocity gradient, velocity at peak, and time since explosion. This gives the following five simulation parameters for each sample in our training data:
\begin{itemize}
    \item Rise time, $t_r$,
    \item Luminosity at maximum light, $L_{\mathrm{max}}$,
    \item Velocity at maximum light, $v_{\mathrm{max}}$,
    \item Velocity gradient, $\Delta v$,
    \item Time since explosion, $t_{\mathrm{exp}}$
\end{itemize}

\cite{magee--24} use an empirical approach to calculate the luminosity $L$ of each \textsc{tardis} simulation based on observed bolometric light curves of SNe~Ia and an assumed rise time of 18.5\,d. Given the increased diversity of our training datasets in this work, here we adopt an alternative approach that is again more directly related to the explosion model predictions. Determining the necessary luminosity for each simulation requires a light curve from which we can calculate the luminosity at the randomly sampled $t_{\textrm{exp}}$. For each model, we interpolate the light curves from HESMA to the appropriate explosion strength parameter discussed in Sect.~\ref{sect:ejecta}. The luminosity of SNe~Ia is proportional to the $^{56}$Ni mass. For our models, we therefore scale the interpolated light curve to match the luminosity expected for the new, scaled $^{56}$Ni mass calculated in Sect.~\ref{sect:ejecta}. To ensure that our models are broadly compatible with the observed luminosities of SNe~Ia we also allow for changes in the rise time, $t_r$. Appendix~\ref{sect:appdx:rise} provides additional discussion on sampling rise times. All interpolated light curves are then stretched to match $t_{r,x}$ and the luminosity at the required phase, $L_x$, is calculated.

\par

Unlike \textsc{tardis}, the radiative transfer codes used to simulate the models discussed in Sect.~\ref{sect:ejecta} do not assume a sharp inner boundary separating optically thick and thin regions. As such, we cannot rely purely on the existing radiative transfer simulations to calculate the appropriate inner boundary velocity, $v_i$, for our \textsc{tardis} simulations. For all models, we determine the velocity at peak and sample a velocity gradient, from which we determine the appropriate $v_{i,x}$ for the sampled $t_{\textrm{exp},x}$ through linear extrapolation. \cite{magee--19} present a number of \textsc{tardis} simulations based on the \cite{fink-2014} pure deflagration models at maximum light. The inner boundary velocities of these models are highly correlated with $N_k$. We therefore use the relation between $v_{\textrm{max}}$ and $N_k$ to interpolate to our selected $N_{k,x}$ and allow for scatter of $\pm$30~per~cent. For all other models, we use an empirical approach based on observed SNe~Ia. \cite{silverman--12b} present \ion{Si}{ii}~$\lambda$6\,355 velocities at maximum light (ranging from $\sim$9\,000 -- 16\,000~km~s$^{-1}$) and velocity gradients (ranging from $\sim$10 -- 500~km~s$^{-1}$~d$^{-1}$) for a sample of SNe~Ia observed as part of the Berkeley Supernova Ia Program \citep{silverman--ia}. Overall, these velocities show significant scatter. No strong correlation is observed with light curve stretch, or indeed between the velocity and velocity gradient. The inner boundary velocity in \textsc{tardis} simulations is typically correlated with, but generally lower than, the \ion{Si}{ii}~$\lambda$6\,355 velocity. We therefore uniformly sample a velocity at peak between 6\,000 -- 13\,000~km~s$^{-1}$. The velocity gradient for all models (including pure deflagrations) is uniformly sampled between 50 -- 250~km~s$^{-1}$~d$^{-1}$ and the resulting $v_{i,x}$ calculated for each model.

\subsection{Training model construction summary}
\label{sect:training_data_summary}
\begin{figure*}
\centering
\includegraphics[width=\textwidth]{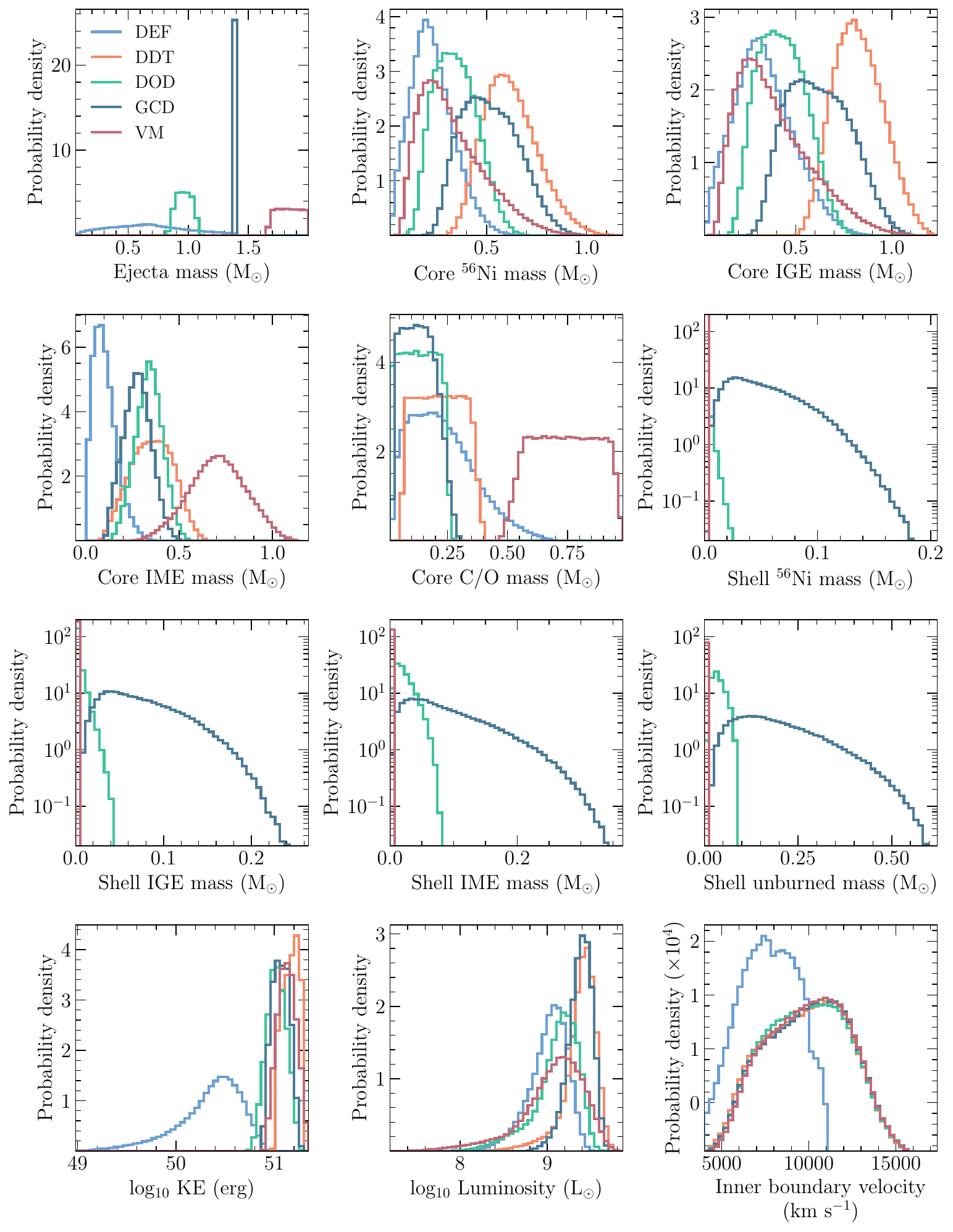}
\caption{Prior distributions for input and derived model parameters. 
}
\label{fig:model_priors}
\centering
\end{figure*}

In summary, each spectrum in our training datasets is generated from a set of 16 input parameters, including 11 parameters defining the structure of the ejecta (Sect.~\ref{sect:ejecta}) and 5 parameters used to determine simulation properties (Sect.~\ref{sect:simulation_parameters}). Input parameters are randomly sampled, generally based on predictions from existing explosion simulations. We note however that these input parameters are generally not independent and typically are highly dependent on the sampled explosion strength parameter. As such, many parameters are highly correlated and some show strong degeneracies. For example, multiple combinations of the velocity at maximum light and velocity gradient could produce the same inner boundary velocity. In this case the velocity at maximum light and velocity gradient are also not used by \textsc{tardis} when simulating spectra, but the inner boundary velocity is (see Sect.~\ref{sect:simulation_parameters}). We therefore caution against over-interpretation of individual model parameters and instead focus on the derived parameters that more directly impact the \textsc{tardis} simulations, including the masses of different elemental groups (to which the abundances are scaled), luminosity, $KE$, and inner boundary velocity. Figure~\ref{fig:model_priors} shows the priors of these input and derived parameters. We also stress that while we refer to the derived masses throughout this work, \textsc{tardis} is not sensitive to material below the photosphere. Values of ejecta mass, $^{56}$Ni mass, etc. are provided simply as comparison points against the explosion simulations upon which the training data is based and to allow reconstruction of the model ejecta. They are only valid within the context of the given explosion scenario and should not be taken as standalone measurements of the \textit{true} ejecta mass, $^{56}$Ni mass, etc.

\par

Given the increased diversity of our models, we generate 300\,000 training spectra for each explosion scenario (c.f. 100\,000 each in \citealt{magee--24}). In total this gives 1\,500\,000 spectra across all five scenarios explored here.

\par

\subsection{Pre-processing}
\label{sect:processing}

We follow a similar pre-processing approach as in \cite{magee--24}, although with an expanded wavelength range. All model spectra are first resampled to 2\,000 log-spaced wavelength bins between 2\,000 -- 10\,000~\AA\, (c.f. 1\,000 log-spaced bins between 3\,000 -- 9\,000~\AA\, in \citealt{magee--24}). We take $\log_{10}$ of the flux in each wavelength bin and transform using the \textsc{standard} scaler in \textsc{sklearn}. Input parameters are also processed by taking $\log_{10}$ and applying the \textsc{standard} scaler. This process is applied to each training dataset independently.

%
%__________________________________________________________________________________________________________________________________________________________________________________________
%__________________________________________________________________________________________________________________________________________________________________________________________
%__________________________________________________________________________________________________________________________________________________________________________________________

\section{Neural networks}
\label{sect:nns}

\subsection{Architecture and training}
\begin{figure*}
\centering
\includegraphics[width=\textwidth]{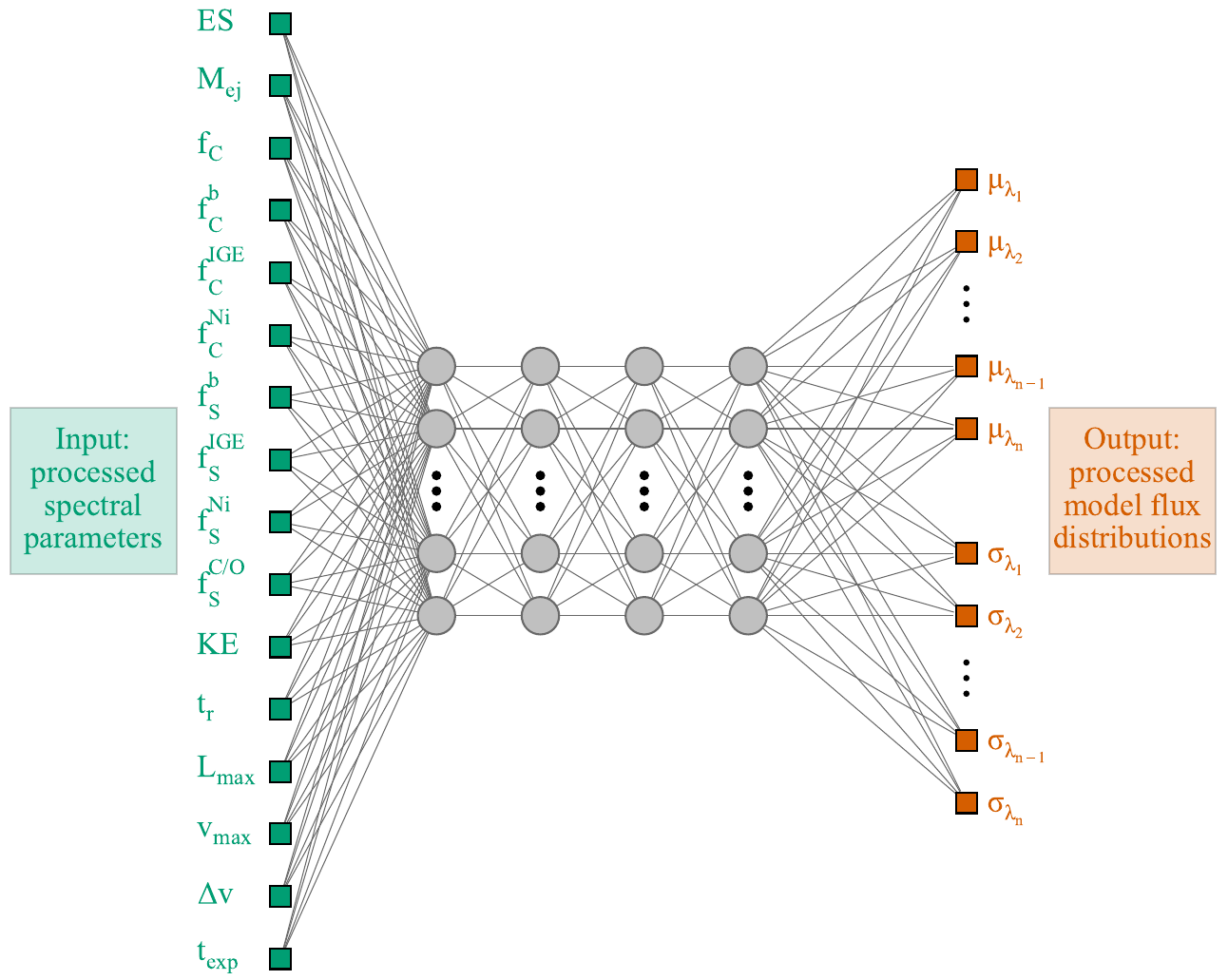}
\caption{Schematic diagram showing an example neural network architecture with four hidden layers. The 16 parameters discussed in Sect.~\ref{sect:training_models} are processed according to Sect.~\ref{sect:processing} and used as inputs for the neural networks. Separate output layers are defined for the mean ($\mu_{\lambda}$) and standard deviation ($\sigma_{\lambda}$) of the (processed) model flux in each wavelength bin, $\lambda$.
}
\label{fig:nn_architecture}
\centering
\end{figure*}

\begin{table}
\centering
\caption{Hyper-parameters and values used during optimisation}\tabularnewline
\label{tab:hyper_param}\tabularnewline
%\resizebox{\columnwidth}{!}{
\begin{tabular}{ll}\hline
\hline
Hyper-parameter     &   Values studied              \tabularnewline
\hline
\hline
Batch size          & 4\,096, \textbf{8\,192}, 16\,384       \tabularnewline
Learning rate       &   $10^{-4}$, $\mathbf{10^{-3}}$          \tabularnewline
Number of layers    &   \textbf{4}, 6, 8                     \tabularnewline
Neurons per layer   &   \textbf{400}, 600, 800               \tabularnewline
Activation function &   \textbf{Softplus}, Sigmoid, LeakyReLU\tabularnewline
Optimizer           &   NAdam, \textbf{AdamW}                \tabularnewline
\hline
\hline
\multicolumn{2}{l}{\textit{Note:} Values given in bold showed the best performance and were} \tabularnewline 
\multicolumn{2}{l}{selected for the final neural networks} \tabularnewline
\end{tabular}
%}
\end{table}

Using the five training datasets discussed in Sect.~\ref{sect:training_models}, we train neural networks specific to each of these explosion scenarios. All neural networks are trained using \textsc{pytorch} \citep{pytorch}. Extensive manual hyper-parameter tuning was performed over the parameters given in Table~\ref{tab:hyper_param}. Figure~\ref{fig:nn_architecture} shows an example of one of the neural network architectures used in this work. 

\par

Each neural network is defined with an input layer of 16 neurons that correspond to the 16 processed parameters discussed in Sect.~\ref{sect:training_models}. The neural networks presented by \cite{magee--24} contain a single output layer corresponding to the processed model flux within each wavelength bin. The flux uncertainty of each predicted spectrum was later estimated based on the typical error of the neural network as a function of two input parameters (time since explosion and inner boundary velocity). Here we attempt to provide a more robust uncertainty estimate that includes all input parameters. We therefore define two output layers and predict the flux distribution within each wavelength bin, $\lambda$, rather than predicting the flux directly \citep{nix--94, sluijterman--23}. The flux distributions are assumed to be Gaussian, with the first and second output layers corresponding to $\mu_{\lambda}$ and $\sigma_{\lambda}$ of the distributions, respectively. A Softplus activation function is also applied to the second output layer to ensure a positive $\sigma_{\lambda}$. Following \cite{sluijterman--23}, during training we implement a warm up period in which only $\mu_{\lambda}$ is learned and $\sigma_{\lambda}$ remains fixed. This ensures the neural networks learn an overall prescription of the training data, rather than ignoring areas with large uncertainties initially and only focusing on those that are well fit \citep{sluijterman--23}. For all neural networks we use a warm up period of 1\,000 epochs. 

\par

To train the neural networks, each dataset of 300\,000 spectra is split into training, validation, and evaluation sets containing 80~per~cent, 10~per~cent, and 10~per~cent of the spectra, respectively. Neither validation nor evaluation sets are seen during training. The former is used to select the best performing model during hyper-parameter optimisation, while the latter is used to evaluate the overall performance of the neural networks in reproducing the training spectra (see Sect.~\ref{sect:nn_performance}). Neural networks are trained for 10\,000 epochs on three Nvidia Quadro RTX 6\,000 GPUs using a Gaussian negative log-likelihood loss function given by
\begin{equation}
\label{eqn:nn_sample_loss}
        \mathcal{L} = \frac{1}{2} \sum \left[ \left( \frac{X_{\lambda} - \mu_{\lambda}}{\sigma_{\lambda}} \right)^2 + \ln (\sigma_{\lambda}^2 ) \right],
\end{equation}
where $X_\lambda$ is the flux of the training spectrum in wavelength bin $\lambda$ and the summation is performed over all wavelength bins and batch samples. During training, we monitor the performance of the neural network on the unseen validation dataset and save the epoch with the lowest validation loss to avoid over-fitting.

\subsection{Performance}
\label{sect:nn_performance}

\begin{figure*}
\centering
\includegraphics[width=\textwidth]{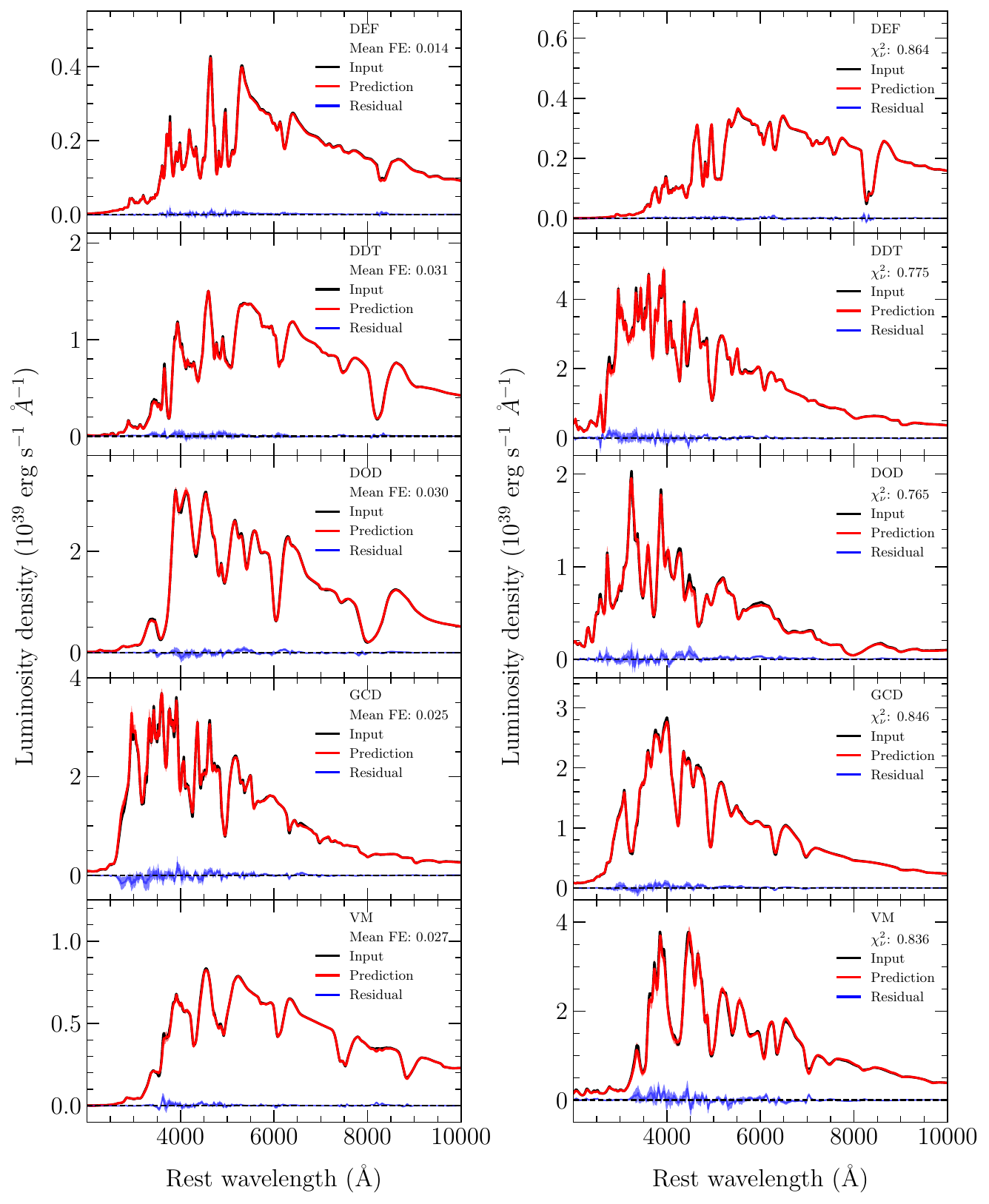}
\caption{Input spectra (black) and neural network predictions (red) for each of the explosion scenarios considered here. The residual between the input and predicted spectra is given in blue. Shaded regions show the 1$\sigma$ uncertainties on the predictions. Panels on the left show spectra with the median mean FE for each explosion scenario, while those spectra with the median $\chi^2_{\nu}$ are shown on the right. 
}
\label{fig:evaluation_spectra}
\centering
\end{figure*}

\begin{table*}
\centering
\caption{Hyperparameters and performance for neural networks}\tabularnewline
\label{tab:best_nns}\tabularnewline
\begin{tabular}{llllllll}\hline
\hline
\tabularnewline[-0.25cm]
Training    & Minimum         & Median    & Median    &  Median           & Median        & Median        & Median         \tabularnewline
dataset     & validation loss    & mean FE   & max FE    &  $\chi^2_{\nu}$   & mean FE (opt) & max FE (opt)  & $\chi^2_{\nu}$ (opt)\tabularnewline

\hline
DEF         & $-$3.509 & 0.014 & 0.078 & 0.864 & 0.009 & 0.048 & 0.852 \tabularnewline
DDT         & $-$2.600 & 0.031 & 0.206 & 0.775 & 0.015 & 0.072 & 0.735 \tabularnewline
DOD         & $-$2.765 & 0.030 & 0.176 & 0.765 & 0.021 & 0.146 & 0.773 \tabularnewline
GCD         & $-$2.621 & 0.025 & 0.132 & 0.846 & 0.017 & 0.080 & 0.845 \tabularnewline
VM          & $-$3.244 & 0.027 & 0.201 & 0.836 & 0.017 & 0.099 & 0.801 \tabularnewline

\hline
\hline
\multicolumn{8}{l}{\textit{Note:} `Opt' values are calculated over the optical wavelength range of $3\,000 \leq \lambda \leq 9000$~\AA.} \tabularnewline 
\end{tabular}
\end{table*}

Table~\ref{tab:hyper_param} gives the selected hyper-parameters for our neural networks. Based on our optimisation we find that these parameters produce the best performance for all validation datasets during training. To determine the performance, we calculate the loss (Eqn.~\ref{eqn:nn_sample_loss}), fractional errors (FEs), and reduced $\chi^2$ ($\chi^2_{\nu}$) for each of the predicted, unprocessed (i.e. flux) spectra.

\par

Figure~\ref{fig:evaluation_spectra} shows the evaluation spectra and corresponding neural network predictions with the median mean FE and median $\chi^2_{\nu}$ across the 30\,000 spectra in each of the evaluation datasets. The neural networks are generally able to reproduce the complex features of the spectra with the appropriate strengths and velocities. For example, Fig.~\ref{fig:evaluation_spectra} shows that our DEF neural network is able to reproduce the narrow features predicted by the pure deflagration explosion scenario, while our DOD neural network matches the broad \ion{Ca}{ii} absorption predicted by double detonations. Typically, the features predicted by the neural networks are smoother than those in the evaluation spectra, particularly around the absorption minima, and some weak features may become blended. The neural network predictions also provide good qualitative agreement with the overall flux scale and spectral shape. 

\par

\begin{figure}
\centering
\includegraphics[width=0.83\columnwidth]{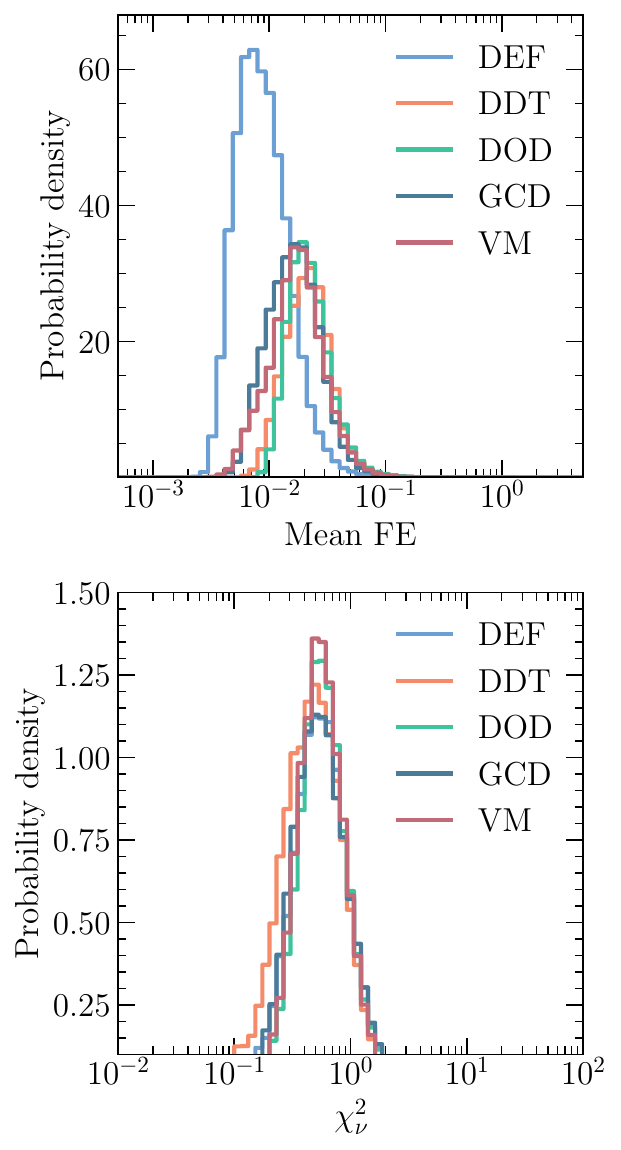}
\caption{Distributions of mean FEs and $\chi^2_{\nu}$ across the 30\,000 spectra for each evaluation dataset.
}
\label{fig:evaluation}
\centering
\end{figure}

In Fig.~\ref{fig:evaluation} we show the distributions of mean FEs and reduced $\chi^2$ for each explosion scenario. The median values for each scenario are also given in Table~\ref{tab:best_nns}. Despite the differences in underlying ejecta structures, input parameters, and spectra, we find that all of our neural networks generally achieve similar levels of performance, with median mean FEs ranging from $\sim$0.014 -- 0.031. We note that the distribution of mean FEs for our DEF neural network is skewed towards lower values (i.e. better approximations of $\mu_{\lambda}$) than those of other models. We speculate that this may arise from the simpler (near-uniform) ejecta structure or lower velocities (and hence less blending) of these models, resulting in the neural network more reliably able to learn the mapping from input parameters to complex spectra. While the predicted $\mu_{\lambda}$ across all models may be discrepant by up to a few per cent, the distributions of $\chi^2_{\nu}$ show that these differences typically fall within the $\sigma_{\lambda}$ predicted by the neural network. Indeed, although the DEF neural network provides a better approximation of $\mu_{\lambda}$, the distributions of $\chi^2_{\nu}$ are similar for all models, peaking around $\chi^2_{\nu} \sim 0.8$. The uncertainties predicted by our neural networks are typically of order a few per cent, which is comparable to the level of Monte Carlo noise within the \textsc{tardis} simulations \citep{kerzendorf--21}. In addition, following \cite{levi--22}, we calibrate the predicted uncertainties from our neural networks. For each wavelength, we bin our 30\,000 evaluation samples based on the predicted variance and calculate the scaling factors $\alpha_{\lambda}$ ($\geq1$) required to ensure that the root mean predicted variance matches the empirical mean squared error in that bin. All predicted neural network uncertainties are then scaled by $\alpha_{\lambda}$, which typically ranges from $\sim$1 -- 1.15 across all wavelengths. This process is performed independently for each training dataset.

\par

For all neural networks, we find that the most prominent discrepancies and largest predicted uncertainties occur at ultraviolet wavelengths. We therefore also calculate the FE and $\chi^2_{\nu}$ for a more limited optical wavelength range (opt) corresponding to that used by the neural networks in \cite{magee--24}, $3\,000 \leq \lambda \leq 9\,000$~\AA. As demonstrated in Table~\ref{tab:best_nns}, the neural networks show considerably improved performances over optical wavelengths, with mean FEs decreasing by $\sim$30 -- 50~per~cent. Despite the increased complexity in the training datasets and neural network architectures, we therefore achieve similar performance over the same wavelength range as the neural networks presented by \cite{magee--24}.

%
%__________________________________________________________________________________________________________________________________________________________________________________________
%__________________________________________________________________________________________________________________________________________________________________________________________
%__________________________________________________________________________________________________________________________________________________________________________________________

\section{Fitting spectral sequences}
\label{sect:fitting}
\subsection{Method}
\label{sect:fit_method}

Our spectral fitting method follows a similar approach to that described by \cite{magee--24}. Briefly, \textsc{riddler} uses the neural networks discussed in Sect.~\ref{sect:nns} in conjunction with \textsc{ultranest} \citep{ultranest} to perform nested sampling and determine the parameters $\theta$ with the highest likelihood for an observed spectral sequence $o$, $\mathcal{L}(\theta|o)$. Given a set of prior distributions $\pi(\theta)$, \textsc{ultranest} draws samples that are used by the neural networks to emulate the radiative transfer simulations and generate synthetic spectra. These spectra are then evaluated against the observations using a Gaussian log-likelihood function, which is summed over all $N$ spectra in the sequence and all wavelengths $\lambda$:
\begin{equation}
\label{eqn:ultanest_likelihood}
    \mathcal{L} = -\frac{1}{2} \sum_N \sum_{\lambda} w \left[ \left( \frac{\mu(\theta) - o}{s(\theta)} \right)^2 + \log(2 \pi s(\theta) ) \right],
\end{equation}
where
\begin{equation}
s(\theta)^2 = \sigma_o^2 + f^2\mu(\theta)^2 + \alpha^2\sigma(\theta)^2.
\end{equation}
The predicted flux and uncertainty from the neural networks are given by $\mu(\theta)$ and $\sigma(\theta)$ respectively, while the observed flux and uncertainty are given by $o$ and $\sigma_o^2$, respectively. As discussed in Sect.~\ref{sect:nn_performance}, the uncertainties predicted by the neural networks are increased by a wavelength-dependent scaling factor $\alpha$. We also include an additional nuisance parameter $f$ to account for systematic differences between explosion models and observations \citep{obrien--24, magee--24}. The weight parameter $w$ allows for independent weighting of each wavelength bin at each epoch and is supplied along with the flux and flux error for each observed spectrum. Similar objective functions have also previously been used for other quantitative spectral comparisons, including those using emulated spectra as in this work \citep{synapps, obrien--21, obrien--24, ogawa--23, harvey--26}.

\par

Using Eqn.~\ref{eqn:ultanest_likelihood}, \textsc{ultranest} attempts to find the best set of parameters that reproduces the observed spectra. Each of the 16 input parameters used to generate the neural network predictions are sampled from the same prior distributions discussed in Sect.~\ref{sect:training_models} and shown in Fig.~\ref{fig:model_priors}. In addition to these parameters, we include the host extinction ($E(B-V)$ and $R_V$) and distance $d$ as optional parameters (fixed values can also be used). If prior constraints are provided, $E(B-V)$ and $R_V$ are sampled from Gaussian distributions. Otherwise, $E(B-V)$ is drawn from an exponential distribution with $\lambda = 0.11$ \citep{stanishev--18} and $R_V = 3.1$. For the distance, we also use a Gaussian prior defined by the mean and standard deviation of the distance modulus. Including the nuisance parameter $f$, this gives a total of 20 parameters to be fit \textit{per spectrum}. The majority of these parameters however are fixed for all spectra in the sequence, with the only exceptions being the velocity at maximum light and velocity gradient. Both parameters are allowed to vary across each spectrum, but we add a constraint that the derived inner boundary velocity decreases over time \citep{magee--24}. The result is a set of up to $18 + 2N$ parameters constrained by fitting a given SN. Again, we stress that degeneracies and correlations exist between many parameters, and focus here on the derived parameters that directly impact \textsc{tardis} simulations (see. Sect.~\ref{sect:training_models}).

\par

The likelihood function given by Eqn.~\ref{eqn:ultanest_likelihood} finds the best overall fit to the data, while the priors outlined in Sect.~\ref{sect:training_models} are motivated by explosion simulations. Together, this approach therefore represents a more unbiased and repeatable method than traditional visual inspection for empirical modelling, and ensures models are physically realistic.

\begin{figure*}
\centering
\includegraphics[width=\textwidth]{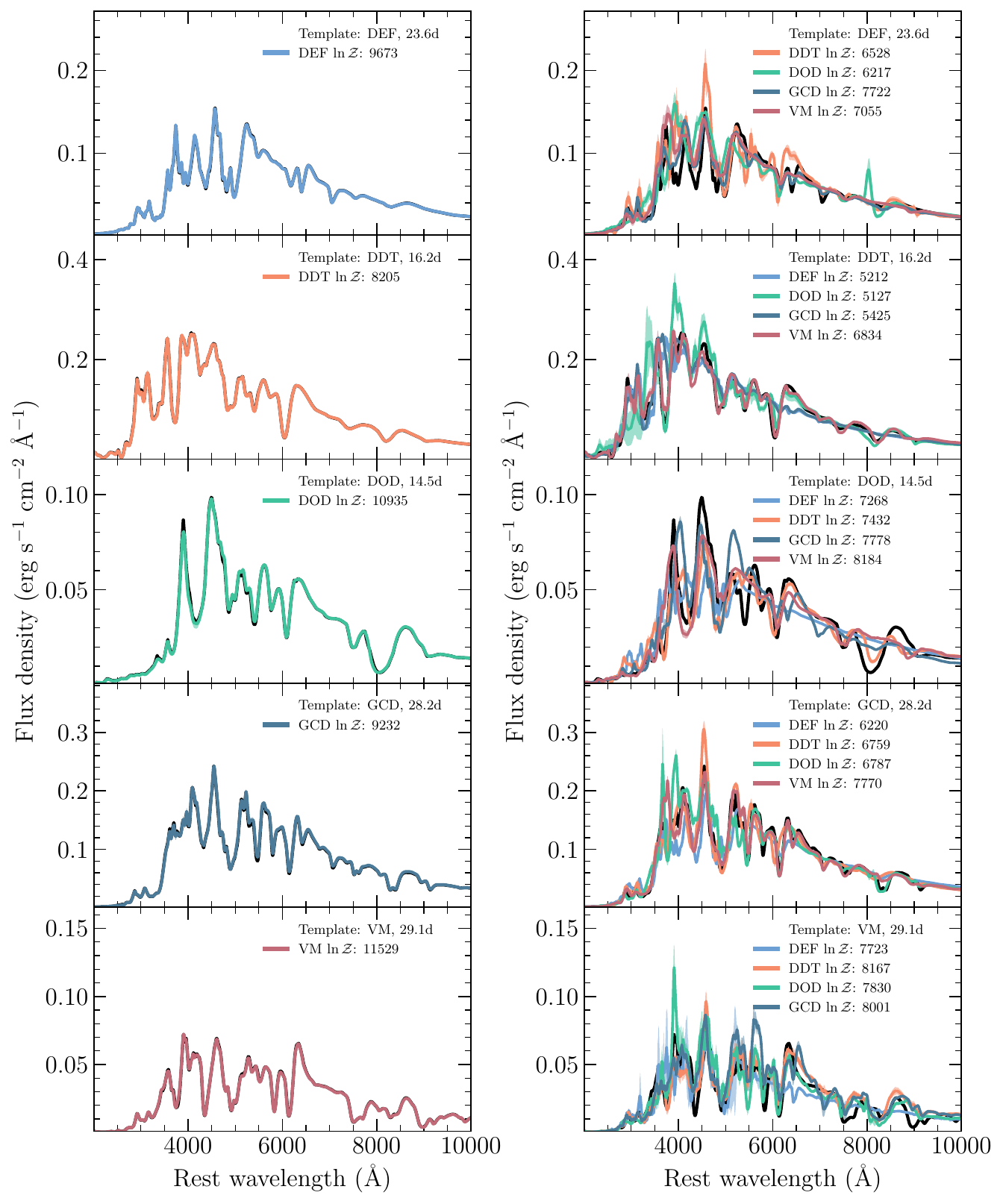}
\caption{ Comparison between template spectra (black) and neural network emulations of the best-fitting parameters determined by \textsc{ultranest}. Fits assuming the same explosion scenario as the template are shown in the left panels and those assuming all other explosion scenarios are shown in the right panels. Each explosion scenario is given by a different colour. Solid lines show the mean flux, while shaded regions show data and emulation uncertainties. The phase of each spectrum relative to explosion and the evidence calculated by \textsc{ultranest} are given in each panel. 
}
\label{fig:ultranest_templates_spectra}
\centering
\end{figure*}

\begin{figure*}
\centering
\includegraphics[width=\textwidth]{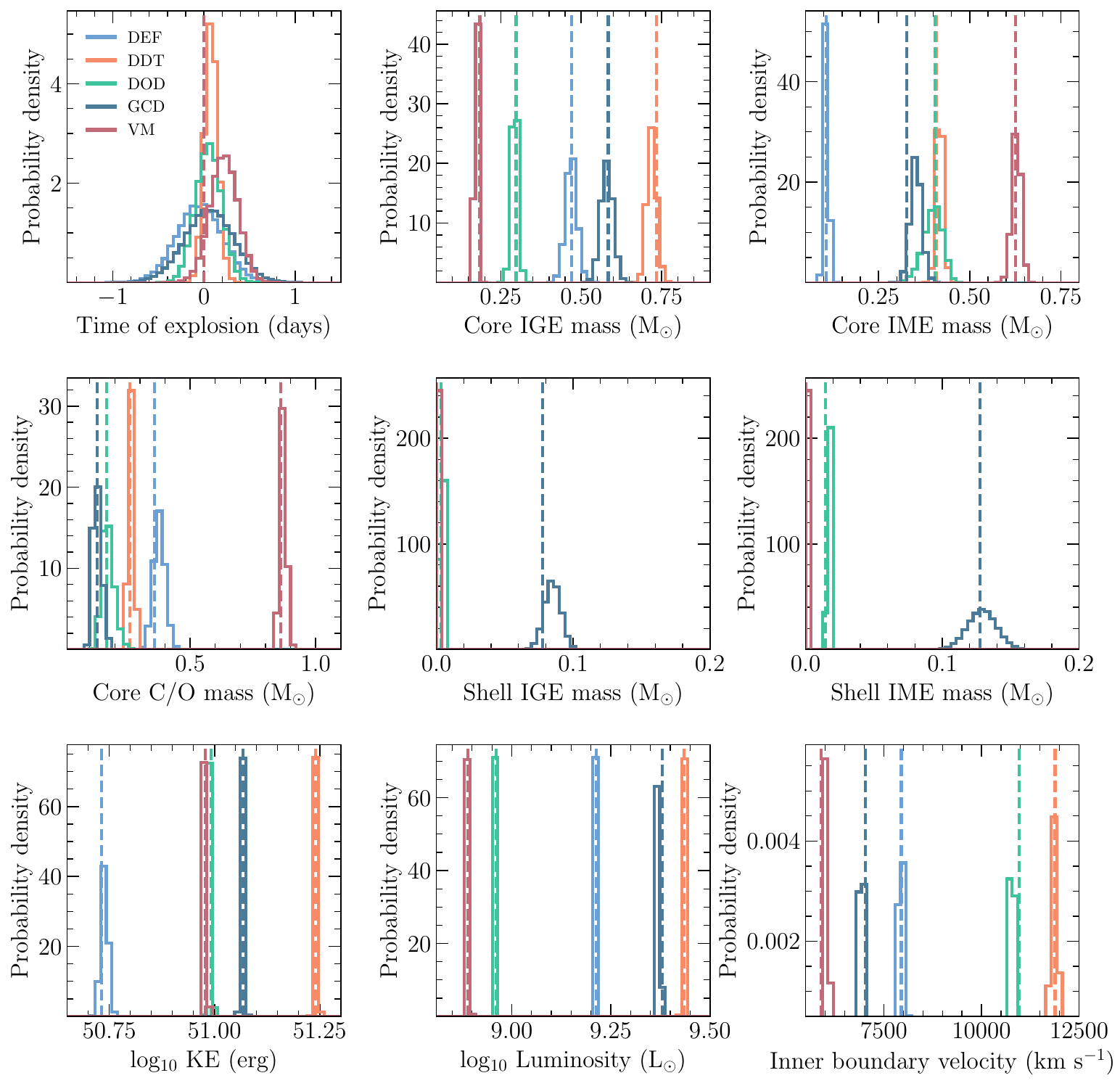}
\caption{ Posterior distributions of model properties for fits shown in the left panels of Fig.~\ref{fig:ultranest_templates_spectra} (i.e. those for which the fitted and template explosion scenarios match). True input values for template spectra are shown as vertical dashed lines.
}
\label{fig:ultranest_templates_params}
\centering
\end{figure*}

\par

Fits are performed independently for each of the neural networks trained on a specific explosion scenario. To determine the relative likelihood between two explosion scenarios, we calculate the Bayes factor (BF) as
\begin{equation}
\label{eqn:bf}
    \ln \mathrm{BF} = \ln \mathcal{Z}_1 - \ln \mathcal{Z}_2,
\end{equation}
where $\mathcal{Z}$ gives the evidence of each explosion scenario and is calculated by \textsc{ultranest} through marginalising the likelihood function (Eqn.~\ref{eqn:ultanest_likelihood}) over the prior distribution $\pi(\theta)$
\begin{equation}
    \mathcal{Z} = \int \mathcal{L}(\theta|o) \pi(\theta) d\theta.
\end{equation}
From Eqn.~\ref{eqn:bf}, $\ln$ BF between 1 -- 3 indicates positive evidence in favour of model 1 over model 2, while $\textgreater$5 indicates very strong evidence in favour of model 1 over model 2 \citep{kass--95}.

%
%__________________________________________________________________________________________________________________________________________________________________________________________
%__________________________________________________________________________________________________________________________________________________________________________________________
%__________________________________________________________________________________________________________________________________________________________________________________________

\subsection{Performance}
\label{sect:fit_performance}

In Fig.~\ref{fig:ultranest_templates_spectra} we show \textsc{riddler} fits to template spectra randomly selected from the evaluation datasets of each explosion scenario. For each fit, we allow the explosion epoch to vary by up to $\pm$5\,d and assume a 2~per~cent flux error. Based on our fits, we find that \textsc{riddler} is able to produce good agreement with the templates, matching the overall flux level, spectral shape, and many prominent spectral features across the full wavelength range. As discussed in Sect.~\ref{sect:nn_performance} however, some weak features are not as well reproduced. For example, Fig.~\ref{fig:ultranest_templates_spectra} shows that our best-fitting DOD model fails to produce the weak \ion{C}{ii}~$\lambda$6\,580 feature present in the template spectrum and the faint structure in the \ion{Ca}{ii}~NIR feature. Nevertheless, we find mean FEs between our neural network reconstructions and templates range from $\sim$1 -- 3~per~cent.

\par

Figure~\ref{fig:ultranest_templates_params} shows the posterior distributions of fitted and derived parameters, along with the true values of the templates. Despite the significant increase in complexity of our models, we find that \textsc{riddler} is generally able to recover the true input parameters of the templates. For all templates, we find no significant offset is required to the time of explosion and masses of different elemental groups within the core are well-recovered. Some parameters, particularly those related to the shell, can have little impact on the spectra and therefore the posteriors may be systematically offset from the true value. For example, our GCD template has 0.15~$M_{\odot}$ of unburned material in the shell, but \textsc{riddler} finds a best-fitting value of $0.12\pm0.01~M_{\odot}$. While the true value therefore does fall within the 3$\sigma$ error range, it is unsurprising that \textsc{riddler} struggles to recover this. Carbon and oxygen produce relatively few strong optical features in thermonuclear SNe spectra and the shell material is likely to be sufficiently low density at this epoch (28.2 days after explosion) that carbon and oxygen in the shell will not significantly affect the spectra. Throughout this paper we therefore quote the 3$\sigma$ error range of our fits. As previously discussed (Sect.~\ref{sect:training_models}) there is a strong degeneracy between the velocity at maximum light and velocity gradient, with different combinations producing the same inner boundary velocity that is ultimately used by \textsc{tardis}. Figure~\ref{fig:ultranest_templates_params} shows however that our \textsc{riddler} fits find good agreement with the inner boundary velocities.

\par

In addition to finding the best-fitting parameters for a given explosion scenario, the purpose of \textsc{riddler} is to find the statistically best-matching explosion scenario. The template spectra shown in Fig.~\ref{fig:ultranest_templates_spectra} are therefore fit with each of the remaining four explosion scenarios. These fits are also shown in Fig.~\ref{fig:ultranest_templates_spectra}. In all cases, based on the evidence and $\ln$ BF, we find that the true explosion scenario is strongly preferred. Figure~\ref{fig:ultranest_templates_spectra} also highlights interesting overlap between different explosion scenarios and where degeneracies could occur when fitting observed SNe~Ia. Unsurprisingly for our DEF template spectrum, most explosion scenarios are not able to reproduce the low velocity features observed and are systematically offset. The DDT template shows remarkable agreement with the best-fitting VM model spectrum. The most prominent differences are that the VM model shows slightly lower velocities, stronger \ion{C}{ii}~$\lambda$6\,580 and \ion{O}{i}~$\lambda$7\,773 features, and slightly weaker \ion{Ca}{ii}~H~\&~K. Interestingly, we find that the VM model requires an earlier explosion epoch by $\sim$4 -- 4.5\,d, therefore SNe~Ia with well-constrained explosion epochs may not be affected. None of the models (including the GCD model, despite also containing a shell in the outer ejecta) are able to match the spectral shape or features, such as the broad \ion{Ca}{ii}~NIR feature, of our DOD template. Similar to the case of the DDT template, the VM model shows some agreement with the GCD template, but this is limited to wavelengths $\gtrsim$4\,000~\AA\, and some features are more noticeably offset. Despite the VM model producing reasonable agreement with both the DDT and GCD templates, Fig.~\ref{fig:ultranest_templates_spectra} shows that the reverse is not true and neither scenario can reproduce the VM template. 

\begin{figure}
\centering
\includegraphics[width=\columnwidth]{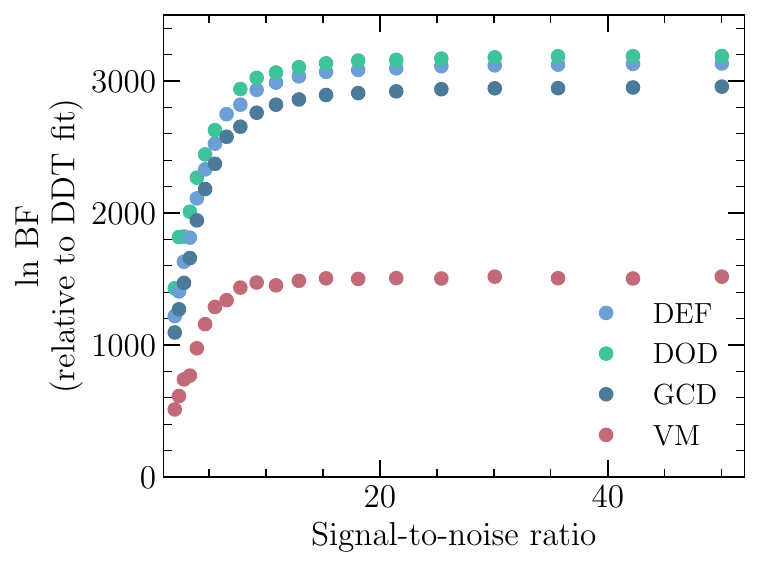}
\caption{Bayes factor relative to the DDT model fit as a function of signal-to-noise ratio. A value of $\log \textrm{BF} \textgreater 5$ indicates that the DDT model is preferred. 
}
\label{fig:snr}
\centering
\end{figure}

\begin{figure}
\centering
\includegraphics[width=\columnwidth]{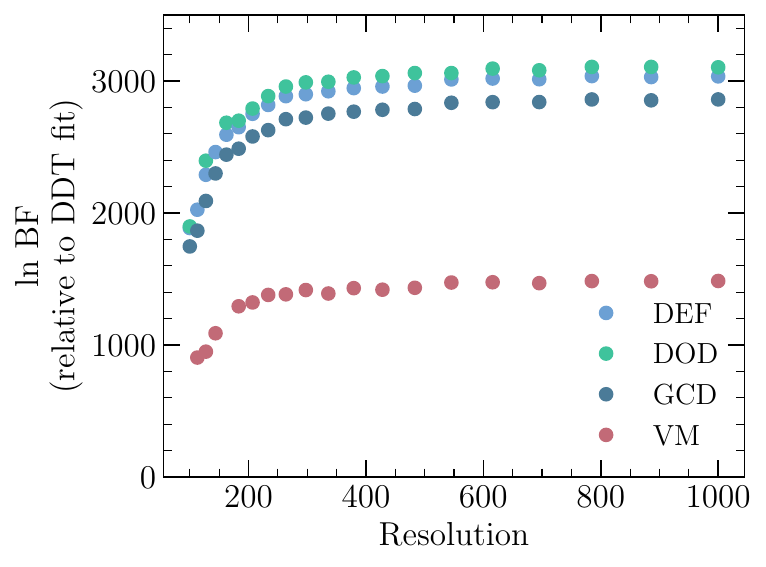}
\caption{Bayes factor relative to the DDT model fit as a function of spectral resolution at 6\,000~\AA. A value of $\log \textrm{BF} \textgreater 5$ indicates that the DDT model is preferred. 
}
\label{fig:resolution}
\centering
\end{figure}
\par

Figure~\ref{fig:ultranest_templates_spectra} demonstrates that the VM model produces reasonable visual agreement with the DDT template, although the DDT model is statistically preferred overall. This raises questions about how the quality of the spectrum being fit affects our results and whether lower quality spectra show differences in best-fitting parameters or indeed preferred model types. To test this, we explore different signal-to-noise ratios and resolutions for the DDT template spectrum shown in Fig.~\ref{fig:ultranest_templates_spectra}. Figure~\ref{fig:snr} shows the resulting $\ln$ BF relative to the DDT model with different levels of Gaussian noise added to the template spectrum. We find that $\ln \textrm{BF} \gg 5$ in all cases, which shows that the DDT model is preferred and \textsc{riddler} is able to recover the correct model type of the template regardless of the signal-to-noise ratio. Figure~\ref{fig:resolution} shows similar results for spectra that have been binned to varying resolutions (measured at 6\,000~\AA, corresponding to $\Delta\lambda = 6$ -- 60~\AA) and assuming a signal-to-noise ratio of 10. While these fits show that the correct model type of the template is recovered in all cases, this is likely not directly applicable to fitting real SNe~Ia and instead may reflect the fact that fitting a DDT template with a DDT model should be able to produce an approximately exact match. This will not be the case with real SNe~Ia -- we do not expect any of our models to produce an exact match. Figures~\ref{fig:snr}~\&~\ref{fig:resolution} do show however that $\ln \textrm{BF}$ is approximately constant for signal-to-noise ratios above $\sim$10 and resolutions above $\sim$300, while below these values $\ln \textrm{BF}$ shows a steady decrease. For real SNe~Ia, we therefore expect that signal-to-noise ratios $\gtrsim$10 and resolutions $\gtrsim$300 ($\Delta\lambda \lesssim 20$~\AA) will have little impact on which model type is preferred overall and below this limit it may not be possible to distinguish between different scenarios. Indeed, we find that this also applies to the best-fitting parameters. Fits below these values show larger uncertainties in best-fitting parameters and in some cases are offset from the true value. We therefore argue that \textsc{riddler} should only be applied to spectra with signal-to-noise ratios $\gtrsim$10 and resolutions $\gtrsim$300.

\par

In summary, we find that \textsc{riddler} is able to recover the correct input parameters and explosion scenario when fitting unseen template spectra. One of the key strengths of \textsc{riddler} is the training data based on predictions from multi-dimensional explosion simulations. Figure~\ref{fig:ultranest_templates_spectra} highlights that due to this physical motivation it is not possible to fit any given observed spectrum with an arbitrary model -- fits are constrained by the explosion physics of each scenario and therefore \textsc{riddler} can provide meaningful insight into the explosion physics of observed SNe~Ia.

%
%__________________________________________________________________________________________________________________________________________________________________________________________
%__________________________________________________________________________________________________________________________________________________________________________________________
%__________________________________________________________________________________________________________________________________________________________________________________________

\section{Application to observations}
\label{sect:application}

\begin{figure}
\centering
\includegraphics[width=0.91\columnwidth]{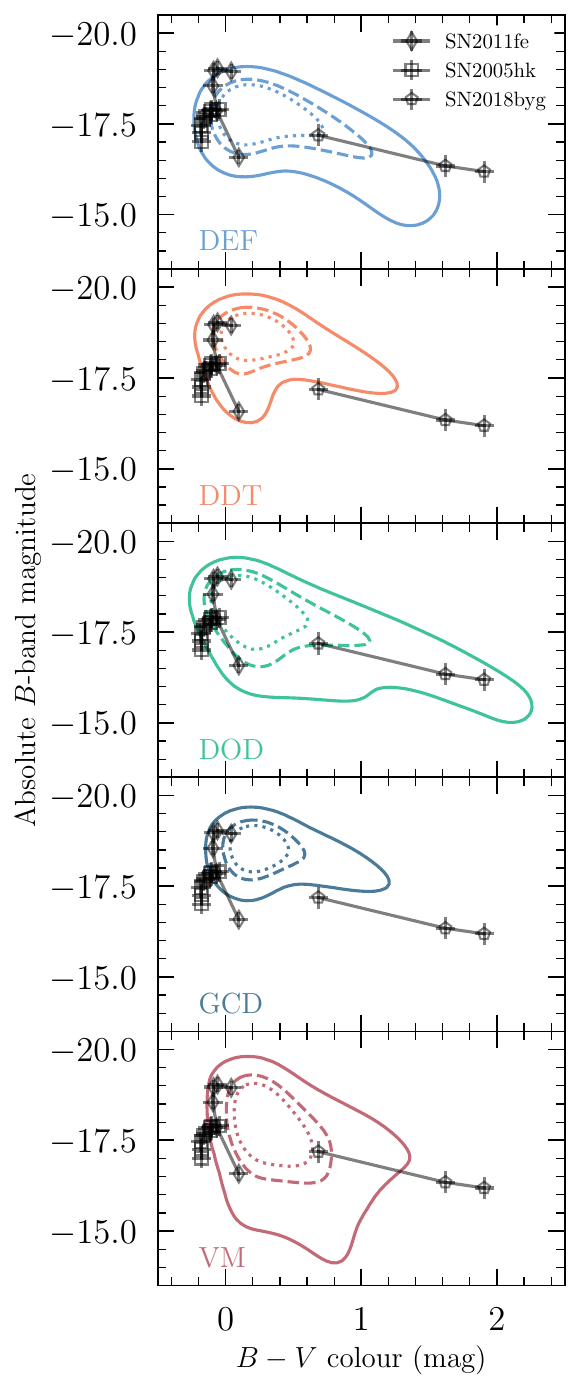}
\caption{Distributions of absolute $B$-band magnitudes and $B-V$ colours for spectra in our training datasets. Contours contain 50, 68, and 95~per~cent of the training samples. Colour and magnitudes of the spectra fit in Sect.~\ref{sect:application} are also shown. Colours have been corrected for Milky Way and host galaxy extinction. Uncertainties include distance modulus and reddening uncertainties.
}
\label{fig:colour_mag}
\centering
\end{figure}

In the following section, we use \textsc{riddler} to fit observations of SNe~2011fe, 2005hk, and 2018byg. These SNe were selected due to the availability of multiple spectra and existing comparisons with explosion models (see Sect.~\ref{sect:previous_fits}). Fits were performed with each explosion scenario independently. Following \cite{magee--24}, due to the photospheric approximation used by \textsc{tardis} wavelengths $\textgreater$6\,500~\AA\, are excluded from the fit, but shown for reference. All spectra were obtained from WISeREP \citep{wiserep}, calibrated to contemporaneous photometry, and corrected for Milky Way extinction. We assume a flat Universe with $H_0 = 70$~km~s$^{-1}$~Mpc$^{-1}$ and $\Omega_M$ = 0.3. Figure~\ref{fig:colour_mag} shows the distributions of absolute $B$-band magnitudes and $B-V$ colours for the training samples and SNe that are fit. We note that Fig.~\ref{fig:colour_mag} contains all spectra in our training datasets and therefore although the magnitude and colour may align with observations, the phase may be incorrect.

\par

We again stress that \textsc{tardis} simulations are not sensitive to material below the photosphere and therefore \textsc{riddler} cannot constrain the entire ejecta. Although we quote the values of ejecta masses, $^{56}$Ni masses, etc., we do not claim to have measured the true values for any of the SNe fit here. These are simply the values of our model for which the material above the photosphere best reproduces the observations and only valid within the context of that specific explosion scenario. They are given here for completeness and to enable comparisons against the explosion simulations upon which our models are based.

\subsection{SN~2011fe}
\label{sect:11fe}
We fit the HST spectra of SN~2011fe presented by \cite{mazzali--14}. Based on the low extinction inferred ($E(B-V) = 0.014$; \citealt{patat--13}), we do not correct for host-galaxy reddening. Following from \cite{shappee--11}, we set a Gaussian prior on the distance modulus of $\mathcal{N}(29.04, 0.19)$. We set a uniform prior on the explosion epoch of MJD = 55\,790.0 -- 55\,797.2

\begin{figure*}
\centering
\includegraphics[width=\textwidth]{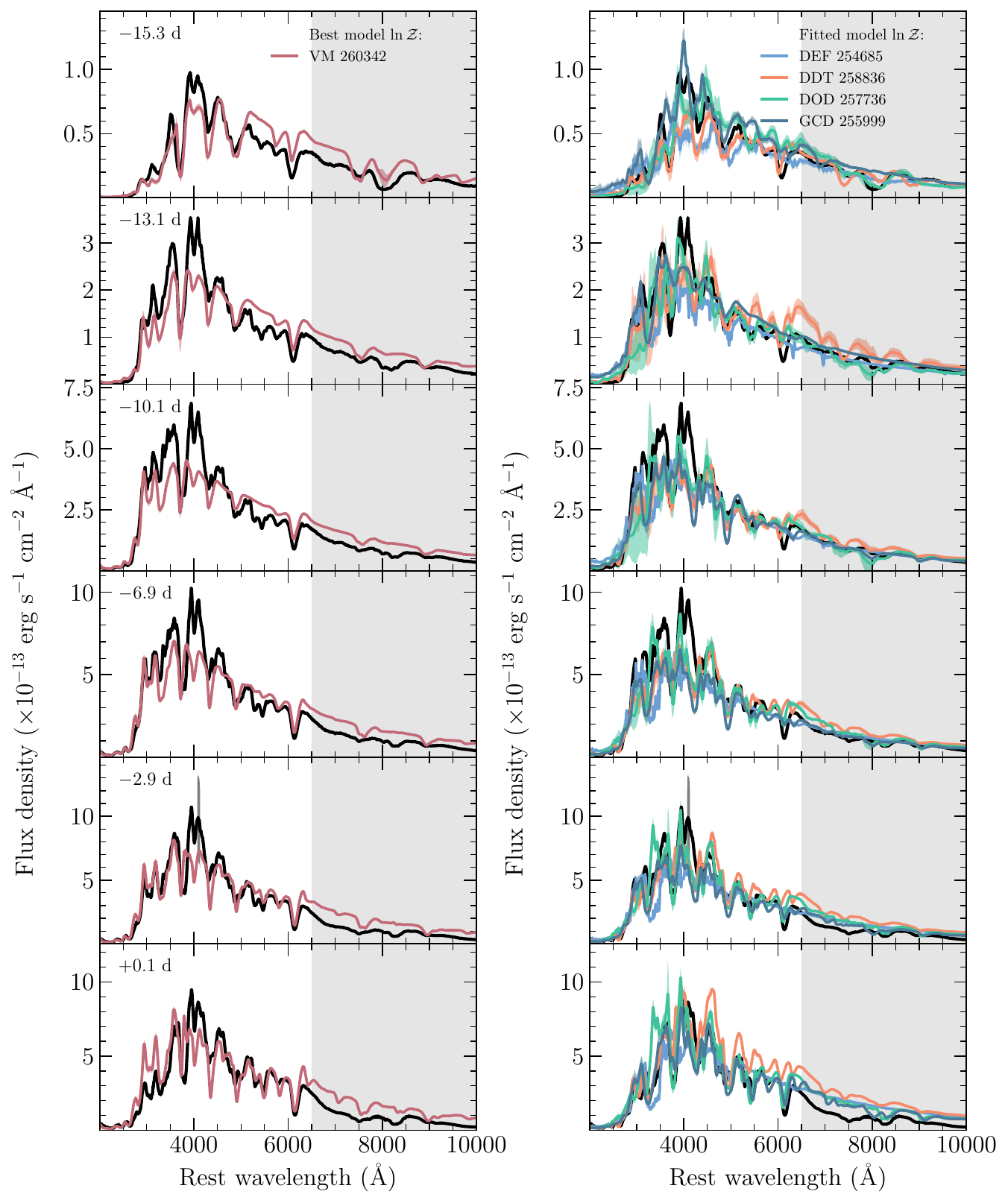}
\caption{As in Fig.~\ref{fig:ultranest_templates_spectra} for SN~2011fe. Grey shaded regions are not included in the fit. Phases are given relative to maximum light.
}
\label{fig:11fe_spectra}
\centering
\end{figure*}

\begin{figure*}
\centering
\includegraphics[width=\textwidth]{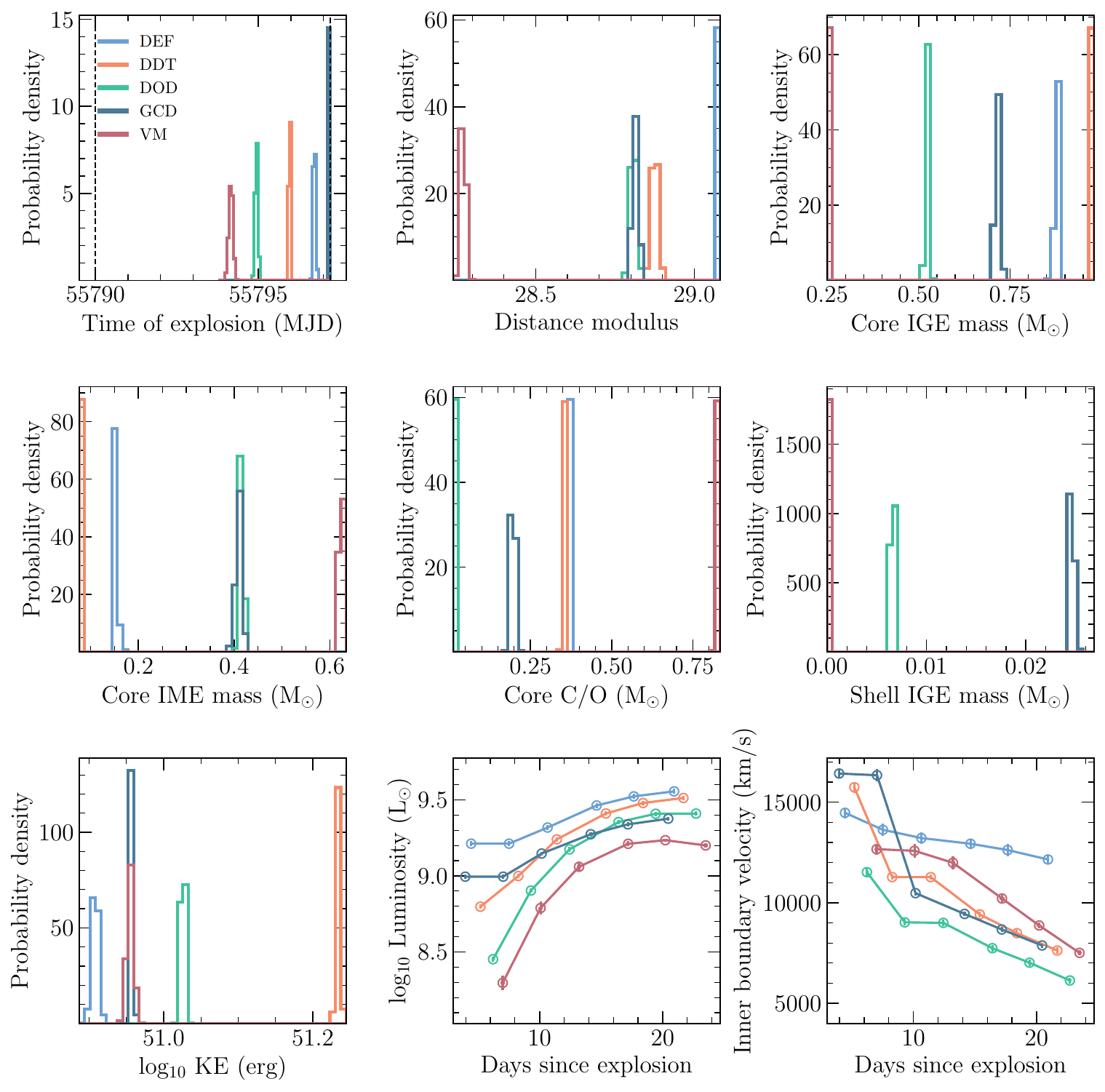}
\caption{As in Fig.~\ref{fig:ultranest_templates_params} for fits to SN~2011fe shown in Fig.~\ref{fig:11fe_spectra}. Black, vertical dashed lines show the assumed boundaries for the time of explosion. 
}
\label{fig:11fe_posteriors}
\centering
\end{figure*}

\par

Based on Fig.~\ref{fig:colour_mag}, all model scenarios show at least some overlap with SN~2011fe in terms of magnitude and colour, but again we note this may not occur at the correct phase. Figure~\ref{fig:11fe_spectra} shows the results of our fits to SN~2011fe. Overall we find that the VM scenario is heavily favoured (based on $\ln$ BF) compared to all other explosion scenarios. At $-$15.3\,d the VM model generally matches the velocities and relative strengths of numerous features, but the overall spectral shape is clearly discrepant with wavelengths $\gtrsim$5\,200~\AA\, showing a systematically higher flux than observed. The best-fitting DEF model fails to reproduce almost all spectral features and indeed has the lowest evidence. Our DOD and GCD models perform somewhat better, reaching broadly similar levels of agreement and producing a few features comparable to the observations, but overall are not good matches to the spectra. The DDT model provides reasonable agreement, again producing numerous features with comparable velocities and relative strengths. Indeed, the DDT model also provides better agreement with the overall spectral shape $\gtrsim$5\,200~\AA, but performs somewhat worse in the near-UV. 

\par

Between $-15.3$ -- $-10.1$\,d, the spectra show only minor changes. The DDT model shows quite significant evolution during this time. Most noticeably, the \ion{Si}{ii} velocities are clearly offset from the observations. At $-$6.9\,d, the \ion{S}{ii}~W feature has begun to develop in the VM model, but is much weaker than observed, while the \ion{Si}{ii}~$\lambda$5\,972 and $\lambda$6\,355 velocities and strengths are now in closer agreement with SN~2011fe. In addition, the VM model also matches the complex blend of \ion{Si}{iii}, \ion{Fe}{iii}, and \ion{Co}{iii} features observed in the near-UV, but the blend around $\sim$3\,300~\AA\, is somewhat too strong. The DEF model again fails to reproduce any spectral features and instead predicts a mostly featureless spectrum. The DDT, DOD, and GCD models all predict the \ion{S}{ii}~W feature, but none are able to match the observed velocity with the DDT being too high and the DOD and GCD too low. Indeed, many of the features produced by the DDT model show velocities that are too high. At $-2.9$\,d, the VM model has developed a weak feature consistent with \ion{C}{ii}~$\lambda$6\,580. This feature is also predicted by the DDT and GCD models, but in all cases is stronger than observed in SN~2011fe. The DOD model now shows generally stronger similarities to SN~2011fe than at previous epochs or indeed the DDT model. 

\par

For our final spectrum at maximum light, the VM model again shows good agreement with many features. As with previous spectra however, the VM model shows only a single absorption feature around $\sim$4\,800~\AA, rather than the complicated blend observed, which has previously been partially attributed to high velocity \ion{Fe}{ii} \citep{branch--06}. Although generally reproducing the spectral features, by this epoch our VM model over-predicts the flux in the near-UV. The DOD and GCD models also generally reproduce SN~2011fe around maximum light. 

\par

Figure~\ref{fig:11fe_posteriors} shows the posterior distributions for our best-fitting input and derived parameters. We find significant variations in parameters across the different explosion scenarios considered here. Depending on the model, the explosion epoch can vary by up to $\sim$3~days. For the VM model we find the earliest explosion epoch of MJD = 55794.15$\pm$0.19, while the GCD model favours a later explosion epoch that falls on our upper limit of MJD = 55\,797.2. The VM model also shows the lowest distance modulus of 28.27$\pm$0.02. Interestingly, we find that the DEF model produces the highest distance modulus and indeed the largest predicted $^{56}$Ni mass. Although it may seem counter-intuitive that our DEF model is intrinsically brightest, this likely results from the need for a stronger explosion that can produce a slower light cure evolution more comparable to that observed in SN~2011fe since most DEF models will evolve much faster. Our DEF model shows $N_k \sim 20$, which is consistent with the need for a brighter and slower-evolving light curve. For the VM model, our best-fitting parameters indicate the ejecta of SN~2011fe is most similar to a VM ejecta containing 0.253$\pm$0.002~$M_{\odot}$ of IGE (of which 0.240$\pm$0.002 are $^{56}$Ni), 0.624$\pm$0.006~$M_{\odot}$ are IME, and the remaining 0.825$\pm$0.009~$M_{\odot}$ are unburned C/O. Both the DOD and GCD models predict small masses of IGE in the outer ejecta shell ($\sim$0.007 and 0.025~$M_{\odot}$, respectively) that are consistent with the lack of strong line blanketing observed in SN~2011fe. 

\par

In summary, we find strong evidence in favour of the VM model ($\ln$BF \textgreater 5) for SN~2011fe. Although many spectroscopic features are reproduced, the agreement is not perfect. Among all of the explosion scenarios considered in this work, our VM training dataset is based on the fewest existing explosion simulations and therefore is the most uncertain and unconstrained, and likely subject to the most biases. Future merger simulations will help to expand the parameter space of models covered by \textsc{riddler} and may find improved agreement.

\subsection{SN~2005hk}
\label{sect:05hk}
For SN~2005hk, we use spectra presented by \cite{phillips--07} and \cite{blondin--12}. Following \cite{phillips--07} we assume a fixed value of $E(B-V) = 0.09$ and draw a distance modulus from $\mathcal{N}(33.46, 0.27)$. A uniform prior on the explosion epoch of MJD = 53\,667.0 -- 53\,673.0 is used.

\begin{figure*}
\centering
\includegraphics[width=\textwidth]{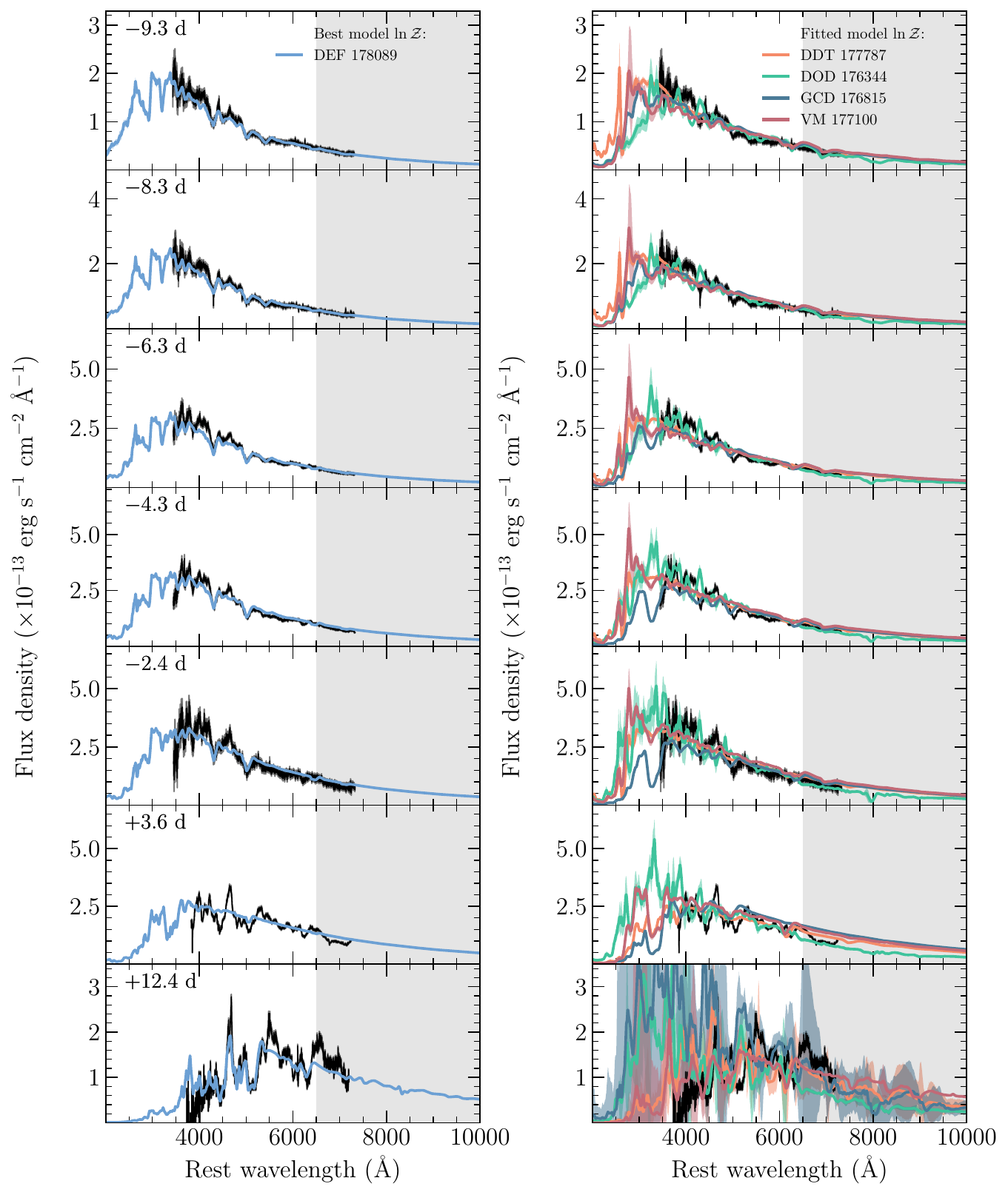}
\caption{As in Fig.~\ref{fig:11fe_spectra} for SN~2005hk.
}
\label{fig:05hk_spectra}
\centering
\end{figure*}

\begin{figure*}
\centering
\includegraphics[width=\textwidth]{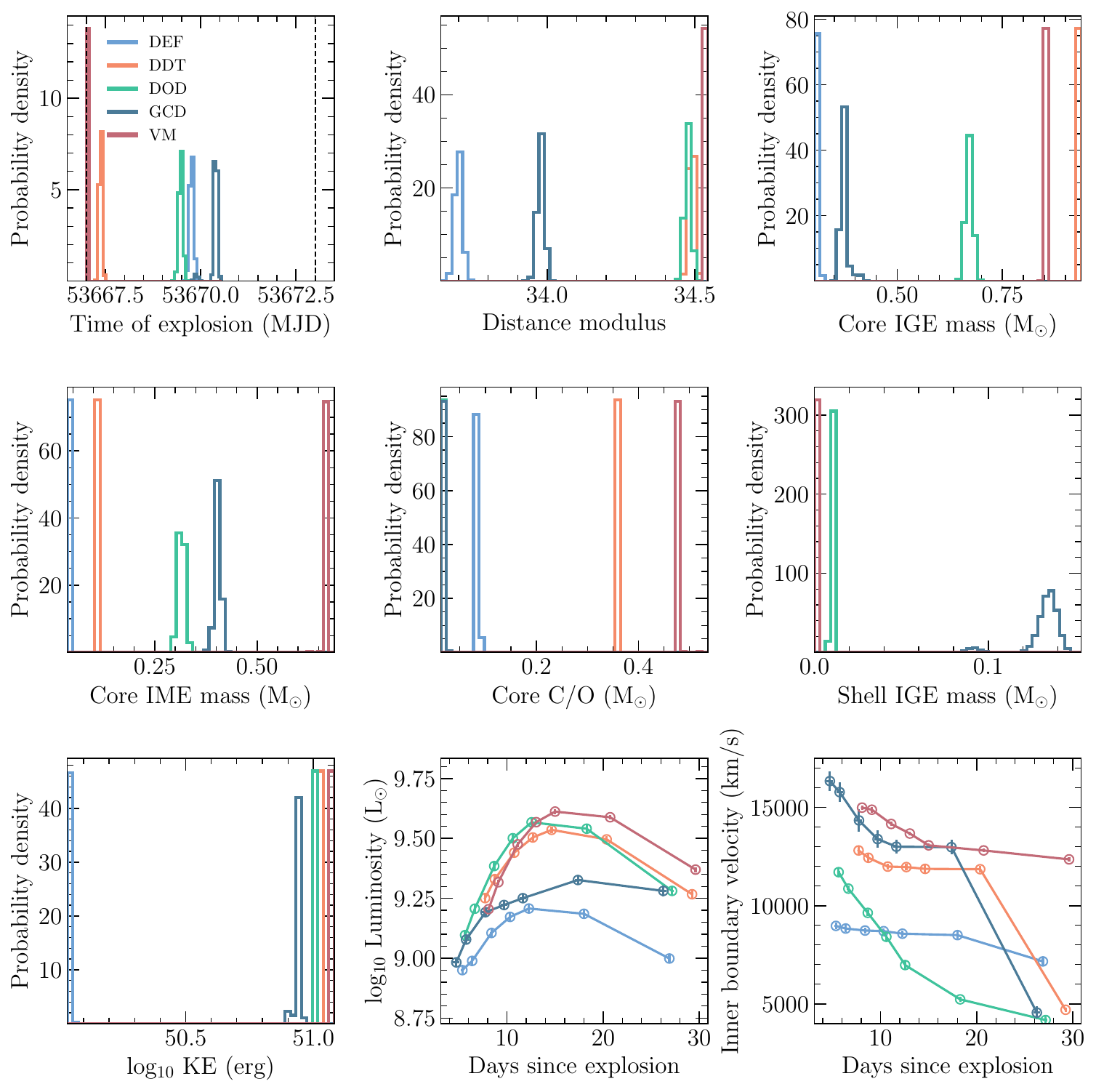}
\caption{As in Fig.~\ref{fig:11fe_posteriors} for SN~2005hk.
}
\label{fig:05hk_posteriors}
\centering
\end{figure*}

As demonstrated by Fig.~\ref{fig:colour_mag}, only the DEF scenario covers the full range of magnitudes and colours observed among the SN~2005hk spectra. Our fits to these spectra are shown in Fig.~\ref{fig:05hk_spectra}. Unsurprisingly, our best-fitting DEF model is able to match most of the features observed in the $-$9.3\,d spectrum, including the strong features due to IGEs. A weak \ion{C}{ii}~$\lambda$6\,580 feature is also predicted, but not clearly observed. At this epoch, the DEF model also provides remarkable agreement with the overall flux level and spectral shape of SN~2005hk. All other models perform significantly worse. The DDT, GCD, and VM models predict generally featureless, but noisy spectra. The DOD model predicts some features, although these are clearly discrepant from SN~2005hk. 

\par

Beginning from $-8.3$\,d and continuing to $-2.4$\,d, SN~2005hk shows an increasingly prominent \ion{Si}{ii}~$\lambda$6\,355 feature that is not reproduced by the DEF model. As discussed previously (Sect.~\ref{sect:fit_performance}), \textsc{riddler} struggles to produce weak spectral features. This likely results from their small contribution to the overall likelihood across the full spectrum. Features due to IMEs are much weaker in SNe~Iax compared to normal SNe~Ia and spectra are instead dominated by IGEs. Therefore it is unsurprising that our fits struggle to replicate the weak IME features and instead produce good matches with the strong IGE features. Alternative weighting schemes, with greater emphasis placed on specific spectral features, may improve the fits. Throughout these phases, SN~2005hk shows some evolution in the spectral features and most are well reproduced by our DEF model. At the same time none of the other models provide any reasonable agreement and instead continue to predict mostly featureless and noisy spectra. 

\par

At $+3.6$\,d the spectrum of SN~2005hk has changed considerably, becoming redder and developing stronger IGE features. These changes are not reflected in our best-fitting DEF model however, which instead shows only a few weak features throughout the optical spectrum. The cause of this failure is unclear. It may result from a difference in light curve evolution, whereby the \textsc{tardis} input luminosity of our model is too high and therefore a higher photospheric velocity is required to produce it. With the inner boundary further into the outer ejecta, this would also explain the lack of strong features observed in the spectrum. At this epoch, all other models remain poor matches. 

\par

By the final $+12.4$\,d spectrum, the DEF model shows better agreement than at $+3.6$\,d. Most of the spectrum in SN~2005hk is now dominated by blended features due to IGEs and this is captured by our DEF model, including the \ion{Si}{ii} and \ion{Fe}{ii} blend around $\sim$6\,300~\AA\, and the broad, flat-bottomed feature at $\sim$5\,200~\AA. While some features are weaker than observed, in general the predicted spectrum shows good agreement with the relative strengths and velocities of most features. In addition to producing poor fits, all other models show significant uncertainties in their predicted spectra. This likely arises from the difference in evolution between SNe~Iax and these models, as \textsc{riddler} tries to force fits into poorly-sampled regions of the parameter space that might better match SNe~Iax and consequently produce larger predicted uncertainties.

\par

Figure~\ref{fig:05hk_posteriors} shows that for our best-fitting DEF model we find an explosion epoch of 53\,669.76$\pm$0.14 and distance modulus of 33.70$\pm$0.04. All other models are higher by $\gtrsim$0.4~mag. The differences in distance modulii reflect differences in the best-fitting $^{56}$Ni masses of each model. Our DEF model contains 0.224$\pm$0.009~$M_{\odot}$ of $^{56}$Ni. Despite some overlap in the priors, none of our other models predict similarly low $^{56}$Ni masses and all are $\gtrsim$0.32~$M_{\odot}$. The DEF model is based on 10.300$\pm$0.033 ignition sparks, with a marginally lower ejecta mass of 0.434$\pm$0.008~$M_{\odot}$ containing 0.310$\pm$0.006~$M_{\odot}$ of IGEs and only 0.038$\pm$0.002~$M_{\odot}$ and 0.086$\pm$0.003~$M_{\odot}$ of IMEs and unburned material, respectively, relative to the canonical N10def model.

\subsection{SN~2018byg}
\label{sect:18byg}
We use the spectra of SN~2018byg presented by \cite{de--19}. From \cite{de--19} we also set $E(B-V) = 0.0$. Based on the redshift of $z = 0.066$, we assume a distance modulus prior of $\mathcal{N}(37.36, 0.30)$. We set explosion epoch limits to MJD = 58\,236 -- 58\,241.

\begin{figure*}
\centering
\includegraphics[width=\textwidth]{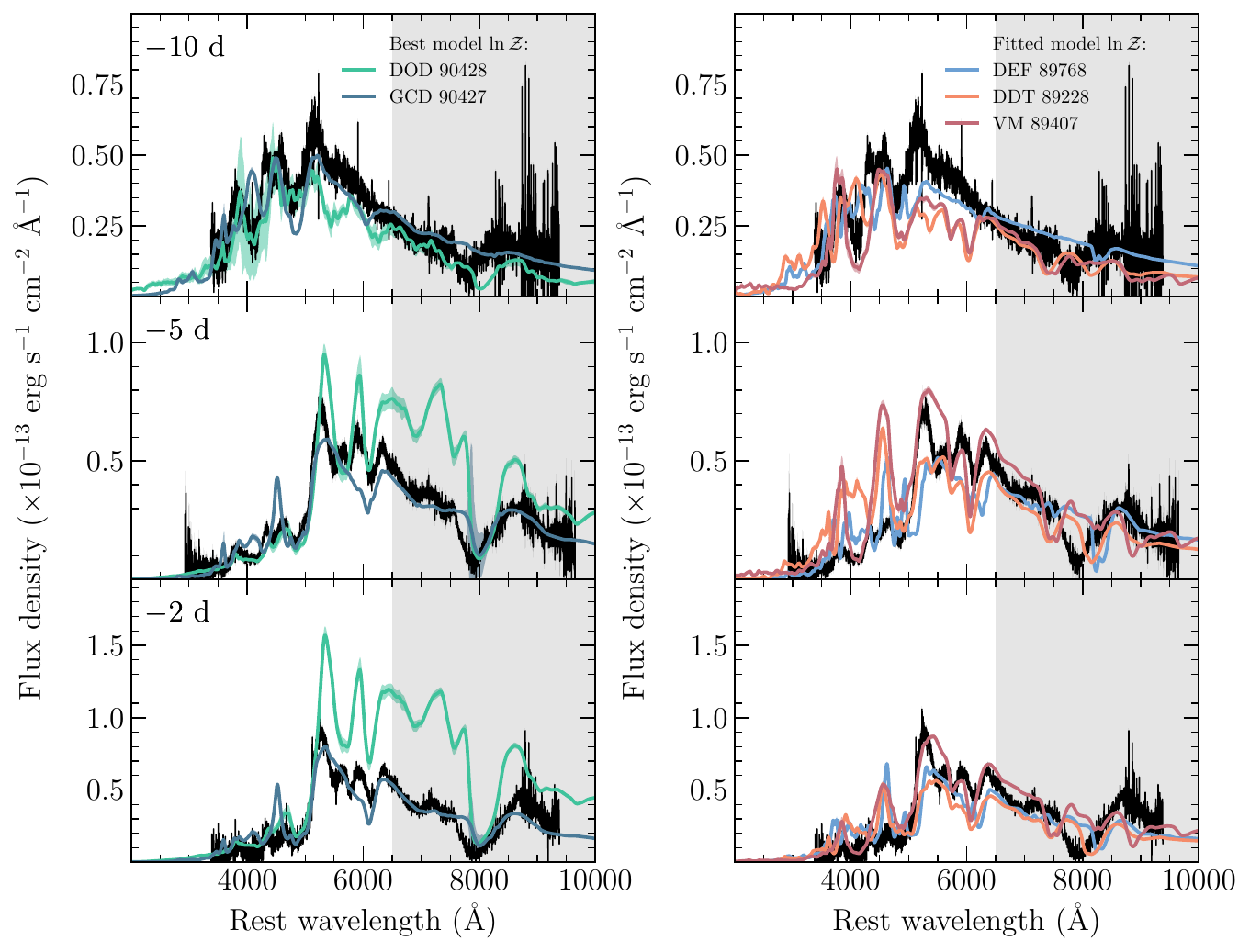}
\caption{As in Fig.~\ref{fig:11fe_spectra} for SN~2018byg.
}
\label{fig:18byg_spectra}
\centering
\end{figure*}

\begin{figure*}
\centering
\includegraphics[width=\textwidth]{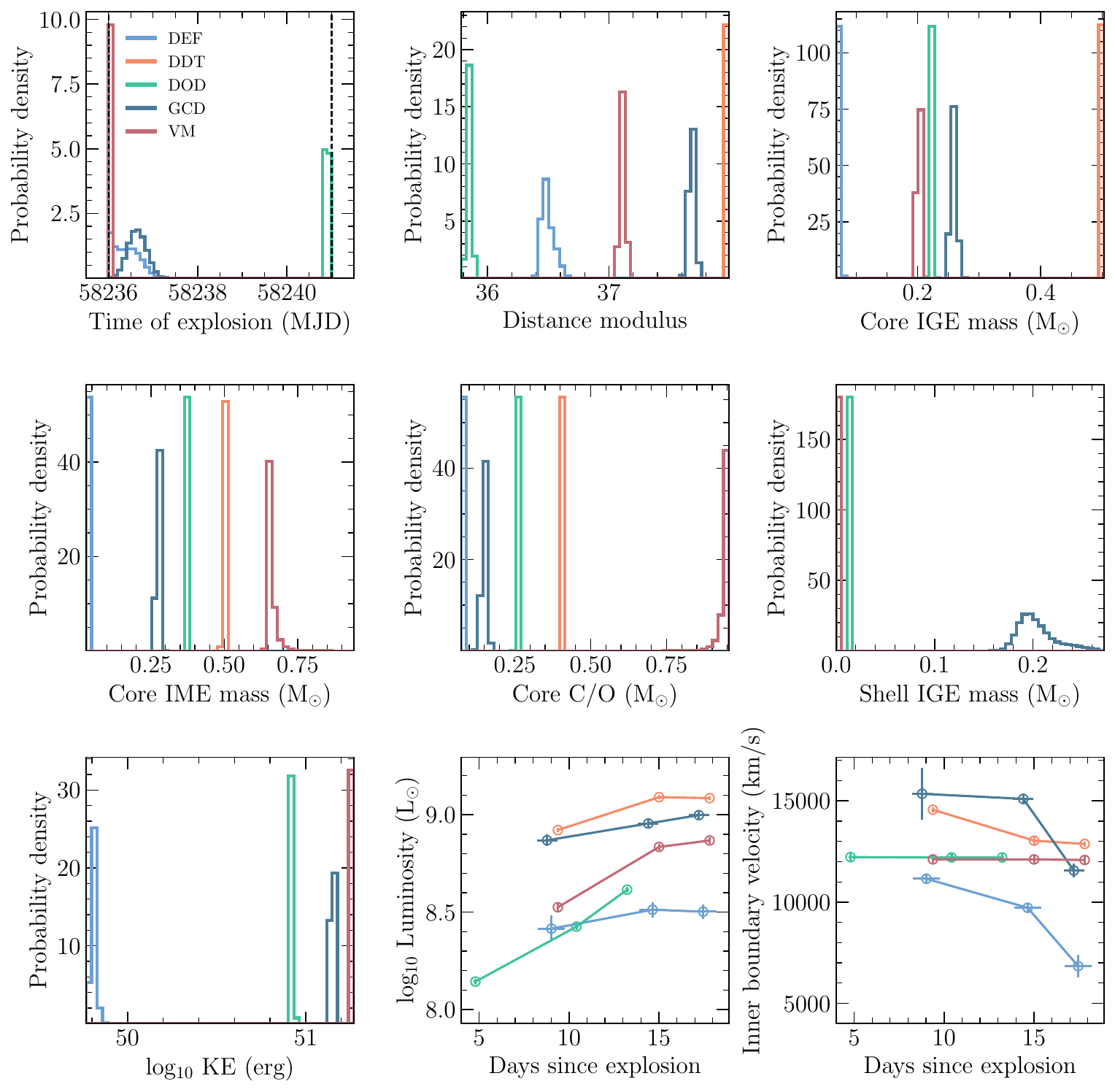}
\caption{As in Fig.~\ref{fig:11fe_spectra} for SN~2018byg.
}
\label{fig:18byg_posteriors}
\centering
\end{figure*}

Figure~\ref{fig:colour_mag} shows that the spectra of SN~2018byg display extremely red $B-V$ colours and most models are significantly bluer. Only the DOD scenario produces similar colours to all SN~2018byg spectra within the 95~per~cent confidence interval. The results of our fits are shown in Fig.~\ref{fig:18byg_spectra}. While we find that the DOD model is favoured overall, the difference in evidence relative to the GCD model ($\ln$~BF $\sim$1) indicates only a slight preference. The $-10$\,d spectrum of SN~2018byg is generally not well reproduced by any model. This spectrum shows strong absorption features at $\sim$4\,100~\AA\, and $\sim$4\,800~\AA, due to \ion{Ti}{ii} and other IGEs, with few other optical features. The DOD model produces both features with approximately the correct velocities and strengths, but a flatter absorption profile from $\sim$4\,800 -- 5\,000~\AA\, and broad features at $\sim$5\,500~\AA\, and $\sim$6\,200~\AA\, that are not observed. The GCD model does not reproduce the \ion{Ti}{ii} feature at $\sim$4\,100~\AA\, but provides better agreement with the feature at $\sim$4\,800~\AA. The spectrum at wavelengths $\gtrsim$5\,200~\AA\, is generally featureless, similar to SN~2018byg, with the exception of a feature at $\sim$6\,000~\AA. Our best-fitting DEF model is also mostly featureless at these wavelengths, but predicts other features at $\lambda \lesssim 5\,200$~\AA\, that are not observed. Both the DDT and VM models predict strong \ion{Si}{ii} features, in disagreement with SN~2018byg. 

\par

By $-$5\,d the spectrum of SN~2018byg is no longer featureless and has developed a few strong features. The DOD model shows good agreement with the sharp decrease in flux below $\lesssim$5\,200~\AA\, and the features from $\sim$3\,800 -- 4\,800~\AA. Beyond $\sim$5\,200~\AA, the overall flux predicted by our DOD is clearly much higher than in SN~2018byg and the features are much stronger. Similar results have previously been shown by \cite{gronow--20} and \cite{collins--22} (see Sect.~\ref{sect:18byg_comp}). The GCD model provides a better match to the overall flux level, but fails to reproduce most spectral features, including predicting what may be a single broad feature from $\sim$5\,400 -- 6\,400~\AA. Other models continue to over-predict the UV flux and produce features that are not observed. At $-$2\,d the level of agreement is similar to the previous $-$5\,d spectrum. The DOD model matches the UV line blanketing and some absorption features, although the flux level is higher than observed. The GCD model matches the flux level, but does not reproduce most absorption features. Likewise none of the other models reproduce the features observed. 

\par

For SN~2018byg, Fig.~\ref{fig:18byg_posteriors} shows that the best-fitting explosion epoch varies by up to $\sim$5\,d, with the DDT and VM models preferring explosion epochs at our lower limit of MJD $\sim$ 58\,236, while our DOD model prefers a value close to our upper limit of MJD $\sim$58\,241. The distance modulus also shows considerable variation of $\sim$3~mag, ranging from 35.85$\pm$0.05 for the DOD model to 38.50$\pm$0.04 for the DDT model. The best-fitting ejecta structure of our DOD model is based on a WD core mass of 0.859$\pm$0.002~$M_{\odot}$ and a shell mass of 0.038$\pm$0.001~$M_{\odot}$, giving a total ejecta mass of 0.897$\pm$0.002~$M_{\odot}$. Within the WD core, this model contains 0.221$\pm$0.005~$M_{\odot}$ of IGE (of which 0.150$\pm$0.004~$M_{\odot}$ are $^{56}$Ni), 0.373$\pm$0.005~$M_{\odot}$ of IME, and 0.265$\pm$0.005~$M_{\odot}$ of unburned C/O. The $^{56}$Ni mass of this model is towards the lower end of our prior distributions (Fig.~\ref{fig:model_priors}), which could explain the over-predicted flux if our models do not cover the appropriate parameter space for SN~2018byg (see Sect.~\ref{sect:18byg_comp}). The strong line blanketing produced in the DOD model results from the burned material in the shell. Approximately half of the mass of the shell is unburned material, while the remainder is comprised of 0.014$\pm$0.001~$M_{\odot}$ of IGE and 0.008$\pm$0.001~$M_{\odot}$ of IMEs. The GCD model predicts a similarly low $^{56}$Ni mass of 0.188$\pm$0.010~$M_{\odot}$ in the core and a large shell mass of 0.716$\pm$0.012, with 0.159$\pm$0.031~$M_{\odot}$ of $^{56}$ Ni in the shell, indicating a significant fraction of burned material at high velocities in the outer ejecta. 

\par

For SN~2018byg, we find comparable levels of agreement with our DOD and GCD models although the DOD model is slightly preferred. Both models predict some amount of burned material in the outer ejecta, resulting in significant line blanketing in agreement with SN~2018byg. While the DOD model is slightly preferred, it generally over-predicts the optical flux. Future models with lower WD masses may improve agreement.

\subsection{Summary}
In summary, we find that \textsc{riddler} is able to produce good qualitative agreement with observations of SNe~2011fe, 2005hk, and 2018byg. While no model is able to match all observed features, we find that SNe~2011fe, 2005hk, and 2018byg are best reproduced by the VM, DEF, and DOD scenarios, respectively. The GCD scenario also provides reasonable agreement with observations of SN~2018byg. These fits show that \textsc{riddler} is able to find intrinsic differences between different SNe~Ia that could be related to intrinsic differences in their explosion physics. Due to our training data, which is based on predictions from realistic explosion simulations, it is not possible to fit any arbitrary model to any set of observations and therefore the fits produce physically meaningful results. Hence our improved version of \textsc{riddler} provides the means for robust and quantitative constraints on the explosion physics of observed samples of SNe~Ia.

%
%__________________________________________________________________________________________________________________________________________________________________________________________
%__________________________________________________________________________________________________________________________________________________________________________________________
%__________________________________________________________________________________________________________________________________________________________________________________________

\section{Discussion}
\label{sect:discussion}

\subsection{Limitations and assumptions}
\label{sect:limits}

While the updated version of \textsc{riddler} presented in this work improves upon that presented by \cite{magee--24}, a number of key limitations and assumptions still remain. Care must therefore be taken when making inferences. Section~\ref{sect:application} discusses how \textsc{tardis}, and therefore \textsc{riddler}, is not sensitive to material below the photosphere. We reiterate this point again here and stress that the masses of various elemental groups within the models should not be taken as measurements of the true masses for a given SN. Rather, the purpose of \textsc{riddler} is to determine the overall preferred explosion scenario for a given SN and best-fitting model parameters for that scenario. As with any comparison between observations and models however, our results depend on both the assumed priors for the model and the likelihood function with which the comparison is performed. Changes to either of these could produce vastly different results (see Sect.~\ref{sect:11fe_comp} and \citealt{magee--24}). Although this problem is not unique to \textsc{riddler}, and indeed is likely more significant for the classical visual inspection approach, it should be considered when performing automated fitting of SNe spectra. With an explicit likelihood function (e.g. Eqn.~\ref{eqn:ultanest_likelihood}), automated fitting is at least easily reproducible and fits to observations can be performed in a systematic way. 

\par

Fits with \textsc{riddler} are based on emulated spectra of radiative transfer simulations. All applications invoking machine learning in similar ways are limited by their training data to some extent and \textsc{riddler} is no different. Nevertheless, using existing explosion simulations as the basis for our radiative transfer simulations comes with advantages and disadvantages. Our training models are more physically realistic than completely ad-hoc, custom models and therefore can provide useful constraints on the explosion physics, but they will almost certainly not be able to provide perfect matches with observations. If models do not predict the correct element at the correct velocity then they will never be able to reproduce certain spectroscopic features. Some SNe~Ia may also simply fall outside of the prior ranges or explosion scenarios considered here. We also note that all of our training data are based on publicly available models from HESMA, but many alternative literature models for the scenarios considered here are available from other groups. Even where our double detonation models, for example, may fail to reproduce certain features or SNe~Ia, models from other groups may fare better (see Sect.~\ref{sect:18byg_comp}).

\par

Indeed, one of the key considerations when fitting with \textsc{riddler} is that it does not find the `correct' explosion scenario for a given SN~Ia. Nor does it quantify whether an explosion scenario provides a `good' match to the observations at all. Formally speaking, with the inclusion of the nuisance parameter $f$, all of our fits are `good' matches in terms of their reduced $\chi^2$ ($\chi_{\nu}^2 \sim 1$). Removing this parameter would result in none of the fits being classed as `good' ($\chi_{\nu}^2 \gg 1$). The same is also true for the underlying models upon which our training data are based and, we suspect, all existing empirical spectral models -- a simple $\chi_{\nu}^2$ would produce values $\gg1$. The evidence values calculated by \textsc{riddler} and \textsc{ultranest} are therefore only informative in a relative sense when considering each explosion scenario for a given SN. Evidences should not be compared across multiple SNe~Ia to determine whether, for example, the delayed detonation scenario provides a better match to one SN or another. Instead, for any given SN, evidences can be used to determine whether the delayed detonation scenario produces a better match than the violent merger scenario, for example. \textsc{riddler} is simply able to identify which model is preferred among those considered in the training data (for a given likelihood and priors), but otherwise does not consider the quality of agreement.

\par

The primary limitation for observations is the requirement of flux calibrated spectra. Fits with \textsc{riddler} are performed in physical flux units and therefore the absolute and relative (i.e. as a function of wavelength) flux calibration will impact the results. Indeed, this limitation is also applicable to all similar spectral comparisons between models and observations. Calibrating an observed spectrum to photometry at the same epoch could follow one of two approaches. The spectrum can be scaled by a single value such that the synthetic magnitudes are on average equal to the photometry. Alternatively, a wavelength dependent scale factor could be applied and the spectrum mangled such that the synthetic magnitudes are exactly equal to the photometry in each band. The former case implicitly assumes that the relative flux calibration of the spectrograph is accurate, which is not necessarily the case. Synthetic magnitudes in the latter case should provide better matches to the photometry in the same band, but interpolating to bands not observed or extrapolating beyond wavelengths covered by photometry could introduce a bias or systematic offset into the calibrated spectrum. Accurately accounting for all of the systematic uncertainties associated with this process is non-trivial. Here we adopt the most simplistic approach of assuming a single scale factor.

\par

As mentioned, some of the limitations highlighted in this section also apply to the classical visual inspection approach. Although automated fitting with \textsc{riddler} does not fully overcome these limitations, it is nevertheless a more robust and repeatable approach than visual inspection that can be readily applied to large numbers of SNe~Ia spectra.

\subsection{Comparison with previous results}
\label{sect:previous_fits}

\subsubsection{SN~2011fe}
\label{sect:11fe_comp}
The proximity and luminosity of SN~2011fe has resulted in an extensive dataset and comprehensive analyses to try and uncover the progenitor and explosion mechanism. Here we compare the results from our \textsc{riddler} fits to existing explosion simulations and empirical models for SN~2011fe.

\renewcommand{\arraystretch}{1.4} % 1.5 times the default height
\begin{table}
\centering
\caption{Impact of UV wavelengths on best-fitting parameters}\tabularnewline
\label{tab:11fe}\tabularnewline
\resizebox{\columnwidth}{!}{
\begin{tabular}{llllll}\hline
\hline
             & \multicolumn{2}{c}{DDT} & \multicolumn{2}{c}{VM} \tabularnewline
Parameter    & Incl. UV         & Excl. UV    &  Incl. UV    & Excl. UV     \tabularnewline
\hline
$\ln f$  & $-$1.218$\pm$0.033 & $-$1.706$\pm$0.032 & $-$1.185$\pm$0.026 & $-$1.630$\pm$0.031 \tabularnewline 
$t_{\textrm{exp}}$ & 55\,795.965$\pm$0.049 & 55\,795.977$\pm$0.090 & 55\,794.157$\pm$0.193 & 55\,794.953$\pm$0.111 \tabularnewline 
$\mu$ & 28.875$\pm$0.031 & 29.052$\pm$0.032 & 28.272$\pm$0.021 & 28.578$\pm$0.043 \tabularnewline 
$M_{C}^{\textrm{IGE}}$ & 0.973$\pm$0.005 & 1.179$\pm$0.025 & 0.253$\pm$0.002 & 0.487$\pm$0.019 \tabularnewline 
$M_{C}^{\textrm{Ni}}$ & 0.633$\pm$0.007 & 0.771$\pm$0.020 & 0.240$\pm$0.002 & 0.453$\pm$0.017 \tabularnewline 
$M_{C}^{\textrm{IME}}$ & 0.078$\pm$0.004 & 0.135$\pm$0.013 & 0.624$\pm$0.006 & 0.470$\pm$0.019 \tabularnewline 
$M_{C}^{\textrm{C/O}}$ & 0.350$\pm$0.002 & 0.086$\pm$0.020 & 0.825$\pm$0.009 & 0.959$\pm$0.017 \tabularnewline 
$\ln KE$ & 51.234$\pm$0.007 & 51.031$\pm$0.002 & 50.955$\pm$0.013 & 51.070$\pm$0.013 \tabularnewline
\hline
\hline
\end{tabular}
}
\end{table}
\renewcommand{\arraystretch}{1.0} % 1.5 times the default height

\par

From our \textsc{riddler} fits we find that the VM model produces the overall best agreement, which could indicate that SN~2011fe is consistent with a sub-Chandrasekhar mass progenitor. The distance modulus resulting from this fit ($\mu = 28.27\pm0.02$) is $\sim4\sigma$ below the mean reported by \cite{shappee--11} and hence our model is somewhat under-luminous compared to typical SNe~Ia. The underlying ejecta structure is based on the merger of a $\sim$0.94~$M_{\odot}$ primary white dwarf with a $\sim$0.77~$M_{\odot}$ secondary, similar to the \cite{kromer--13b} model but with a slightly higher $^{56}$Ni mass (0.24~$M_{\odot}$ c.f. 0.18~$M_{\odot}$). Mergers of $\sim$0.9~$M_{\odot}$ white dwarfs have generally been argued to be consistent with under-luminous SN~1991bg-like SNe~Ia \citep{91bg--disc, pakmor--10} as opposed to normal SNe~Ia. Indeed, our VM model shows some features in disagreement with SN~2011fe that are more commonly observed in fainter SNe~Ia, such as the strong \ion{Si}{ii}~$\lambda$5\,972 feature at maximum light. Mergers involving higher mass primaries ($\sim$1.0 -- 1.1~$M_{\odot}$) have instead been favoured for normal SNe~Ia, but still show some disagreement \citep{pakmor-2012}.

\par

\begin{figure*}
\centering
\includegraphics[width=\textwidth]{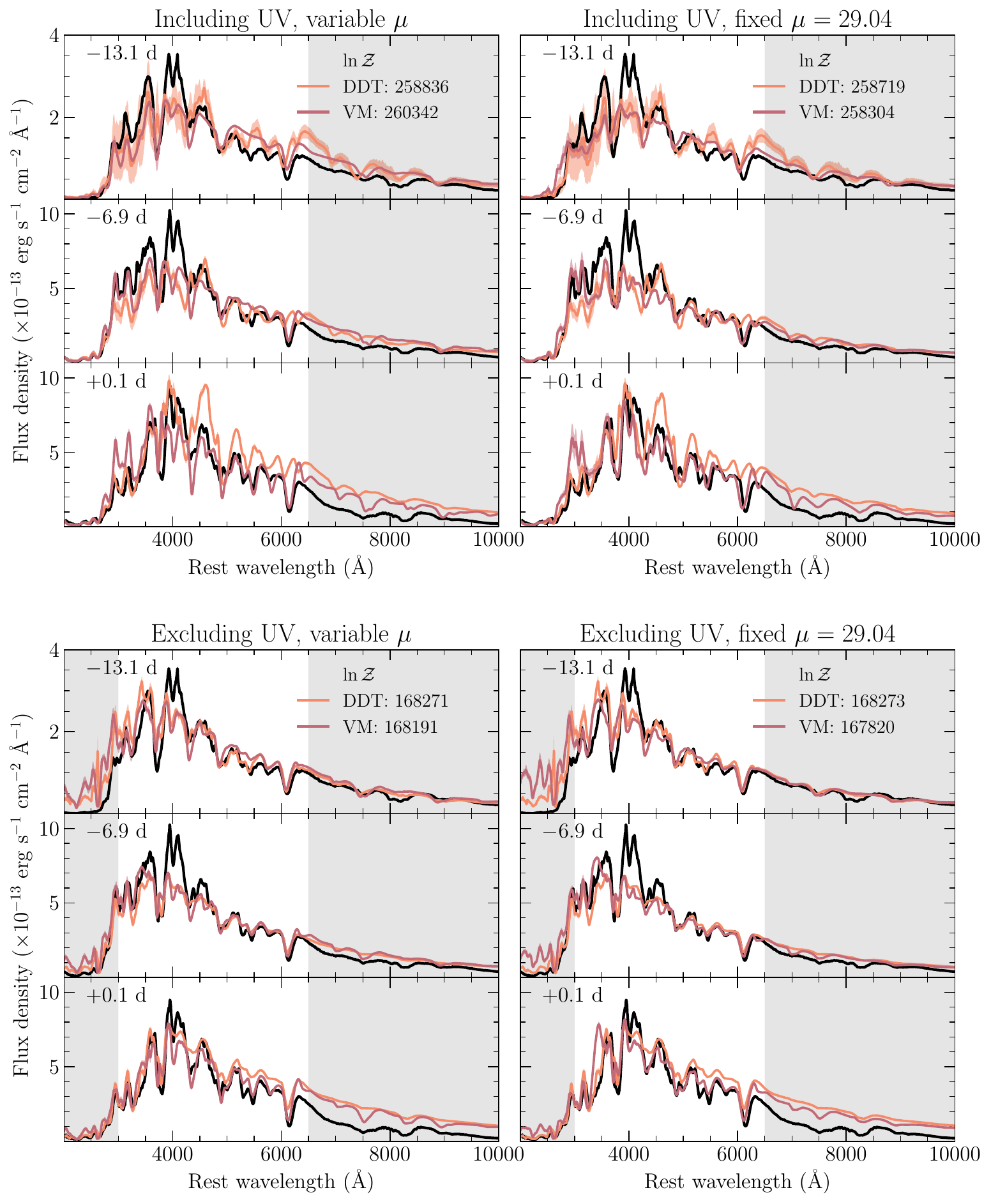}
\caption{Variations in spectral fits to SN~2011fe for the DDT and VM models under four different sets of assumptions. Fits are shown for cases in which the distance modulus $\mu$ is allowed to vary or fixed at $\mu$ = 29.04~mag and the UV is either included or excluded from the likelihood calculation. Grey shaded regions are not included in the fit. Phases are given relative to maximum light. 
}
\label{fig:11fe_fit_variations}
\centering
\end{figure*}

Various arguments for and against sub-Chandrasekhar mass progenitors for SN~2011fe have been presented in the literature (e.g. \citealt{mazzali--15, dimitriadis--17, kerzendorf--17, shappee--17}). Although our fiducial fit favours such progenitors, the model clearly has some short comings. As discussed in Sect.~\ref{sect:limits}, the assumed priors and likelihood function can significantly impact the results of fits. We therefore perform additional fits in which we either fit for the distance $\mu$ (assuming a Gaussian prior $\mathcal{N}(29.04, 0.19)$) or use a fixed $\mu = 29.04$ and either include or exclude the UV ($\lambda \textless 3\,000$~\AA). The fit allowing $\mu$ to vary and including the UV represents our fiducial case. The best-fitting parameters for both models when varying $\mu$ and either including or excluding the UV are shown in Table~\ref{tab:11fe}. All four fits for both models are shown in Fig.~\ref{fig:11fe_fit_variations}.

\par

The resulting models display significant variation. In all but our fiducial case, the DDT scenario would be preferred over the VM scenario, based on $\ln$ BF. With a fixed $\mu = 29.04$ and including the UV we find that the resulting VM model is closer to that expected for a normal SN~Ia. Specifically the ejecta is similar to the merger of a $\sim$1.04~$M_{\odot}$ primary and $\sim$0.86~$M_{\odot}$ secondary, and predicts $0.426\pm0.006$~$M_{\odot}$ of $^{56}$Ni. This is comparable to the $^{56}$Ni mass estimated from the bolometric light curve ($\sim$0.44~$M_{\odot}$; \citealt{pereira--13}). This model generally has weaker \ion{Si}{ii}~$\lambda$5\,972 throughout its evolution, in better agreement with SN~2011fe, and better matches the \ion{Si}{ii} and \ion{Ca}{ii} blend around $\sim$3\,800~\AA. Overall, some features are shifted to higher velocities, the UV flux is generally higher than observed, and the $-$15.3\,d spectrum is not as well matched. This could indicate that the outer ejecta (composition and/or ionisation state) of the fiducial case is favoured at early times, while the inner ejecta of the more luminous model is a better match around maximum. Fixing $\mu = 29.04$ has little impact on the DDT model while the significant decrease in the fit at $-15.3$\,d and the need for a larger systematic uncertainty ($f$) results in a lower evidence for the VM model. Excluding the UV region from our fit also results in a somewhat more luminous VM model and larger distance modulus ($\mu = 28.58\pm0.04$) relative to our fiducial case. As expected, this model shows weaker \ion{Si}{ii}~$\lambda$5\,972 that is in closer agreement with SN~2011fe, but unsurprisingly tends to strongly over-predict the flux below $\lesssim$2\,600~\AA. The DDT model excluding the UV shows a significant increase in the optical fit quality. Although the features are somewhat weaker than observed around maximum light, the velocities and relative strengths throughout the evolution are comparable to those of SN~2011fe. For the DDT model, removing the UV results in a larger inner boundary velocity at all epochs, which may explain the weaker features at maximum light in particular.

\par

Our results show that the best-fitting DDT models are based on ejecta structures arising from $N_k \sim 10$ -- 13, with $^{56}$Ni masses ranging from 0.63 -- 0.77~$M_{\odot}$ depending on whether the UV is included or excluded. \cite{sim--13} show that \cite{seitenzahl--13} DDT models with intermediate $N_k$ ($\sim$40) provide reasonable agreement with normal SNe~Ia, but the \ion{Si}{ii}~$\lambda$6\,355 velocities are too high and the spectra are redder than observed. Weaker explosions ($N_k \sim 1\,600$) result in lower velocities and simultaneously colours that are even redder. Stronger explosions ($N_k \sim1$ -- 3) however show somewhat broader and weaker \ion{Si}{ii}~$\lambda$6\,355. The preference for a relatively low $N_k$ ($\sim 10$) in our model likely arises from trying to balance both the spectral features and continuum shape. \cite{sim--13} also show that stronger explosions ($N_k \sim 3$ -- 5) display pronounced viewing angle variations (up to $\sim$0.5~mag in the $U$- and $B$-bands), with viewing angles producing fainter luminosities providing greater similarities to normal SNe~Ia. Our best-fitting model has a lower $^{56}$Ni mass than the canonical N10 model (0.939~$M_{\odot}$), which is qualitatively similar to the model being projected along similar viewing angles.

\par

The differences between the best-fitting models in these four scenarios highlight the importance of an explicit likelihood function, something which is not captured with visual inspection. Indeed, for all but one of these scenarios the DDT model is favoured over the VM model statistically, but one could argue that the VM model provides comparable or better visual agreement at some epochs (particularly around maximum). Again we note that we take an agnostic approach of weighting every wavelength ($\leq6\,500$~\AA) and phase equally. With log-spaced wavelength bins however, approximately half of our bins occur in the region $\sim2\,000$ -- 4\,500~\AA\, with the remainder spanning a larger wavelength range of $\sim$4\,500 -- 10\,000~\AA. These bins were selected to ensure that all spectral features are equally resolved in velocity space, but will also result in a larger contribution of smaller wavelengths to the likelihood. Similarly, if a spectral sequence contains multiple spectra at early times and fewer at later times (or vice versa) then the fit will be biased towards specific phases and may not necessarily reflect the full ejecta. The modeller employing a classical visual inspection approach may be aware of these limitations and adjust their model accordingly, but again this is not explicit and difficult to replicate. Similar results may be obtained through automated fitting if the wavelength and phase weights are also adjusted in an attempt to remove these biases. This is an arbitrary decision however and it is not clear what (if any) alternative weights should be used. 

\par

These limitations are apparent when comparing our results to the SN~2011fe fits presented by \cite{mazzali--14}. Our DDT models struggle to replicate both the UV and optical features of SN~2011fe simultaneously and our model including the UV is clearly not a good fit beyond the first few days immediately after explosion. \cite{mazzali--14} present empirical models based on the density profile of the one-dimensional WS15DD1 delayed detonation model \citep{iwamoto--99}, assuming a fixed $\mu = 29.04$, and fitting for the best-matching composition. \cite{mazzali--14} show that the relatively large amount of material at high velocities in the WS15DD1 model results in UV features with velocities that are too high. Instead, \cite{mazzali--14} favour a model in which there is a lower density tail towards high velocities. Our DDT models including the UV show qualitatively similar results. Although the overall flux level is (naturally) in better agreement when the UV is included, the velocities are somewhat higher. Conversely, when the UV is excluded the best-matching model favours a lower kinetic energy ($\sim1.1\times10^{51}$~erg c.f. $1.7\times10^{51}$~erg) that provides better agreement and reduces the density of the outer ejecta. 
\par

SN~2011fe was also previously modelled with an earlier version of \textsc{riddler} by \cite{magee--24}. This version included only a single realisation of a delayed detonation explosion, represented by N100 (0.604~$M_{\odot}$ of $^{56}$Ni), and is most directly comparable to our models in which the UV is excluded and $\mu = 29.04$. Compared to the \cite{magee--24} N100 model, the $N_k \sim 10$ models favoured here produce better agreement with the early spectra. In particular, many of the optical features from $\sim4\,500$ -- 6\,500~\AA\, are produced with the correct velocities and relative strengths. At $-$6.9\,d, the \cite{magee--24} model continues to over-predict the flux around $\sim$3\,000 -- 3\,500~\AA, while our model now shows good agreement. At the same time, the \cite{magee--24} model reproduces the absorption features around $\sim$4\,200 -- 4\,500~\AA\, that are weaker in our model. Finally, as mentioned, the features in our model at maximum light are also weaker than observed in SN~2011fe, while those in \cite{magee--24} show good agreement.

\par

Overall, our \textsc{riddler} fits to SN~2011fe highlight the benefits and challenges associated with automated fitting and where caution should be applied. None of the models considered here are simultaneously able to reproduce all of the UV and optical features. Careful consideration must be given to the priors and likelihood function when applied to fitting SNe~Ia, and these should be made explicitly clear when fitting. This is not possible with a classical visual inspection approach.

\subsubsection{SN~2005hk}
\label{sect:05hk_comp}
\cite{kromer-13} present comparisons between the \cite{fink-2014} N5def ($N_k = 5)$ pure deflagration model and observations of SN~2005hk. \cite{kromer-13} find that the N5def model broadly reproduces the observed features of SN~2005hk, including the peak absolute magnitude, colours at maximum light, and low velocities. From our fits with \textsc{riddler}, we find that a model based on $N_k \sim 10$ is preferred, which has higher ejecta, IGEs, and $^{56}$Ni masses than N5def. The increased $^{56}$Ni mass of our best-fitting model (and therefore higher luminosity) compared to N5def is compensated by the higher distance modulus found by our fits ($\mu = 33.70\pm0.04$ c.f. $\mu = 33.46\pm0.27$). 

\par

\cite{kromer-13} show that at $-$6.3\,d, the N5def model is somewhat fainter and redder than SN~2005hk. Conversely, our \textsc{riddler} fits show closer agreement with the spectrum flux level and shape, and reproduce the strong features observed. Neither N5def nor our model however produce significant \ion{Si}{ii}~$\lambda$6\,355 absorption and instead over-predict the \ion{C}{ii}~$\lambda$6\,580 feature. By $-$3.3\,d, \cite{kromer-13} show that the N5def model now provides better agreement with the spectrum flux and shape of SN~2005hk, in addition to many of the spectral features. At this phase, our \textsc{riddler} fits provide good agreement with the velocities of features in SN~2005hk, but this requires a lower kinetic energy for our model of $\log KE = 50.04\pm0.01$, while for the \cite{fink-2014} N5def and N10def models $\log KE = 50.13$ and $50.29$, respectively. As previously discussed (Sect.~\ref{sect:05hk}), our best-fitting DEF spectrum does not reproduce the observed spectral features at $+$3.6\,d and instead is mostly featureless. We speculate that this is due to differences in the luminosity evolution of our model and SN~2005hk. Specifically, the luminosity of our model may be too high at this epoch, forcing the photosphere too far into the outer ejecta and the appearance of only weak features. The \textsc{artis} models presented by \cite{kromer-13} do not contain a similar artificial photosphere in this way and instead are able to broadly reproduce observations of SN~2005hk around this time. At later epochs, N5def and our best-fitting model show similar levels of agreement, reproducing (or not) many of the same features. 

\par

SN~2005hk has also been subject to multiple empirical modelling analyses that provide useful comparison points for our \textsc{riddler} models. \cite{05hk--400days} model the spectra of SN~2005hk based on the W7 deflagration model \citep{nomoto-w7} scaled to a lower $^{56}$Ni mass (0.18~$M_{\odot}$) and kinetic energy (0.30$\times10^{51}$~erg). Unlike N5def and our model, the SN~2005hk model presented by \cite{05hk--400days} is able to match the \ion{Si}{ii}~$\lambda$6\,355 feature observed at early times, likely due to the increased silicon mass (0.074~$M_{\odot}$). \cite{barna--18} model the spectra of SN~2005hk using customised density and composition profiles based on the \cite{fink-2014} deflagration models. \cite{barna--18} find that SN~2005hk is best reproduced by assuming a steeper density profile in the outer ejecta of the N5def model, which is qualitatively similar to the lower kinetic energy found by our fitting. Overall, the models show broadly similar levels of agreement.

\par

\cite{fink-2014} argue that the N5def and N10def models are promising candidates for SNe~Iax, including SN~2005hk. Our fits support this claim, but suggest that some differences may improve agreement further. In particular, our fits suggest a brighter model that simultaneously has a lower kinetic energy is in better agreement with SN~2005hk than the standard \cite{fink-2014} models. The \cite{lach--22} pure deflagration models test various parameters affecting the initial ignition of the deflagration, including the size of the offset from the centre of the white dwarf and the central density. Models with an offset of 10~km and central densities of 4$\times10^9$, 5$\times10^9$, and $6\times10^9$~g~cm$^{-3}$ show similar $^{56}$Ni masses (0.092, 0.085, and 0.090~$M_{\odot}$ respectively), but kinetic energies that can differ by $\sim$40~per~cent (0.68$\times10^{50}$, 0.75$\times10^{50}$, and 0.97$\times10^{50}$~erg, respectively). Models ignited with larger offsets from the white dwarf centre show smaller variations in their kinetic energies and predict overall smaller $^{56}$Ni masses. An offset of 10~km is the lowest value considered by \cite{lach--22} therefore we speculate that even smaller ignition offsets may show larger variations in their predicted ejecta as other parameters are varied and could produce pure deflagrations similar to those that appear to be favoured by our fits.

\subsubsection{SN~2018byg}
\label{sect:18byg_comp}
Our double detonation models are based on the suite presented by \cite{gronow--21}. \cite{collins--22} generate synthetic observables of these models and perform comparisons with multiple SNe~Ia, including the SN~2018byg spectra presented in Fig.~\ref{fig:18byg_spectra}. \cite{collins--22} show that the \cite{gronow--21} models display significant viewing angle variation, due to the location at which the helium shell is ignited. None of the models are able to match all of the features observed in SN~2018byg however, regardless of viewing angle. Models with relatively low mass white dwarfs and helium shells ($M_C = 0.8~M_{\odot}$, $M_S = 0.03$ -- $0.05~M_{\odot}$) provide somewhat reasonable agreement with the spectrum at $-10$\,d when viewed offset ($\theta \gtrsim 130\degree$) from the shell ignition point. These viewing angles experience minimal line blanketing and approximately match the luminosity and colour, but do not reproduce many of the spectral features. At later epochs, such viewing angles clearly do not reproduce the strong flux suppression for wavelengths $\lesssim$5\,000~\AA. When viewed closer to the ignition point ($\theta \lesssim 90\degree$), the early spectra instead show line blanketing that is too strong and do not match the colour of SN~2018byg at $-10$\,d. Qualitatively, the strong line blanketing observed at $-$5 -- $-$2\,d in these models is similar to that seen in SN~2018byg, but many of the spectral features are too strong and the flux level and colour are not well reproduced. Unsurprisingly, our best-fitting DOD model for SN~2018byg shows many of the same shortcomings as the \cite{gronow--21} models, despite the increased variation. In particular, \cite{collins--22} show that these models have a similarly high flux at longer wavelengths and later epochs as that observed in Fig.~\ref{fig:18byg_spectra}. We note that differences in ionisation and excitation treatment can result in less significant viewing angle effects \citep{collins--25}.

\par

Our best-fitting DOD model is based on an ejecta structure with $M_C \sim 0.859~M_{\odot}$ and $M_S \sim 0.038$, which is similar to the models studied by \cite{collins--22}. \cite{de--19} also compare observations of SN~2018byg to the 1D double detonations presented by \cite{polin--19}. \cite{de--19} find that a model with $M_C = 0.76~M_{\odot}$ and $M_S = 0.15~M_{\odot}$ produces the best agreement with the observations, in particular the $r$-band evolution. The large helium shell mass of this model however produces a strong early flux excess in the $r$-band light curve, with an initial decline of $\sim$0.4~mag before the light curve begins to rise again. Such a decline is not observed in SN~2018byg therefore arbitrary mixing is applied to the ejecta before the radiative transfer simulations. Spectroscopically, this model produces better agreement with SN~2018byg than the \cite{gronow--21} models or indeed our best-fitting DOD model. The best-matching core and shell masses found by \cite{de--19} are outside our prior range for the double detonation scenario (see Sect.~\ref{sect:dedet}) therefore it is unsurprising that we find different results. Nevertheless, strong arguments have been made in the literature that SN~2018byg is a promising candidate for a double detonation explosion. Our results with \textsc{riddler} are consistent with this interpretation. Despite the multiple scenarios included in our training data, we find that the double detonation scenario is the favoured model. This shows that even if \textsc{riddler} is not able to provide an exact match to a given SN, it is still sensitive to intrinsic properties of the ejecta and capable of providing constraints on the explosion mechanism.

%
%__________________________________________________________________________________________________________________________________________________________________________________________
%__________________________________________________________________________________________________________________________________________________________________________________________
%__________________________________________________________________________________________________________________________________________________________________________________________

\section{Conclusions}
\label{sect:conclusions}
\textsc{riddler} is a machine learning based method for automated, quantitative, and repeatable fitting of SNe~Ia spectral time series up to shortly after maximum light \citep{magee--24}. In this work, we presented an updated version of \textsc{riddler}, rewritten in \textsc{pytorch} with a significantly expanded training dataset, broader wavelength coverage, and more robust uncertainty estimates.

\par

Training data for \textsc{riddler} is based on predictions from existing literature explosion simulations. Here we added a wider range of explosion scenarios to \textsc{riddler}, including pure deflagrations \citep{fink-2014}, delayed detonations \citep{seitenzahl--13}, double detonations \citep{gronow--21}, gravitationally confined detonations \citep{lach--22b}, and violent mergers \citep{pakmor--10, pakmor--11, kromer--13b}. For each scenario we considered a broad set of input parameters and generated 300\,000 synthetic spectra with \textsc{tardis} \citep{tardis} that were used to train neural networks. The neural networks act as emulators of the \textsc{tardis} radiative transfer simulations, predicting fluxes and flux uncertainties from 2\,000 -- 10\,000~\AA\, in a fraction of the time. Our emulators show typical accuracies of $\sim$1 -- 3~per~cent. Nested sampling is performed with \textsc{ultranest} \citep{ultranest} and the trained emulators to determine the best-fitting set of parameters for each explosion scenario. Comparing the overall likelihoods of different explosion scenarios also allows for a quantitative determination of the best-matching scenario for a given SN~Ia.

\par

Using template spectra generated following the same procedure as our training data, but not seen during training, we showed that \textsc{riddler} is able to accurately recover the correct input parameters and explosion mechanism of the template. We tested the impact of the signal-to-noise ratio and spectral resolution on our fits and found that signal-to-noise ratios $\gtrsim$10 and resolutions $\gtrsim$300 (at 6\,000~\AA) are required for robust results. Above these limits, fits showed no significant variation in the best-fitting parameters or explosion scenarios. 

\par

We applied \textsc{riddler} to fit observations of the normal SN~Ia SN~2011fe \citep{11fe--nature}, the SN~Iax SN~2005hk \citep{phillips--07}, and the peculiar, line-blanketed SN~Ia, SN~2018byg \citep{de--19}. We showed that the violent merger scenario produces the overall best agreement with SN~2011fe, but this requires a sub-luminous model and a distance modulus that is $\sim4\sigma$ below that reported by \cite{shappee--11} ($\mu = 28.27\pm0.02$ c.f. 29.04$\pm$0.19). Alternatively, using a fixed distance modulus of $\mu = 29.04$ and/or excluding the UV region from the fit favours the delayed detonation scenario overall and produces results comparable to previous analyses. Our fits to SN~2005hk are also consistent with previous works. Specifically, we find that the pure deflagration scenario produces the best agreement and indicates a somewhat lower kinetic energy for a given $^{56}$Ni mass is favoured. We argue that pure deflagration explosion simulations ignited closer to the centre of the white dwarf may produce similar ejecta structures to those favoured by our fits. For SN~2018byg, we find that the double detonation scenario broadly reproduces the strong line-blanketing observed, but our models show some disagreement with features at early times and the flux level around maximum light. 

\par

All fits performed with \textsc{riddler} are only valid given the assumed priors and likelihood function. Changes to either of these can significantly impact the best-fitting parameters, as demonstrated by our fits to SN~2011fe. Some explosion scenarios, in particular double detonations and violent mergers, are highly asymmetric, while our model spectra are based on angle-averaged ejecta properties. Although we include variation in ejecta properties in an effort to mimic these variations, the use of multi-dimensional models may impact the conclusions. In addition, although \textsc{riddler} provides a quantitative means with which to perform model comparisons, it does not measure the overall goodness of fit. Therefore the best-fitting model may in fact represent an overall poor fit to the data. Indeed, \textsc{riddler} does not find the `correct' explosion scenario for a given SN~Ia, but merely allows one to quantitatively determine which scenario among the training data provides the best match under the priors and likelihood function used. Typically, similar fits to SNe~Ia spectra are performed manually, using qualitative visual inspection to judge the fit quality. While automated fitting comes with a number of assumptions and limitations, these are generally also applicable to qualitative fitting. Automated fitting nevertheless has the benefit that the priors and likelihood function can be made explicit, and large samples of SNe~Ia can be treated in a systematic and reproducible way. 

\par

Automated fitting of SNe~Ia spectra similar to that shown here will play an increasingly important role within the coming years. Thousands of SNe~Ia spectra already exist, far exceeding our capability to model them individually with a qualitative approach. The time and computational expense of this method means that we must move to an automated approach in order to fully exploit the possibilities of these large datasets. In a companion paper, \citealt{magee--26b}, we use \textsc{riddler} to fit spectra of cosmological SNe~Ia observed as part of the Zwicky Transient Facility survey \citep{bellm--14, bellm--19, rigault--25}. Systematically applying \textsc{riddler} to new and upcoming surveys, such a TiDES on 4MOST \citep{tides, frohmaier--25}, will provide new constraints on thermonuclear explosion physics that have so far been out of reach.

%
%__________________________________________________________________________________________________________________________________________________________________________________________
%__________________________________________________________________________________________________________________________________________________________________________________________
%__________________________________________________________________________________________________________________________________________________________________________________________

\section*{Acknowledgements}

We thank the referee for their constructive comments that have helped to improve this work. We thank Conor Byrne, Madeleine Ginolin, Lisa Kelsey, Tom Killestein, Joe Lyman, Miika Pursiainen, and Danny Steeghs for useful discussion. MRM acknowledges a Warwick Astrophysics prize post-doctoral fellowship made possible thanks to a generous philanthropic donation. Computing facilities were provided by the Scientific Computing Research Technology Platform of the University of Warwick and the Queen's University Belfast HPC Kelvin cluster. This research made use of \textsc{Tardis}, a community-developed software package for spectral synthesis in supernovae \citep{tardis}. The development of \textsc{Tardis} received support from the Google Summer of Code initiative and from ESA's Summer of Code in Space program. \textsc{Tardis} makes extensive use of Astropy and PyNE. This work made use of the Heidelberg Supernova Model Archive (HESMA), https://hesma.h-its.org. We derive posterior probability distributions and the Bayesian
evidence with the nested sampling Monte Carlo algorithm
MLFriends (Buchner, 2014; 2019) using the
UltraNest\footnote{\url{https://johannesbuchner.github.io/UltraNest/}} package (Buchner 2021).

%%%%%%%%%%%%%%%%%%%%%%%%%%%%%%%%%%%%%%%%%%%%%%%%%%
\section*{Data Availability}

\texttt{riddler} is publicly available on GitHub\footnote{\href{https://github.com/MarkMageeAstro/riddler}{https://github.com/MarkMageeAstro/riddler}}.



\bibliographystyle{mnras}
\bibliography{mnras_template}

@ARTICLE{ashall--16,
       author = {{Ashall}, C. and {Mazzali}, P.~A. and {Pian}, E. and {James}, P.~A.},
        title = "{Abundance stratification in Type Ia supernovae - V. SN 1986G bridging the gap between normal and subluminous SNe Ia}",
      journal = {\mnras},
         year = 2016,
        month = dec,
       volume = {463},
       number = {2},
        pages = {1891-1906},
          doi = {10.1093/mnras/stw2114},
archivePrefix = {arXiv},
       eprint = {1608.05244},
 primaryClass = {astro-ph.HE},
       adsurl = {https://ui.adsabs.harvard.edu/abs/2016MNRAS.463.1891A}
}

@ARTICLE{badenas-agusti--24,
       author = {{Badenas-Agusti}, Mariona and {Via{\~n}a}, Javier and {Vanderburg}, Andrew and {Blouin}, Simon and {Dufour}, Patrick and {Xu}, Siyi and {Sha}, Lizhou},
        title = "{cecilia: a machine learning-based pipeline for measuring metal abundances of helium-rich polluted white dwarfs}",
      journal = {\mnras},
         year = 2024,
        month = apr,
       volume = {529},
       number = {2},
        pages = {1688-1714},
          doi = {10.1093/mnras/stae421},
archivePrefix = {arXiv},
       eprint = {2402.05176},
 primaryClass = {astro-ph.IM},
       adsurl = {https://ui.adsabs.harvard.edu/abs/2024MNRAS.529.1688B}
}

@ARTICLE{badenas-augusti--25,
       author = {{Badenas-Agusti}, Mariona and {Xu}, Siyi and {Vanderburg}, Andrew and {De}, Kishalay and {Dufour}, Patrick and {Rogers}, Laura K. and {Hoyos}, Susana and {Blouin}, Simon and {Via{\~n}a}, Javier and {Bonsor}, Amy and {Zuckerman}, Ben},
        title = "{A machine-learning compositional study of exoplanetary material accreted onto five helium-atmosphere white dwarfs with cecilia}",
      journal = {\mnras},
         year = 2025,
        month = jun,
       volume = {540},
       number = {1},
        pages = {746-773},
          doi = {10.1093/mnras/staf777},
archivePrefix = {arXiv},
       eprint = {2505.06228},
 primaryClass = {astro-ph.EP},
       adsurl = {https://ui.adsabs.harvard.edu/abs/2025MNRAS.540..746B}
}

@ARTICLE{barna--18,
   author = {{Barna}, B. and {Szalai}, T. and {Kerzendorf}, W.~E. and {Kromer}, M. and 
	{Sim}, S.~A. and {Magee}, M.~R. and {Leibundgut}, B.},
    title = "{Type Iax supernovae as a few-parameter family}",
  journal = {\mnras},
archivePrefix = "arXiv",
   eprint = {1808.00448},
 primaryClass = "astro-ph.SR",
     year = 2018,
    month = nov,
   volume = 480,
    pages = {3609-3627},
      doi = {10.1093/mnras/sty2065},
   adsurl = {http://adsabs.harvard.edu/abs/2018MNRAS.480.3609B}
}

@ARTICLE{baron--12,
       author = {{Baron}, E. and {H{\"o}flich}, P. and {Krisciunas}, K. and {Dominguez}, I. and {Khokhlov}, A.~M. and {Phillips}, M.~M. and {Suntzeff}, N. and {Wang}, L.},
        title = "{A Physical Model for SN 2001ay, a Normal, Bright, Extremely Slow Declining Type Ia Supernova}",
      journal = {\apj},
         year = 2012,
        month = jul,
       volume = {753},
       number = {2},
          eid = {105},
        pages = {105},
          doi = {10.1088/0004-637X/753/2/105},
archivePrefix = {arXiv},
       eprint = {1205.0814},
 primaryClass = {astro-ph.SR},
       adsurl = {https://ui.adsabs.harvard.edu/abs/2012ApJ...753..105B}
}

@INPROCEEDINGS{bellm--14,
   author = {{Bellm}, E.},
    title = "{The Zwicky Transient Facility}",
booktitle = {The Third Hot-wiring the Transient Universe Workshop},
     year = 2014,
archivePrefix = "arXiv",
   eprint = {1410.8185},
 primaryClass = "astro-ph.IM",
   editor = {{Wozniak}, P.~R. and {Graham}, M.~J. and {Mahabal}, A.~A. and 
	{Seaman}, R.},
    pages = {27-33},
   adsurl = {http://adsabs.harvard.edu/abs/2014htu..conf...27B}
}

@ARTICLE{bellm--19,
       author = {{Bellm}, Eric C. and {Kulkarni}, Shrinivas R. and {Graham}, Matthew J. and {Dekany}, Richard and {Smith}, Roger M. and {Riddle}, Reed and {Masci}, Frank J. and {Helou}, George and {Prince}, Thomas A. and {Adams}, Scott M. and {Barbarino}, C. and {Barlow}, Tom and {Bauer}, James and {Beck}, Ron and {Belicki}, Justin and {Biswas}, Rahul and {Blagorodnova}, Nadejda and {Bodewits}, Dennis and {Bolin}, Bryce and {Brinnel}, Valery and {Brooke}, Tim and {Bue}, Brian and {Bulla}, Mattia and {Burruss}, Rick and {Cenko}, S. Bradley and {Chang}, Chan-Kao and {Connolly}, Andrew and {Coughlin}, Michael and {Cromer}, John and {Cunningham}, Virginia and {De}, Kishalay and {Delacroix}, Alex and {Desai}, Vandana and {Duev}, Dmitry A. and {Eadie}, Gwendolyn and {Farnham}, Tony L. and {Feeney}, Michael and {Feindt}, Ulrich and {Flynn}, David and {Franckowiak}, Anna and {Frederick}, S. and {Fremling}, C. and {Gal-Yam}, Avishay and {Gezari}, Suvi and {Giomi}, Matteo and {Goldstein}, Daniel A. and {Golkhou}, V. Zach and {Goobar}, Ariel and {Groom}, Steven and {Hacopians}, Eugean and {Hale}, David and {Henning}, John and {Ho}, Anna Y.~Q. and {Hover}, David and {Howell}, Justin and {Hung}, Tiara and {Huppenkothen}, Daniela and {Imel}, David and {Ip}, Wing-Huen and {Ivezi{\'c}}, {\v{Z}}eljko and {Jackson}, Edward and {Jones}, Lynne and {Juric}, Mario and {Kasliwal}, Mansi M. and {Kaspi}, S. and {Kaye}, Stephen and {Kelley}, Michael S.~P. and {Kowalski}, Marek and {Kramer}, Emily and {Kupfer}, Thomas and {Landry}, Walter and {Laher}, Russ R. and {Lee}, Chien-De and {Lin}, Hsing Wen and {Lin}, Zhong-Yi and {Lunnan}, Ragnhild and {Giomi}, Matteo and {Mahabal}, Ashish and {Mao}, Peter and {Miller}, Adam A. and {Monkewitz}, Serge and {Murphy}, Patrick and {Ngeow}, Chow-Choong and {Nordin}, Jakob and {Nugent}, Peter and {Ofek}, Eran and {Patterson}, Maria T. and {Penprase}, Bryan and {Porter}, Michael and {Rauch}, Ludwig and {Rebbapragada}, Umaa and {Reiley}, Dan and {Rigault}, Mickael and {Rodriguez}, Hector and {van Roestel}, Jan and {Rusholme}, Ben and {van Santen}, Jakob and {Schulze}, S. and {Shupe}, David L. and {Singer}, Leo P. and {Soumagnac}, Maayane T. and {Stein}, Robert and {Surace}, Jason and {Sollerman}, Jesper and {Szkody}, Paula and {Taddia}, F. and {Terek}, Scott and {Van Sistine}, Angela and {van Velzen}, Sjoert and {Vestrand}, W. Thomas and {Walters}, Richard and {Ward}, Charlotte and {Ye}, Quan-Zhi and {Yu}, Po-Chieh and {Yan}, Lin and {Zolkower}, Jeffry},
        title = "{The Zwicky Transient Facility: System Overview, Performance, and First Results}",
      journal = {\pasp},
         year = 2019,
        month = jan,
       volume = {131},
       number = {995},
        pages = {018002},
          doi = {10.1088/1538-3873/aaecbe},
archivePrefix = {arXiv},
       eprint = {1902.01932},
 primaryClass = {astro-ph.IM},
       adsurl = {https://ui.adsabs.harvard.edu/abs/2019PASP..131a8002B}
}

@ARTICLE{bildsten--07,
   author = {{Bildsten}, L. and {Shen}, K.~J. and {Weinberg}, N.~N. and {Nelemans}, G.
	},
    title = "{Faint Thermonuclear Supernovae from AM Canum Venaticorum Binaries}",
  journal = {\apjl},
   eprint = {astro-ph/0703578},
     year = 2007,
    month = jun,
   volume = 662,
    pages = {L95-L98},
      doi = {10.1086/519489},
   adsurl = {http://adsabs.harvard.edu/abs/2007ApJ...662L..95B}
}

@ARTICLE{blondin--11,
       author = {{Blondin}, St{\'e}phane and {Kasen}, Daniel and {R{\"o}pke}, Friedrich K. and {Kirshner}, Robert P. and {Mandel}, Kaisey S.},
        title = "{Confronting 2D delayed-detonation models with light curves and spectra of Type Ia supernovae}",
      journal = {\mnras},
         year = 2011,
        month = oct,
       volume = {417},
       number = {2},
        pages = {1280-1302},
          doi = {10.1111/j.1365-2966.2011.19345.x},
archivePrefix = {arXiv},
       eprint = {1107.0009},
 primaryClass = {astro-ph.HE},
       adsurl = {https://ui.adsabs.harvard.edu/abs/2011MNRAS.417.1280B}
}

@ARTICLE{blondin--12,
   author = {{Blondin}, S. and {Matheson}, T. and {Kirshner}, R.~P. and {Mandel}, K.~S. and 
	{Berlind}, P. and {Calkins}, M. and {Challis}, P. and {Garnavich}, P.~M. and 
	{Jha}, S.~W. and {Modjaz}, M. and {Riess}, A.~G. and {Schmidt}, B.~P.
	},
    title = "{The Spectroscopic Diversity of Type Ia Supernovae}",
  journal = {\aj},
archivePrefix = "arXiv",
   eprint = {1203.4832},
 primaryClass = "astro-ph.SR",
     year = 2012,
    month = may,
   volume = 143,
      eid = {126},
    pages = {126},
      doi = {10.1088/0004-6256/143/5/126},
   adsurl = {http://adsabs.harvard.edu/abs/2012AJ....143..126B}
}

@ARTICLE{blondin--13,
       author = {{Blondin}, St{\'e}phane and {Dessart}, Luc and {Hillier}, D. John and {Khokhlov}, Alexei M.},
        title = "{One-dimensional delayed-detonation models of Type Ia supernovae: confrontation to observations at bolometric maximum}",
      journal = {\mnras},
         year = 2013,
        month = mar,
       volume = {429},
       number = {3},
        pages = {2127-2142},
          doi = {10.1093/mnras/sts484},
archivePrefix = {arXiv},
       eprint = {1211.5892},
 primaryClass = {astro-ph.SR},
       adsurl = {https://ui.adsabs.harvard.edu/abs/2013MNRAS.429.2127B}
}

@ARTICLE{blondin--22,
       author = {{Blondin}, St{\'e}phane and {Blinnikov}, Sergei and {Callan}, Fionntan P. and {Collins}, Christine E. and {Dessart}, Luc and {Even}, Wesley and {Fl{\"o}rs}, Andreas and {Fullard}, Andrew G. and {Hillier}, D. John and {Jerkstrand}, Anders and {Kasen}, Daniel and {Katz}, Boaz and {Kerzendorf}, Wolfgang and {Kozyreva}, Alexandra and {O'Brien}, Jack and {P{\'a}ssaro}, Ezequiel A. and {Roth}, Nathaniel and {Shen}, Ken J. and {Shingles}, Luke and {Sim}, Stuart A. and {Singhal}, Jaladh and {Smith}, Isaac G. and {Sorokina}, Elena and {Utrobin}, Victor P. and {Vogl}, Christian and {Williamson}, Marc and {Wollaeger}, Ryan and {Woosley}, Stan E. and {Wygoda}, Nahliel},
        title = "{StaNdaRT: a repository of standardised test models and outputs for supernova radiative transfer}",
      journal = {\aap},
         year = 2022,
        month = dec,
       volume = {668},
          eid = {A163},
        pages = {A163},
          doi = {10.1051/0004-6361/202244134},
archivePrefix = {arXiv},
       eprint = {2209.11671},
 primaryClass = {astro-ph.SR},
       adsurl = {https://ui.adsabs.harvard.edu/abs/2022A&A...668A.163B}
}

@ARTICLE{branch--06,
       author = {{Branch}, David and {Dang}, Leeann Chau and {Hall}, Nicholas and {Ketchum}, Wesley and {Melakayil}, Mercy and {Parrent}, Jerod and {Troxel}, M.~A. and {Casebeer}, D. and {Jeffery}, David J. and {Baron}, E.},
        title = "{Comparative Direct Analysis of Type Ia Supernova Spectra. II. Maximum Light}",
      journal = {\pasp},
         year = 2006,
        month = apr,
       volume = {118},
       number = {842},
        pages = {560-571},
          doi = {10.1086/502778},
archivePrefix = {arXiv},
       eprint = {astro-ph/0601048},
 primaryClass = {astro-ph},
       adsurl = {https://ui.adsabs.harvard.edu/abs/2006PASP..118..560B}
}

@ARTICLE{ultranest,
       author = {{Buchner}, Johannes},
        title = "{UltraNest - a robust, general purpose Bayesian inference engine}",
      journal = {The Journal of Open Source Software},
         year = 2021,
        month = apr,
       volume = {6},
       number = {60},
          eid = {3001},
        pages = {3001},
          doi = {10.21105/joss.03001},
archivePrefix = {arXiv},
       eprint = {2101.09604},
 primaryClass = {stat.CO},
       adsurl = {https://ui.adsabs.harvard.edu/abs/2021JOSS....6.3001B}
}

@ARTICLE{bulla--16,
       author = {{Bulla}, M. and {Sim}, S.~A. and {Kromer}, M. and {Seitenzahl}, I.~R. and {Fink}, M. and {Ciaraldi-Schoolmann}, F. and {R{\"o}pke}, F.~K. and {Hillebrandt}, W. and {Pakmor}, R. and {Ruiter}, A.~J. and {Taubenberger}, S.},
        title = "{Predicting polarization signatures for double-detonation and delayed-detonation models of Type Ia supernovae}",
      journal = {\mnras},
         year = 2016,
        month = oct,
       volume = {462},
       number = {1},
        pages = {1039-1056},
          doi = {10.1093/mnras/stw1733},
archivePrefix = {arXiv},
       eprint = {1607.04081},
 primaryClass = {astro-ph.HE},
       adsurl = {https://ui.adsabs.harvard.edu/abs/2016MNRAS.462.1039B}
}

@ARTICLE{camacho-neves--23,
       author = {{Camacho-Neves}, Yssavo and {Jha}, Saurabh W. and {Barna}, Barnabas and {Dai}, Mi and {Filippenko}, Alexei V. and {Foley}, Ryan J. and {Hosseinzadeh}, Griffin and {Howell}, D. Andrew and {Johansson}, Joel and {Kelly}, Patrick L. and {Kerzendorf}, Wolfgang E. and {Kwok}, Lindsey A. and {Larison}, Conor and {Magee}, Mark R. and {McCully}, Curtis and {O'Brien}, John T. and {Pan}, Yen-Chen and {Pandya}, Viraj and {Singhal}, Jaladh and {Stahl}, Benjamin E. and {Szalai}, Tam{\'a}s and {Wieber}, Meredith and {Williamson}, Marc},
        title = "{Over 500 Days in the Life of the Photosphere of the Type Iax Supernova SN 2014dt}",
      journal = {\apj},
         year = 2023,
        month = jul,
       volume = {951},
       number = {1},
          eid = {67},
        pages = {67},
          doi = {10.3847/1538-4357/acd558},
archivePrefix = {arXiv},
       eprint = {2302.03105},
 primaryClass = {astro-ph.HE},
       adsurl = {https://ui.adsabs.harvard.edu/abs/2023ApJ...951...67C}
}

@ARTICLE{chen--20,
       author = {{Chen}, Xingzhuo and {Hu}, Lei and {Wang}, Lifan},
        title = "{Artificial Intelligence-Assisted Inversion (AIAI) of Synthetic Type Ia Supernova Spectra}",
      journal = {\apjs},
         year = 2020,
        month = sep,
       volume = {250},
       number = {1},
          eid = {12},
        pages = {12},
          doi = {10.3847/1538-4365/ab9a3b},
archivePrefix = {arXiv},
       eprint = {1911.05209},
 primaryClass = {astro-ph.HE},
       adsurl = {https://ui.adsabs.harvard.edu/abs/2020ApJS..250...12C}
}

@ARTICLE{chen--24,
       author = {{Chen}, Xingzhuo and {Wang}, Lifan and {Hu}, Lei and {Brown}, Peter J.},
        title = "{Artificial Intelligence Assisted Inversion (AIAI): Quantifying the Spectral Features of $^{56}$Ni of Type Ia Supernovae}",
      journal = {\apj},
         year = 2024,
        month = feb,
       volume = {962},
       number = {2},
          eid = {125},
        pages = {125},
          doi = {10.3847/1538-4357/ad0a33},
archivePrefix = {arXiv},
       eprint = {2210.15892},
 primaryClass = {astro-ph.HE},
       adsurl = {https://ui.adsabs.harvard.edu/abs/2024ApJ...962..125C}
}

@ARTICLE{collins--22,
       author = {{Collins}, Christine E. and {Gronow}, Sabrina and {Sim}, Stuart A. and {R{\"o}pke}, Friedrich K.},
        title = "{Double detonations: variations in Type Ia supernovae due to different core and He shell masses - II. Synthetic observables}",
      journal = {\mnras},
         year = 2022,
        month = dec,
       volume = {517},
       number = {4},
        pages = {5289-5302},
          doi = {10.1093/mnras/stac2665},
archivePrefix = {arXiv},
       eprint = {2209.04305},
 primaryClass = {astro-ph.SR},
       adsurl = {https://ui.adsabs.harvard.edu/abs/2022MNRAS.517.5289C}
}

@ARTICLE{collins--25,
       author = {{Collins}, Christine E. and {Shingles}, Luke J. and {Sim}, Stuart A. and {Callan}, Fionntan P. and {Gronow}, Sabrina and {Hillebrandt}, Wolfgang and {Kromer}, Markus and {Pakmor}, R{\"u}diger and {R{\"o}pke}, Friedrich K.},
        title = "{Non-LTE radiative transfer simulations: improved agreement of the double detonation with normal Type Ia supernovae}",
      journal = {\mnras},
         year = 2025,
        month = apr,
       volume = {538},
       number = {3},
        pages = {1289-1300},
          doi = {10.1093/mnras/staf261},
archivePrefix = {arXiv},
       eprint = {2411.11643},
 primaryClass = {astro-ph.SR},
       adsurl = {https://ui.adsabs.harvard.edu/abs/2025MNRAS.538.1289C}
}

@ARTICLE{czekala--15,
       author = {{Czekala}, Ian and {Andrews}, Sean M. and {Mandel}, Kaisey S. and {Hogg}, David W. and {Green}, Gregory M.},
        title = "{Constructing a Flexible Likelihood Function for Spectroscopic Inference}",
      journal = {\apj},
         year = 2015,
        month = oct,
       volume = {812},
       number = {2},
          eid = {128},
        pages = {128},
          doi = {10.1088/0004-637X/812/2/128},
archivePrefix = {arXiv},
       eprint = {1412.5177},
 primaryClass = {astro-ph.SR},
       adsurl = {https://ui.adsabs.harvard.edu/abs/2015ApJ...812..128C}
}

@ARTICLE{de--19,
       author = {{De}, Kishalay and {Kasliwal}, Mansi M. and {Polin}, Abigail and
         {Nugent}, Peter E. and {Bildsten}, Lars and {Adams}, Scott M. and
         {Bellm}, Eric C. and {Blagorodnova}, Nadia and {Burdge}, Kevin B. and
         {Cannella}, Christopher and {Cenko}, S. Bradley and
         {Dekany}, Richard G. and {Feeney}, Michael and {Hale}, David and
         {Fremling}, U. Christoffer and {Graham}, Matthew J. and
         {Ho}, Anna Y.~Q. and {Jencson}, Jacob E. and {Kulkarni}, S.~R. and
         {Laher}, Russ R. and {Masci}, Frank J. and {Miller}, Adam A. and
         {Patterson}, Maria T. and {Rebbapragada}, Umaa and {Riddle}, Reed L. and
         {Shupe}, David L. and {Smith}, Roger M.},
        title = "{ZTF 18aaqeasu (SN2018byg): A Massive Helium-shell Double Detonation on a Sub-Chandrasekhar-mass White Dwarf}",
      journal = {\apjl},
         year = 2019,
        month = mar,
       volume = {873},
       number = {2},
          eid = {L18},
        pages = {L18},
          doi = {10.3847/2041-8213/ab0aec},
archivePrefix = {arXiv},
       eprint = {1901.00874},
 primaryClass = {astro-ph.HE},
       adsurl = {https://ui.adsabs.harvard.edu/abs/2019ApJ...873L..18D}
}

@ARTICLE{dessart--14a,
   author = {{Dessart}, L. and {Blondin}, S. and {Hillier}, D.~J. and {Khokhlov}, A.
	},
    title = "{Constraints on the explosion mechanism and progenitors of Type Ia supernovae}",
  journal = {\mnras},
archivePrefix = "arXiv",
   eprint = {1310.7747},
 primaryClass = "astro-ph.SR",
     year = 2014,
    month = jun,
   volume = 441,
    pages = {532-550},
      doi = {10.1093/mnras/stu598},
   adsurl = {http://adsabs.harvard.edu/abs/2014MNRAS.441..532D}
}

@ARTICLE{dimitriadis--17,
       author = {{Dimitriadis}, G. and {Sullivan}, M. and {Kerzendorf}, W. and {Ruiter}, A.~J. and {Seitenzahl}, I.~R. and {Taubenberger}, S. and {Doran}, G.~B. and {Gal-Yam}, A. and {Laher}, R.~R. and {Maguire}, K. and {Nugent}, P. and {Ofek}, E.~O. and {Surace}, J.},
        title = "{The late-time light curve of the Type Ia supernova SN 2011fe}",
      journal = {\mnras},
         year = 2017,
        month = jul,
       volume = {468},
       number = {4},
        pages = {3798-3812},
          doi = {10.1093/mnras/stx683},
archivePrefix = {arXiv},
       eprint = {1701.07267},
 primaryClass = {astro-ph.HE},
       adsurl = {https://ui.adsabs.harvard.edu/abs/2017MNRAS.468.3798D}
}

@ARTICLE{dimitriadis--23,
       author = {{Dimitriadis}, Georgios and {Maguire}, Kate and {Karambelkar}, Viraj R. and {Lebron}, Ryan J. and {Liu}, Chang and {Kozyreva}, Alexandra and {Miller}, Adam A. and {Ridden-Harper}, Ryan and {Anderson}, Joseph P. and {Chen}, Ting-Wan and {Coughlin}, Michael and {Della Valle}, Massimo and {Drake}, Andrew and {Galbany}, Llu{\'\i}s and {Gromadzki}, Mariusz and {Groom}, Steven L. and {Guti{\'e}rrez}, Claudia P. and {Ihanec}, Nada and {Inserra}, Cosimo and {Johansson}, Joel and {M{\"u}ller-Bravo}, Tom{\'a}s E. and {Nicholl}, Matt and {Polin}, Abigail and {Rusholme}, Ben and {Schulze}, Steve and {Sollerman}, Jesper and {Srivastav}, Shubham and {Taggart}, Kirsty and {Wang}, Qinan and {Yang}, Yi and {Young}, David R.},
        title = "{SN 2021zny: an early flux excess combined with late-time oxygen emission suggests a double white dwarf merger event}",
      journal = {\mnras},
         year = 2023,
        month = may,
       volume = {521},
       number = {1},
        pages = {1162-1183},
          doi = {10.1093/mnras/stad536},
archivePrefix = {arXiv},
       eprint = {2302.08228},
 primaryClass = {astro-ph.HE},
       adsurl = {https://ui.adsabs.harvard.edu/abs/2023MNRAS.521.1162D}
}

@ARTICLE{dong--22,
       author = {{Dong}, Yize and {Valenti}, Stefano and {Polin}, Abigail and {Boyle}, Aoife and {Fl{\"o}rs}, Andreas and {Vogl}, Christian and {Kerzendorf}, Wolfgang E. and {Sand}, David J. and {Jha}, Saurabh W. and {Wyrzykowski}, {\L}ukasz and {Bostroem}, K. Azalee and {Pearson}, Jeniveve and {McCully}, Curtis and {Andrews}, Jennifer E. and {Benetti}, Stefano and {Blondin}, St{\'e}phane and {Galbany}, L. and {Gromadzki}, Mariusz and {Hosseinzadeh}, Griffin and {Howell}, D. Andrew and {Inserra}, Cosimo and {Jencson}, Jacob E. and {Lundquist}, Michael and {Lyman}, J.~D. and {Magee}, Mark and {Maguire}, Kate and {Meza}, Nicolas and {Srivastav}, Shubham and {Taubenberger}, Stefan and {Terwel}, J.~H. and {Wyatt}, Samuel and {Young}, D.~R.},
        title = "{SN 2016dsg: A Thermonuclear Explosion Involving a Thick Helium Shell}",
      journal = {\apj},
         year = 2022,
        month = aug,
       volume = {934},
       number = {2},
          eid = {102},
        pages = {102},
          doi = {10.3847/1538-4357/ac75eb},
archivePrefix = {arXiv},
       eprint = {2206.07065},
 primaryClass = {astro-ph.SR},
       adsurl = {https://ui.adsabs.harvard.edu/abs/2022ApJ...934..102D}
}

@ARTICLE{ergon--18,
       author = {{Ergon}, M. and {Fransson}, C. and {Jerkstrand}, A. and {Kozma}, C. and {Kromer}, M. and {Spricer}, K.},
        title = "{Monte-Carlo methods for NLTE spectral synthesis of supernovae}",
      journal = {\aap},
         year = 2018,
        month = dec,
       volume = {620},
          eid = {A156},
        pages = {A156},
          doi = {10.1051/0004-6361/201833043},
archivePrefix = {arXiv},
       eprint = {1810.07165},
 primaryClass = {astro-ph.SR},
       adsurl = {https://ui.adsabs.harvard.edu/abs/2018A&A...620A.156E}
}

@ARTICLE{91t--disc,
   author = {{Filippenko}, A.~V. and {Richmond}, M.~W. and {Matheson}, T. and 
	{Shields}, J.~C. and {Burbidge}, E.~M. and {Cohen}, R.~D. and 
	{Dickinson}, M. and {Malkan}, M.~A. and {Nelson}, B. and {Pietz}, J. and 
	{Schlegel}, D. and {Schmeer}, P. and {Spinrad}, H. and {Steidel}, C.~C. and 
	{Tran}, H.~D. and {Wren}, W.},
    title = "{The peculiar Type IA SN 1991T - Detonation of a white dwarf?}",
  journal = {\apjl},
     year = 1992,
    month = jan,
   volume = 384,
    pages = {L15-L18},
      doi = {10.1086/186252},
   adsurl = {http://adsabs.harvard.edu/abs/1992ApJ...384L..15F}
}

@ARTICLE{91bg--disc,
   author = {{Filippenko}, A.~V. and {Richmond}, M.~W. and {Branch}, D. and 
	{Gaskell}, M. and {Herbst}, W. and {Ford}, C.~H. and {Treffers}, R.~R. and 
	{Matheson}, T. and {Ho}, L.~C. and {Dey}, A. and {Sargent}, W.~L.~W. and 
	{Small}, T.~A. and {van Breugel}, W.~J.~M.},
    title = "{The subluminous, spectroscopically peculiar type IA supernova 1991bg in the elliptical galaxy NGC 4374}",
  journal = {\aj},
     year = 1992,
    month = oct,
   volume = 104,
    pages = {1543-1556},
      doi = {10.1086/116339},
   adsurl = {http://adsabs.harvard.edu/abs/1992AJ....104.1543F}
}

@ARTICLE{fink--10,
   author = {{Fink}, M. and {R{\"o}pke}, F.~K. and {Hillebrandt}, W. and 
	{Seitenzahl}, I.~R. and {Sim}, S.~A. and {Kromer}, M.},
    title = "{Double-detonation sub-Chandrasekhar supernovae: can minimum helium shell masses detonate the core?}",
  journal = {\aap},
archivePrefix = "arXiv",
   eprint = {1002.2173},
 primaryClass = "astro-ph.SR",
     year = 2010,
    month = may,
   volume = 514,
      eid = {A53},
    pages = {A53},
      doi = {10.1051/0004-6361/200913892},
   adsurl = {http://adsabs.harvard.edu/abs/2010A\%26A...514A..53F}
}

@ARTICLE{fink-2014,
   author = {{Fink}, M. and {Kromer}, M. and {Seitenzahl}, I.~R. and {Ciaraldi-Schoolmann}, F. and 
	{R{\"o}pke}, F.~K. and {Sim}, S.~A. and {Pakmor}, R. and {Ruiter}, A.~J. and 
	{Hillebrandt}, W.},
    title = "{Three-dimensional pure deflagration models with nucleosynthesis and synthetic observables for Type Ia supernovae}",
  journal = {\mnras},
archivePrefix = "arXiv",
   eprint = {1308.3257},
 primaryClass = "astro-ph.SR",
     year = 2014,
    month = feb,
   volume = 438,
    pages = {1762-1783},
      doi = {10.1093/mnras/stt2315},
   adsurl = {http://adsabs.harvard.edu/abs/2014MNRAS.438.1762F}
}

@ARTICLE{frohmaier--25,
       author = {{Frohmaier}, C. and {Vincenzi}, M. and {Sullivan}, M. and {H{\"o}nig}, S.~F. and {Smith}, M. and {Addison}, H. and {Collett}, T. and {Dimitriadis}, G. and {Ellis}, R.~S. and {Gandhi}, P. and {Graur}, O. and {Hook}, I. and {Kelsey}, L. and {Kim}, Y.-L. and {Lidman}, C. and {Maguire}, K. and {Makrygianni}, L. and {Martin}, B. and {M{\"o}ller}, A. and {Nichol}, R.~C. and {Nicholl}, M. and {Schady}, P. and {Simmons}, B.~D. and {Smartt}, S.~J. and {Tempel}, E. and {Wiseman}, P. and {the LSST Dark Energy Science Collaboration}},
        title = "{TiDES: The 4MOST Time Domain Extragalactic Survey}",
      journal = {\apj},
         year = 2025,
        month = oct,
       volume = {992},
       number = {1},
          eid = {158},
        pages = {158},
          doi = {10.3847/1538-4357/adff4e},
archivePrefix = {arXiv},
       eprint = {2501.16311},
 primaryClass = {astro-ph.HE},
       adsurl = {https://ui.adsabs.harvard.edu/abs/2025ApJ...992..158F}
}

@ARTICLE{galbanu--19,
       author = {{Galbany}, L. and {Ashall}, C. and {H{\"o}flich}, P. and {Gonz{\'a}lez-Gait{\'a}n}, S. and {Taubenberger}, S. and {Stritzinger}, M. and {Hsiao}, E.~Y. and {Mazzali}, P. and {Baron}, E. and {Blondin}, S. and {Bose}, S. and {Bulla}, M. and {Burke}, J.~F. and {Burns}, C.~R. and {Cartier}, R. and {Chen}, P. and {Della Valle}, M. and {Diamond}, T.~R. and {Guti{\'e}rrez}, C.~P. and {Harmanen}, J. and {Hiramatsu}, D. and {Holoien}, T.~W.-S. and {Hosseinzadeh}, G. and {Howell}, D. Andrew and {Huang}, Y. and {Inserra}, C. and {de Jaeger}, T. and {Jha}, S.~W. and {Kangas}, T. and {Kromer}, M. and {Lyman}, J.~D. and {Maguire}, K. and {Marion}, G. Howie and {Milisavljevic}, D. and {Prentice}, S.~J. and {Razza}, A. and {Reynolds}, T.~M. and {Sand}, D.~J. and {Shappee}, B.~J. and {Shekhar}, R. and {Smartt}, S.~J. and {Stassun}, K.~G. and {Sullivan}, M. and {Valenti}, S. and {Villanueva}, S. and {Wang}, X. and {Wheeler}, J. Craig and {Zhai}, Q. and {Zhang}, J.},
        title = "{Evidence for a Chandrasekhar-mass explosion in the Ca-strong 1991bg-like type Ia supernova 2016hnk}",
      journal = {\aap},
         year = 2019,
        month = oct,
       volume = {630},
          eid = {A76},
        pages = {A76},
          doi = {10.1051/0004-6361/201935537},
archivePrefix = {arXiv},
       eprint = {1904.10034},
 primaryClass = {astro-ph.SR},
       adsurl = {https://ui.adsabs.harvard.edu/abs/2019A&A...630A..76G}
}

@ARTICLE{gamezo--03,
       author = {{Gamezo}, Vadim N. and {Khokhlov}, Alexei M. and {Oran}, Elaine S. and {Chtchelkanova}, Almadena Y. and {Rosenberg}, Robert O.},
        title = "{Thermonuclear Supernovae: Simulations of the Deflagration Stage and Their Implications}",
      journal = {Science},
         year = 2003,
        month = jan,
       volume = {299},
       number = {5603},
        pages = {77-81},
          doi = {10.1126/science.299.5603.77},
archivePrefix = {arXiv},
       eprint = {astro-ph/0212054},
 primaryClass = {astro-ph},
       adsurl = {https://ui.adsabs.harvard.edu/abs/2003Sci...299...77G}
}

@ARTICLE{gamezo--04,
       author = {{Gamezo}, Vadim N. and {Khokhlov}, Alexei M. and {Oran}, Elaine S.},
        title = "{Deflagrations and Detonations in Thermonuclear Supernovae}",
      journal = {\prl},
         year = 2004,
        month = may,
       volume = {92},
       number = {21},
          eid = {211102},
        pages = {211102},
          doi = {10.1103/PhysRevLett.92.211102},
archivePrefix = {arXiv},
       eprint = {astro-ph/0406101},
 primaryClass = {astro-ph},
       adsurl = {https://ui.adsabs.harvard.edu/abs/2004PhRvL..92u1102G}
}

@ARTICLE{gamezo--05,
       author = {{Gamezo}, Vadim N. and {Khokhlov}, Alexei M. and {Oran}, Elaine S.},
        title = "{Three-dimensional Delayed-Detonation Model of Type Ia Supernovae}",
      journal = {\apj},
         year = 2005,
        month = apr,
       volume = {623},
       number = {1},
        pages = {337-346},
          doi = {10.1086/428767},
archivePrefix = {arXiv},
       eprint = {astro-ph/0409598},
 primaryClass = {astro-ph},
       adsurl = {https://ui.adsabs.harvard.edu/abs/2005ApJ...623..337G}
}

@ARTICLE{gronow--20,
       author = {{Gronow}, Sabrina and {Collins}, Christine and {Ohlmann}, Sebastian T. and
         {Pakmor}, R{\"u}diger and {Kromer}, Markus and {Seitenzahl}, Ivo R. and
         {Sim}, Stuart A. and {R{\"o}pke}, Friedrich K.},
        title = "{SNe Ia from double detonations: Impact of core-shell mixing on the carbon ignition mechanism}",
      journal = {\aap},
         year = 2020,
        month = mar,
       volume = {635},
          eid = {A169},
        pages = {A169},
          doi = {10.1051/0004-6361/201936494},
archivePrefix = {arXiv},
       eprint = {2002.00981},
 primaryClass = {astro-ph.SR},
       adsurl = {https://ui.adsabs.harvard.edu/abs/2020A&A...635A.169G}
}

@ARTICLE{gronow--21,
       author = {{Gronow}, Sabrina and {Collins}, Christine E. and {Sim}, Stuart A. and {R{\"o}pke}, Friedrich K.},
        title = "{Double detonations of sub-M$_{Ch}$ CO white dwarfs: variations in Type Ia supernovae due to different core and He shell masses}",
      journal = {\aap},
         year = 2021,
        month = may,
       volume = {649},
          eid = {A155},
        pages = {A155},
          doi = {10.1051/0004-6361/202039954},
archivePrefix = {arXiv},
       eprint = {2102.06719},
 primaryClass = {astro-ph.SR},
       adsurl = {https://ui.adsabs.harvard.edu/abs/2021A&A...649A.155G}
}

@ARTICLE{gronow--21b,
       author = {{Gronow}, Sabrina and {C{\^o}t{\'e}}, Benoit and {Lach}, Florian and {Seitenzahl}, Ivo R. and {Collins}, Christine E. and {Sim}, Stuart A. and {R{\"o}pke}, Friedrich K.},
        title = "{Metallicity-dependent nucleosynthetic yields of Type Ia supernovae originating from double detonations of sub-M$_{Ch}$ white dwarfs}",
      journal = {\aap},
         year = 2021,
        month = dec,
       volume = {656},
          eid = {A94},
        pages = {A94},
          doi = {10.1051/0004-6361/202140881},
archivePrefix = {arXiv},
       eprint = {2103.14050},
 primaryClass = {astro-ph.HE},
       adsurl = {https://ui.adsabs.harvard.edu/abs/2021A&A...656A..94G}
}

@ARTICLE{hachinger--09,
   author = {{Hachinger}, S. and {Mazzali}, P.~A. and {Taubenberger}, S. and 
	{Pakmor}, R. and {Hillebrandt}, W.},
    title = "{Spectral analysis of the 91bg-like Type Ia SN 2005bl: low luminosity, low velocities, incomplete burning}",
  journal = {\mnras},
archivePrefix = "arXiv",
   eprint = {0907.2542},
 primaryClass = "astro-ph.HE",
     year = 2009,
    month = nov,
   volume = 399,
    pages = {1238-1254},
      doi = {10.1111/j.1365-2966.2009.15403.x},
   adsurl = {http://adsabs.harvard.edu/abs/2009MNRAS.399.1238H}
}

@ARTICLE{harvey--23,
       author = {{Harvey}, L. and {Maguire}, K. and {Magee}, M.~R. and {Bulla}, M. and {Dhawan}, S. and {Schulze}, S. and {Sollerman}, J. and {Deckers}, M. and {Dimitriadis}, G. and {Reusch}, S. and {Smith}, M. and {Terwel}, J. and {Coughlin}, M.~W. and {Masci}, F. and {Purdum}, J. and {Reedy}, A. and {Robert}, E. and {Wold}, A.},
        title = "{Early-time spectroscopic modelling of the transitional Type Ia Supernova 2021rhu with TARDIS}",
      journal = {\mnras},
         year = 2023,
        month = jul,
       volume = {522},
       number = {3},
        pages = {4444-4467},
          doi = {10.1093/mnras/stad1226},
archivePrefix = {arXiv},
       eprint = {2304.10129},
 primaryClass = {astro-ph.HE},
       adsurl = {https://ui.adsabs.harvard.edu/abs/2023MNRAS.522.4444H}
}

@ARTICLE{harvey--26,
       author = {{Harvey}, L. and {Maguire}, K. and {Holas}, A. and {Anderson}, J.~P. and {Chen}, T.-W. and {Galbany}, L. and {Gonz{\'a}lez-Gait{\'a}n}, S. and {Gromadzki}, M. and {M{\"u}ller-Bravo}, T.~E. and {Pignata}, G. and {Seitenzahl}, I.~R.},
        title = "{Exploring the origins of high-velocity features in SNe Ia with the spectral synthesis code TARDIS}",
      journal = {\aap},
         year = 2026,
        month = feb,
       volume = {707},
          eid = {A57},
        pages = {A57},
          doi = {10.1051/0004-6361/202555042},
archivePrefix = {arXiv},
       eprint = {2512.13791},
 primaryClass = {astro-ph.HE},
       adsurl = {https://ui.adsabs.harvard.edu/abs/2026A&A...707A..57H}
}

@ARTICLE{heringer--17,
       author = {{Heringer}, E. and {van Kerkwijk}, M.~H. and {Sim}, S.~A. and {Kerzendorf}, W.~E.},
        title = "{Spectral Sequences of Type Ia Supernovae. I. Connecting Normal and Subluminous SNe Ia and the Presence of Unburned Carbon}",
      journal = {\apj},
         year = 2017,
        month = sep,
       volume = {846},
       number = {1},
          eid = {15},
        pages = {15},
          doi = {10.3847/1538-4357/aa8309},
archivePrefix = {arXiv},
       eprint = {1707.08572},
 primaryClass = {astro-ph.HE},
       adsurl = {https://ui.adsabs.harvard.edu/abs/2017ApJ...846...15H}
}

@ARTICLE{heringer--19,
       author = {{Heringer}, E. and {van Kerkwijk}, M.~H. and {Sim}, S.~A. and {Kerzendorf}, W.~E. and {Graham}, Melissa L.},
        title = "{Spectral Sequences of Type Ia Supernovae. II. Carbon as a Diagnostic Tool for Explosion Mechanisms}",
      journal = {\apj},
         year = 2019,
        month = feb,
       volume = {871},
       number = {2},
          eid = {250},
        pages = {250},
          doi = {10.3847/1538-4357/aafad5},
archivePrefix = {arXiv},
       eprint = {1902.01904},
 primaryClass = {astro-ph.HE},
       adsurl = {https://ui.adsabs.harvard.edu/abs/2019ApJ...871..250H}
}

@ARTICLE{hillier--2012,
   author = {{Hillier}, D.~J. and {Dessart}, L.},
    title = "{Time-dependent radiative transfer calculations for supernovae}",
  journal = {\mnras},
archivePrefix = "arXiv",
   eprint = {1204.0527},
 primaryClass = "astro-ph.SR",
     year = 2012,
    month = jul,
   volume = 424,
    pages = {252-271},
      doi = {10.1111/j.1365-2966.2012.21192.x},
   adsurl = {http://adsabs.harvard.edu/abs/2012MNRAS.424..252H}
}

@ARTICLE{hoeflich-95,
   author = {{Hoeflich}, P. and {Khokhlov}, A.~M. and {Wheeler}, J.~C.},
    title = "{Delayed detonation models for normal and subluminous type IA sueprnovae: Absolute brightness, light curves, and molecule formation}",
  journal = {\apj},
     year = 1995,
    month = may,
   volume = 444,
    pages = {831-847},
      doi = {10.1086/175656},
   adsurl = {http://adsabs.harvard.edu/abs/1995ApJ...444..831H}
}

@INPROCEEDINGS{hoeflich--2003a,
   author = {{H{\"o}flich}, P.},
    title = "{ALI in Rapidly Expanding Envelopes}",
booktitle = {Stellar Atmosphere Modeling},
     year = 2003,
   series = {Astronomical Society of the Pacific Conference Series},
   volume = 288,
   editor = {{Hubeny}, I. and {Mihalas}, D. and {Werner}, K.},
    month = jan,
    pages = {185},
   adsurl = {http://adsabs.harvard.edu/abs/2003ASPC..288..185H}
}

@ARTICLE{hoeflich--2017,
   author = {{Hoeflich}, P. and {Hsiao}, E.~Y. and {Ashall}, C. and {Burns}, C.~R. and 
	{Diamond}, T.~R. and {Phillips}, M.~M. and {Sand}, D. and {Stritzinger}, M.~D. and 
	{Suntzeff}, N. and {Contreras}, C. and {Krisciunas}, K. and 
	{Morrell}, N. and {Wang}, L.},
    title = "{Light and Color Curve Properties of Type Ia Supernovae: Theory Versus Observations}",
  journal = {\apj},
archivePrefix = "arXiv",
   eprint = {1707.05350},
 primaryClass = "astro-ph.SR",
     year = 2017,
    month = sep,
   volume = 846,
      eid = {58},
    pages = {58},
      doi = {10.3847/1538-4357/aa84b2},
   adsurl = {http://adsabs.harvard.edu/abs/2017ApJ...846...58H}
}

@ARTICLE{iben--84,
   author = {{Iben}, Jr., I. and {Tutukov}, A.~V.},
    title = "{Supernovae of type I as end products of the evolution of binaries with components of moderate initial mass (M not greater than about 9 solar masses)}",
  journal = {\apjs},
     year = 1984,
    month = feb,
   volume = 54,
    pages = {335-372},
      doi = {10.1086/190932},
   adsurl = {http://adsabs.harvard.edu/abs/1984ApJS...54..335I}
}

@ARTICLE{iwamoto--99,
       author = {{Iwamoto}, Koichi and {Brachwitz}, Franziska and {Nomoto}, Ken'ICHI and {Kishimoto}, Nobuhiro and {Umeda}, Hideyuki and {Hix}, W. Raphael and {Thielemann}, Friedrich-Karl},
        title = "{Nucleosynthesis in Chandrasekhar Mass Models for Type IA Supernovae and Constraints on Progenitor Systems and Burning-Front Propagation}",
      journal = {\apjs},
         year = 1999,
        month = dec,
       volume = {125},
       number = {2},
        pages = {439-462},
          doi = {10.1086/313278},
archivePrefix = {arXiv},
       eprint = {astro-ph/0002337},
 primaryClass = {astro-ph},
       adsurl = {https://ui.adsabs.harvard.edu/abs/1999ApJS..125..439I}
}

@ARTICLE{jacobson-galan--20,
       author = {{Jacobson-Gal{\'a}n}, Wynn V. and {Polin}, Abigail and {Foley}, Ryan J. and {Dimitriadis}, Georgios and {Kilpatrick}, Charles D. and {Margutti}, Raffaella and {Coulter}, David A. and {Jha}, Saurabh W. and {Jones}, David O. and {Kirshner}, Robert P. and {Pan}, Yen-Chen and {Piro}, Anthony L. and {Rest}, Armin and {Rojas-Bravo}, C{\'e}sar},
        title = "{Ca hnk: The Calcium-rich Transient Supernova 2016hnk from a Helium Shell Detonation of a Sub-Chandrasekhar White Dwarf}",
      journal = {\apj},
         year = 2020,
        month = jun,
       volume = {896},
       number = {2},
          eid = {165},
        pages = {165},
          doi = {10.3847/1538-4357/ab94b8},
archivePrefix = {arXiv},
       eprint = {1910.05436},
 primaryClass = {astro-ph.HE},
       adsurl = {https://ui.adsabs.harvard.edu/abs/2020ApJ...896..165J}
}

@ARTICLE{jiang--2017,
   author = {{Jiang}, J.-A. and {Doi}, M. and {Maeda}, K. and {Shigeyama}, T. and 
	{Nomoto}, K. and {Yasuda}, N. and {Jha}, S.~W. and {Tanaka}, M. and 
	{Morokuma}, T. and {Tominaga}, N. and {Ivezi{\'c}}, {\v Z}. and 
	{Ruiz-Lapuente}, P. and {Stritzinger}, M.~D. and {Mazzali}, P.~A. and 
	{Ashall}, C. and {Mould}, J. and {Baade}, D. and {Suzuki}, N. and 
	{Connolly}, A.~J. and {Patat}, F. and {Wang}, L. and {Yoachim}, P. and 
	{Jones}, D. and {Furusawa}, H. and {Miyazaki}, S.},
    title = "{A hybrid type Ia supernova with an early flash triggered by helium-shell detonation}",
  journal = {\nat},
archivePrefix = "arXiv",
   eprint = {1710.01824},
 primaryClass = "astro-ph.HE",
     year = 2017,
    month = oct,
   volume = 550,
    pages = {80-83},
      doi = {10.1038/nature23908},
   adsurl = {http://adsabs.harvard.edu/abs/2017Natur.550...80J}
}

@ARTICLE{kass--95,
       author = {{Kass}, Robert E. and {Raftery}, Adrian E.},
        title = "{Bayes factors}",
      journal = {Journal of the American Statistical Association},
         year = "1995",
       volume = {90},
       number = {430},
        pages = {773-795},
}

@ARTICLE{kasen-06b,
   author = {{Kasen}, D. and {Thomas}, R.~C. and {Nugent}, P.},
    title = "{Time-dependent Monte Carlo Radiative Transfer Calculations for Three-dimensional Supernova Spectra, Light Curves, and Polarization}",
  journal = {\apj},
   eprint = {astro-ph/0606111},
     year = 2006,
    month = nov,
   volume = 651,
    pages = {366-380},
      doi = {10.1086/506190},
   adsurl = {http://adsabs.harvard.edu/abs/2006ApJ...651..366K}
}

@ARTICLE{kasen--07b,
       author = {{Kasen}, Daniel and {Plewa}, Tomasz},
        title = "{Detonating Failed Deflagration Model of Thermonuclear Supernovae. II. Comparison to Observations}",
      journal = {\apj},
         year = 2007,
        month = jun,
       volume = {662},
       number = {1},
        pages = {459-471},
          doi = {10.1086/516834},
archivePrefix = {arXiv},
       eprint = {astro-ph/0612198},
 primaryClass = {astro-ph},
       adsurl = {https://ui.adsabs.harvard.edu/abs/2007ApJ...662..459K}
}

@ARTICLE{tardis,
   author = {{Kerzendorf}, W.~E. and {Sim}, S.~A.},
    title = "{A spectral synthesis code for rapid modelling of supernovae}",
  journal = {\mnras},
archivePrefix = "arXiv",
   eprint = {1401.5469},
 primaryClass = "astro-ph.SR",
     year = 2014,
    month = may,
   volume = 440,
    pages = {387-404},
      doi = {10.1093/mnras/stu055},
   adsurl = {http://adsabs.harvard.edu/abs/2014MNRAS.440..387K}
}

@ARTICLE{kerzendorf--17,
       author = {{Kerzendorf}, W.~E. and {McCully}, C. and {Taubenberger}, S. and {Jerkstrand}, A. and {Seitenzahl}, I. and {Ruiter}, A.~J. and {Spyromilio}, J. and {Long}, K.~S. and {Fransson}, C.},
        title = "{Extremely late photometry of the nearby SN 2011fe}",
      journal = {\mnras},
         year = 2017,
        month = dec,
       volume = {472},
       number = {3},
        pages = {2534-2542},
          doi = {10.1093/mnras/stx1923},
archivePrefix = {arXiv},
       eprint = {1706.01460},
 primaryClass = {astro-ph.SR},
       adsurl = {https://ui.adsabs.harvard.edu/abs/2017MNRAS.472.2534K}
}

@ARTICLE{kerzendorf--21,
       author = {{Kerzendorf}, Wolfgang E. and {Vogl}, Christian and {Buchner}, Johannes and {Contardo}, Gabriella and {Williamson}, Marc and {van der Smagt}, Patrick},
        title = "{Dalek: A Deep Learning Emulator for TARDIS}",
      journal = {\apjl},
         year = 2021,
        month = apr,
       volume = {910},
       number = {2},
          eid = {L23},
        pages = {L23},
          doi = {10.3847/2041-8213/abeb1b},
archivePrefix = {arXiv},
       eprint = {2007.01868},
 primaryClass = {astro-ph.IM},
       adsurl = {https://ui.adsabs.harvard.edu/abs/2021ApJ...910L..23K}
}

@ARTICLE{khokhlov--89,
       author = {{Khokhlov}, A.~M.},
        title = "{The structure of detonation waves in supernovae}",
      journal = {\mnras},
         year = 1989,
        month = aug,
       volume = {239},
        pages = {785-808},
          doi = {10.1093/mnras/239.3.785},
       adsurl = {https://ui.adsabs.harvard.edu/abs/1989MNRAS.239..785K}
}

@ARTICLE{khokhlov--91b,
   author = {{Khokhlov}, A.~M.},
    title = "{Nucleosynthesis in delayed detonation models of Type IA supernovae}",
  journal = {\aap},
     year = 1991,
    month = may,
   volume = 245,
    pages = {L25-L28},
   adsurl = {http://adsabs.harvard.edu/abs/1991A\%26A...245L..25K}
}

@ARTICLE{khokhlov--91a,
   author = {{Khokhlov}, A.~M.},
    title = "{Delayed detonation model for type IA supernovae}",
  journal = {\aap},
     year = 1991,
    month = may,
   volume = 245,
    pages = {114-128},
   adsurl = {http://adsabs.harvard.edu/abs/1991A\%26A...245..114K}
}

@ARTICLE{artis,
   author = {{Kromer}, M. and {Sim}, S.~A.},
    title = "{Time-dependent three-dimensional spectrum synthesis for Type Ia supernovae}",
  journal = {\mnras},
archivePrefix = "arXiv",
   eprint = {0906.3152},
 primaryClass = "astro-ph.HE",
     year = 2009,
    month = oct,
   volume = 398,
    pages = {1809-1826},
      doi = {10.1111/j.1365-2966.2009.15256.x},
   adsurl = {http://adsabs.harvard.edu/abs/2009MNRAS.398.1809K}
}

@ARTICLE{kromer--10,
   author = {{Kromer}, M. and {Sim}, S.~A. and {Fink}, M. and {R{\"o}pke}, F.~K. and 
	{Seitenzahl}, I.~R. and {Hillebrandt}, W.},
    title = "{Double-detonation Sub-Chandrasekhar Supernovae: Synthetic Observables for Minimum Helium Shell Mass Models}",
  journal = {\apj},
archivePrefix = "arXiv",
   eprint = {1006.4489},
 primaryClass = "astro-ph.HE",
     year = 2010,
    month = aug,
   volume = 719,
    pages = {1067-1082},
      doi = {10.1088/0004-637X/719/2/1067},
   adsurl = {http://adsabs.harvard.edu/abs/2010ApJ...719.1067K}
}

@ARTICLE{kromer-13,
   author = {{Kromer}, M. and {Fink}, M. and {Stanishev}, V. and {Taubenberger}, S. and 
	{Ciaraldi-Schoolman}, F. and {Pakmor}, R. and {R{\"o}pke}, F.~K. and 
	{Ruiter}, A.~J. and {Seitenzahl}, I.~R. and {Sim}, S.~A. and 
	{Blanc}, G. and {Elias-Rosa}, N. and {Hillebrandt}, W.},
    title = "{3D deflagration simulations leaving bound remnants: a model for 2002cx-like Type Ia supernovae}",
  journal = {\mnras},
archivePrefix = "arXiv",
   eprint = {1210.5243},
 primaryClass = "astro-ph.HE",
     year = 2013,
    month = mar,
   volume = 429,
    pages = {2287-2297},
      doi = {10.1093/mnras/sts498},
   adsurl = {http://adsabs.harvard.edu/abs/2013MNRAS.429.2287K}
}

@ARTICLE{kromer--13b,
       author = {{Kromer}, M. and {Pakmor}, R. and {Taubenberger}, S. and {Pignata}, G. and {Fink}, M. and {R{\"o}pke}, F.~K. and {Seitenzahl}, I.~R. and {Sim}, S.~A. and {Hillebrandt}, W.},
        title = "{SN 2010lp{\textemdash}a Type Ia Supernova from a Violent Merger of Two Carbon-Oxygen White Dwarfs}",
      journal = {\apjl},
         year = 2013,
        month = nov,
       volume = {778},
       number = {1},
          eid = {L18},
        pages = {L18},
          doi = {10.1088/2041-8205/778/1/L18},
archivePrefix = {arXiv},
       eprint = {1311.0310},
 primaryClass = {astro-ph.HE},
       adsurl = {https://ui.adsabs.harvard.edu/abs/2013ApJ...778L..18K}
}

@ARTICLE{hesma,
       author = {{Kromer}, M. and {Ohlmann}, S. and {R{\"o}pke}, F.~K.},
        title = "{Simulating the observed diversity of Type Ia supernovae . Introducing a model data base}",
      journal = {\memsai},
         year = 2017,
        month = jan,
       volume = {88},
        pages = {312},
          doi = {10.48550/arXiv.1706.09879},
archivePrefix = {arXiv},
       eprint = {1706.09879},
 primaryClass = {astro-ph.HE},
       adsurl = {https://ui.adsabs.harvard.edu/abs/2017MmSAI..88..312K}
}

@ARTICLE{lach--22,
       author = {{Lach}, F. and {Callan}, F.~P. and {Bubeck}, D. and {R{\"o}pke}, F.~K. and {Sim}, S.~A. and {Schrauth}, M. and {Ohlmann}, S.~T. and {Kromer}, M.},
        title = "{Type Iax supernovae from deflagrations in Chandrasekhar mass white dwarfs}",
      journal = {\aap},
         year = 2022,
        month = feb,
       volume = {658},
          eid = {A179},
        pages = {A179},
          doi = {10.1051/0004-6361/202141453},
archivePrefix = {arXiv},
       eprint = {2109.02926},
 primaryClass = {astro-ph.SR},
       adsurl = {https://ui.adsabs.harvard.edu/abs/2022A&A...658A.179L}
}

@ARTICLE{lach--22b,
       author = {{Lach}, F. and {Callan}, F.~P. and {Sim}, S.~A. and {R{\"o}pke}, F.~K.},
        title = "{Models of pulsationally assisted gravitationally confined detonations with different ignition conditions}",
      journal = {\aap},
         year = 2022,
        month = mar,
       volume = {659},
          eid = {A27},
        pages = {A27},
          doi = {10.1051/0004-6361/202142194},
archivePrefix = {arXiv},
       eprint = {2111.14394},
 primaryClass = {astro-ph.SR},
       adsurl = {https://ui.adsabs.harvard.edu/abs/2022A&A...659A..27L}
}

@ARTICLE{levi--22,
       author = {{Levi}, Dan and {Gispan}, Liran and {Giladi}, Niv and {Fetaya}, Ethan},
        title = "{Evaluating and Calibrating Uncertainty Prediction in Regression Tasks}",
      journal = {Sensors},
         year = 2022,
        month = jul,
       volume = {22},
       number = {15},
        pages = {5540},
          doi = {10.3390/s22155540},
       adsurl = {https://ui.adsabs.harvard.edu/abs/2022Senso..22.5540L}
}

@ARTICLE{91t--like,
   author = {{Li}, W.~D. and {Qiu}, Y.~L. and {Qiao}, Q.~Y. and {Zhu}, X.~H. and 
	{Hu}, J.~Y. and {Richmond}, M.~W. and {Filippenko}, A.~V. and 
	{Treffers}, R.~R. and {Peng}, C.~Y. and {Leonard}, D.~C.},
    title = "{The Type IA Supernova 1997BR in ESO 576-G40}",
  journal = {\aj},
   eprint = {astro-ph/9903466},
     year = 1999,
    month = jun,
   volume = 117,
    pages = {2709-2724},
      doi = {10.1086/300895},
   adsurl = {http://adsabs.harvard.edu/abs/1999AJ....117.2709L}
}

@ARTICLE{liu--23,
       author = {{Liu}, Chang and {Miller}, Adam A. and {Polin}, Abigail and {Nugent}, Anya E. and {De}, Kishalay and {Nugent}, Peter E. and {Schulze}, Steve and {Gal-Yam}, Avishay and {Fremling}, Christoffer and {Anand}, Shreya and {Andreoni}, Igor and {Blanchard}, Peter and {Brink}, Thomas G. and {Dhawan}, Suhail and {Filippenko}, Alexei V. and {Maguire}, Kate and {Schweyer}, Tassilo and {Sears}, Huei and {Sharma}, Yashvi and {Graham}, Matthew J. and {Groom}, Steven L. and {Hale}, David and {Kasliwal}, Mansi M. and {Masci}, Frank J. and {Purdum}, Josiah and {Racine}, Benjamin and {Sollerman}, Jesper and {Kulkarni}, Shrinivas R.},
        title = "{SN 2020jgb: A Peculiar Type Ia Supernova Triggered by a Helium-shell Detonation in a Star-forming Galaxy}",
      journal = {\apj},
         year = 2023,
        month = apr,
       volume = {946},
       number = {2},
          eid = {83},
        pages = {83},
          doi = {10.3847/1538-4357/acbb5e},
archivePrefix = {arXiv},
       eprint = {2209.04463},
 primaryClass = {astro-ph.HE},
       adsurl = {https://ui.adsabs.harvard.edu/abs/2023ApJ...946...83L}
}

@ARTICLE{liu--23b,
       author = {{Liu}, Zheng-Wei and {R{\"o}pke}, Friedrich K. and {Han}, Zhanwen},
        title = "{Type Ia Supernova Explosions in Binary Systems: A Review}",
      journal = {Research in Astronomy and Astrophysics},
         year = 2023,
        month = aug,
       volume = {23},
       number = {8},
          eid = {082001},
        pages = {082001},
          doi = {10.1088/1674-4527/acd89e},
archivePrefix = {arXiv},
       eprint = {2305.13305},
 primaryClass = {astro-ph.HE},
       adsurl = {https://ui.adsabs.harvard.edu/abs/2023RAA....23h2001L}
}

@ARTICLE{livne--90,
   author = {{Livne}, E.},
    title = "{Successive detonations in accreting white dwarfs as an alternative mechanism for type I supernovae}",
  journal = {\apjl},
     year = 1990,
    month = may,
   volume = 354,
    pages = {L53-L55},
      doi = {10.1086/185721},
   adsurl = {http://adsabs.harvard.edu/abs/1990ApJ...354L..53L}
}

@ARTICLE{livne--91,
   author = {{Livne}, E. and {Glasner}, A.~S.},
    title = "{Numerical simulations of off-center detonations in helium shells}",
  journal = {\apj},
     year = 1991,
    month = mar,
   volume = 370,
    pages = {272-281},
      doi = {10.1086/169813},
   adsurl = {http://adsabs.harvard.edu/abs/1991ApJ...370..272L}
}

@ARTICLE{15h,
   author = {{Magee}, M.~R. and {Kotak}, R. and {Sim}, S.~A. and {Kromer}, M. and 
	{Rabinowitz}, D. and {Smartt}, S.~J. and {Baltay}, C. and {Campbell}, H.~C. and 
	{Chen}, T.-W. and {Fink}, M. and {Gal-Yam}, A. and {Galbany}, L. and 
	{Hillebrandt}, W. and {Inserra}, C. and {Kankare}, E. and {Le Guillou}, L. and 
	{Lyman}, J.~D. and {Maguire}, K. and {Pakmor}, R. and {R{\"o}pke}, F.~K. and 
	{Ruiter}, A.~J. and {Seitenzahl}, I.~R. and {Sullivan}, M. and 
	{Valenti}, S. and {Young}, D.~R.},
    title = "{The type Iax supernova, SN 2015H. A white dwarf deflagration candidate}",
  journal = {\aap},
archivePrefix = "arXiv",
   eprint = {1603.04728},
 primaryClass = "astro-ph.HE",
     year = 2016,
    month = apr,
   volume = 589,
      eid = {A89},
    pages = {A89},
      doi = {10.1051/0004-6361/201528036},
   adsurl = {http://adsabs.harvard.edu/abs/2016A\%26A...589A..89M}
}

@ARTICLE{12bwh,
   author = {{Magee}, M.~R. and {Kotak}, R. and {Sim}, S.~A. and {Wright}, D. and 
	{Smartt}, S.~J. and {Berger}, E. and {Chornock}, R. and {Foley}, R.~J. and 
	{Howell}, D.~A. and {Kaiser}, N. and {Magnier}, E.~A. and {Wainscoat}, R. and 
	{Waters}, C.},
    title = "{Growing evidence that SNe Iax are not a one-parameter family. The case of PS1-12bwh}",
  journal = {\aap},
archivePrefix = "arXiv",
   eprint = {1701.05459},
 primaryClass = "astro-ph.HE",
     year = 2017,
    month = may,
   volume = 601,
      eid = {A62},
    pages = {A62},
      doi = {10.1051/0004-6361/201629643},
   adsurl = {http://adsabs.harvard.edu/abs/2017A%26A...601A..62M}
}

@ARTICLE{magee--19,
       author = {{Magee}, M.~R. and {Sim}, S.~A. and {Kotak}, R. and {Maguire}, K. and
         {Boyle}, A.},
        title = "{Detecting the signatures of helium in type Iax supernovae}",
      journal = {\aap},
         year = 2019,
        month = feb,
       volume = {622},
          eid = {A102},
        pages = {A102},
          doi = {10.1051/0004-6361/201834420},
archivePrefix = {arXiv},
       eprint = {1812.08695},
 primaryClass = {astro-ph.HE},
       adsurl = {https://ui.adsabs.harvard.edu/abs/2019A&A...622A.102M}
}

@ARTICLE{magee--24,
       author = {{Magee}, M.~R. and {Siebenaler}, L. and {Maguire}, K. and {Ackley}, K. and {Killestein}, T.},
        title = "{Quantitative modelling of type Ia supernovae spectral time series: constraining the explosion physics}",
      journal = {\mnras},
         year = 2024,
        month = jul,
       volume = {531},
       number = {3},
        pages = {3042-3068},
          doi = {10.1093/mnras/stae1233},
archivePrefix = {arXiv},
       eprint = {2403.16889},
 primaryClass = {astro-ph.HE},
       adsurl = {https://ui.adsabs.harvard.edu/abs/2024MNRAS.531.3042M}
}

@ARTICLE{magee--25,
       author = {{Magee}, M.~R. and {Killestein}, T.~L. and {Pursiainen}, M. and {Godson}, B. and {Jarvis}, D. and {Jim{\'e}nez-Palau}, C. and {Lyman}, J.~D. and {Steeghs}, D. and {Warwick}, B. and {Anderson}, J.~P. and {Butterley}, T. and {Chen}, T.-W. and {Dhillon}, V.~S. and {Galbany}, L. and {Gonz{\'a}lez-Gait{\'a}n}, S. and {Gromadzki}, M. and {Inserra}, C. and {Kelsey}, L. and {Kumar}, A. and {Leloudas}, G. and {Mattila}, S. and {Moran}, S. and {M{\"u}ller-Bravo}, T.~E. and {Noysena}, K. and {Ramsay}, G. and {Srivastav}, S. and {Starling}, R. and {Wilson}, R.~W. and {Young}, D.~R. and {Ackley}, K. and {Breton}, R.~P. and {Casares Vel{\'a}zquez}, J. and {Dyer}, M.~J. and {Galloway}, D.~K. and {Kankare}, E. and {Kotak}, R. and {Nuttall}, L.~K. and {O'Neill}, D. and {Pessi}, P. and {Pollacco}, D. and {Ulaczyk}, K. and {Yaron}, O.},
        title = "{SN 2024bfu, SN 2025qe, and the early light curves of type Iax supernovae}",
      journal = {\mnras},
         year = 2025,
        month = nov,
       volume = {543},
       number = {4},
        pages = {3731-3753},
          doi = {10.1093/mnras/staf1675},
archivePrefix = {arXiv},
       eprint = {2506.02118},
 primaryClass = {astro-ph.HE},
       adsurl = {https://ui.adsabs.harvard.edu/abs/2025MNRAS.543.3731M}
}

@ARTICLE{magee--26b,
       author = {{Magee}, M.~R.},
        title = "{Quantitative modelling of type Ia supernovae spectral time series III: Implications for type Ia supernovae standardisation in cosmology}",
      journal = {arXiv e-prints},
         year = 2026,
        month = apr,
          eid = {arXiv:2604.22928},
        pages = {arXiv:2604.22928},
          doi = {10.48550/arXiv.2604.22928},
archivePrefix = {arXiv},
       eprint = {2604.22928},
 primaryClass = {astro-ph.HE},
       adsurl = {https://ui.adsabs.harvard.edu/abs/2026arXiv260422928M}
}

@ARTICLE{obs--05hk--08a,
   author = {{McCully}, C. and {Jha}, S.~W. and {Foley}, R.~J. and {Chornock}, R. and 
	{Holtzman}, J.~A. and {Balam}, D.~D. and {Branch}, D. and {Filippenko}, A.~V. and 
	{Frieman}, J. and {Fynbo}, J. and {Galbany}, L. and {Ganeshalingam}, M. and 
	{Garnavich}, P.~M. and {Graham}, M.~L. and {Hsiao}, E.~Y. and 
	{Leloudas}, G. and {Leonard}, D.~C. and {Li}, W. and {Riess}, A.~G. and 
	{Sako}, M. and {Schneider}, D.~P. and {Silverman}, J.~M. and 
	{Sollerman}, J. and {Steele}, T.~N. and {Thomas}, R.~C. and 
	{Wheeler}, J.~C. and {Zheng}, C.},
    title = "{Hubble Space Telescope and Ground-based Observations of the Type Iax Supernovae SN 2005hk and SN 2008A}",
  journal = {\apj},
archivePrefix = "arXiv",
   eprint = {1309.4457},
 primaryClass = "astro-ph.SR",
     year = 2014,
    month = may,
   volume = 786,
      eid = {134},
    pages = {134},
      doi = {10.1088/0004-637X/786/2/134},
   adsurl = {http://adsabs.harvard.edu/abs/2014ApJ...786..134M}
}

@ARTICLE{maguire--23,
       author = {{Maguire}, Kate and {Magee}, Mark R. and {Leloudas}, Giorgos and {Miller}, Adam A. and {Dimitriadis}, Georgios and {Pursiainen}, Miika and {Bulla}, Mattia and {De}, Kishalay and {Gal-Yam}, Avishay and {Perley}, Daniel A. and {Fremling}, Christoffer and {Karambelkar}, Viraj R. and {Nordin}, Jakob and {Reusch}, Simeon and {Schulze}, Steve and {Sollerman}, Jesper and {Terreran}, Giacomo and {Yang(杨轶)}, Yi and {Bellm}, Eric C. and {Groom}, Steven L. and {Kasliwal}, Mansi M. and {Kulkarni}, Shrinivas R. and {Lacroix}, Leander and {Masci}, Frank J. and {Purdum}, Josiah N. and {Sharma}, Yashvi and {Smith}, Roger},
        title = "{SN 2020udy: an SN Iax with strict limits on interaction consistent with a helium-star companion}",
      journal = {\mnras},
         year = 2023,
        month = oct,
       volume = {525},
       number = {1},
        pages = {1210-1228},
          doi = {10.1093/mnras/stad2316},
archivePrefix = {arXiv},
       eprint = {2304.12361},
 primaryClass = {astro-ph.HE},
       adsurl = {https://ui.adsabs.harvard.edu/abs/2023MNRAS.525.1210M}
}

@ARTICLE{mazzali--lucy--93,
   author = {{Mazzali}, P.~A. and {Lucy}, L.~B.},
    title = "{The application of Monte Carlo methods to the synthesis of early-time supernovae spectra}",
  journal = {\aap},
     year = 1993,
    month = nov,
   volume = 279,
    pages = {447-456},
   adsurl = {http://adsabs.harvard.edu/abs/1993A\%26A...279..447M}
}

@ARTICLE{mazzali--07,
   author = {{Mazzali}, P.~A. and {R{\"o}pke}, F.~K. and {Benetti}, S. and 
	{Hillebrandt}, W.},
    title = "{A Common Explosion Mechanism for Type Ia Supernovae}",
  journal = {Science},
   eprint = {astro-ph/0702351},
     year = 2007,
    month = feb,
   volume = 315,
    pages = {825},
      doi = {10.1126/science.1136259},
   adsurl = {http://adsabs.harvard.edu/abs/2007Sci...315..825M}
}

@ARTICLE{mazzali--08,
   author = {{Mazzali}, P.~A. and {Sauer}, D.~N. and {Pastorello}, A. and 
	{Benetti}, S. and {Hillebrandt}, W.},
    title = "{Abundance stratification in Type Ia supernovae - II. The rapidly declining, spectroscopically normal SN2004eo}",
  journal = {\mnras},
archivePrefix = "arXiv",
   eprint = {0803.1383},
     year = 2008,
    month = jun,
   volume = 386,
    pages = {1897-1906},
      doi = {10.1111/j.1365-2966.2008.13199.x},
   adsurl = {http://adsabs.harvard.edu/abs/2008MNRAS.386.1897M}
}

@ARTICLE{mazzali--14,
   author = {{Mazzali}, P.~A. and {Sullivan}, M. and {Hachinger}, S. and 
	{Ellis}, R.~S. and {Nugent}, P.~E. and {Howell}, D.~A. and {Gal-Yam}, A. and 
	{Maguire}, K. and {Cooke}, J. and {Thomas}, R. and {Nomoto}, K. and 
	{Walker}, E.~S.},
    title = "{Hubble Space Telescope spectra of the Type Ia supernova SN 2011fe: a tail of low-density, high-velocity material with Z $\lt$ Z$_{⊙}$}",
  journal = {\mnras},
archivePrefix = "arXiv",
   eprint = {1305.2356},
     year = 2014,
    month = apr,
   volume = 439,
    pages = {1959-1979},
      doi = {10.1093/mnras/stu077},
   adsurl = {http://adsabs.harvard.edu/abs/2014MNRAS.439.1959M}
}

@ARTICLE{mazzali--15,
       author = {{Mazzali}, P.~A. and {Sullivan}, M. and {Filippenko}, A.~V. and {Garnavich}, P.~M. and {Clubb}, K.~I. and {Maguire}, K. and {Pan}, Y.-C. and {Shappee}, B. and {Silverman}, J.~M. and {Benetti}, S. and {Hachinger}, S. and {Nomoto}, K. and {Pian}, E.},
        title = "{Nebular spectra and abundance tomography of the Type Ia supernova SN 2011fe: a normal SN Ia with a stable Fe core}",
      journal = {\mnras},
         year = 2015,
        month = jul,
       volume = {450},
       number = {3},
        pages = {2631-2643},
          doi = {10.1093/mnras/stv761},
archivePrefix = {arXiv},
       eprint = {1504.04857},
 primaryClass = {astro-ph.HE},
       adsurl = {https://ui.adsabs.harvard.edu/abs/2015MNRAS.450.2631M}
}

@ARTICLE{meakin--09,
       author = {{Meakin}, Casey A. and {Seitenzahl}, Ivo and {Townsley}, Dean and
         {Jordan}, George C., IV and {Truran}, James and {Lamb}, Don},
        title = "{Study of the Detonation Phase in the Gravitationally Confined Detonation Model of Type Ia Supernovae}",
      journal = {\apj},
         year = "2009",
        month = "Mar",
       volume = {693},
       number = {2},
        pages = {1188-1208},
          doi = {10.1088/0004-637X/693/2/1188},
archivePrefix = {arXiv},
       eprint = {0806.4972},
 primaryClass = {astro-ph},
       adsurl = {https://ui.adsabs.harvard.edu/abs/2009ApJ...693.1188M}
}

@ARTICLE{miller--20a,
       author = {{Miller}, A.~A. and {Yao}, Y. and {Bulla}, M. and {Pankow}, C. and {Bellm}, E.~C. and {Cenko}, S.~B. and {Dekany}, R. and {Fremling}, C. and {Graham}, M.~J. and {Kupfer}, T. and {Laher}, R.~R. and {Mahabal}, A.~A. and {Masci}, F.~J. and {Nugent}, P.~E. and {Riddle}, R. and {Rusholme}, B. and {Smith}, R.~M. and {Shupe}, D.~L. and {van Roestel}, J. and {Kulkarni}, S.~R.},
        title = "{ZTF Early Observations of Type Ia Supernovae. II. First Light, the Initial Rise, and Time to Reach Maximum Brightness}",
      journal = {\apj},
         year = 2020,
        month = oct,
       volume = {902},
       number = {1},
          eid = {47},
        pages = {47},
          doi = {10.3847/1538-4357/abb13b},
archivePrefix = {arXiv},
       eprint = {2001.00598},
 primaryClass = {astro-ph.HE},
       adsurl = {https://ui.adsabs.harvard.edu/abs/2020ApJ...902...47M}
}

@ARTICLE{moll--13,
       author = {{Moll}, R. and {Woosley}, S.~E.},
        title = "{Multi-dimensional Models for Double Detonation in Sub-Chandrasekhar Mass White Dwarfs}",
      journal = {\apj},
         year = "2013",
        month = "Sep",
       volume = {774},
       number = {2},
          eid = {137},
        pages = {137},
          doi = {10.1088/0004-637X/774/2/137},
archivePrefix = {arXiv},
       eprint = {1303.0324},
 primaryClass = {astro-ph.HE},
       adsurl = {https://ui.adsabs.harvard.edu/abs/2013ApJ...774..137M}
}

@ARTICLE{ness--15,
       author = {{Ness}, M. and {Hogg}, David W. and {Rix}, H.-W. and {Ho}, Anna. Y.~Q. and {Zasowski}, G.},
        title = "{The Cannon: A data-driven approach to Stellar Label Determination}",
      journal = {\apj},
         year = 2015,
        month = jul,
       volume = {808},
       number = {1},
          eid = {16},
        pages = {16},
          doi = {10.1088/0004-637X/808/1/16},
archivePrefix = {arXiv},
       eprint = {1501.07604},
 primaryClass = {astro-ph.SR},
       adsurl = {https://ui.adsabs.harvard.edu/abs/2015ApJ...808...16N}
}

@ARTICLE{niemeyer--97,
   author = {{Niemeyer}, J.~C. and {Woosley}, S.~E.},
    title = "{The Thermonuclear Explosion of Chandrasekhar Mass White Dwarfs}",
  journal = {\apj},
   eprint = {astro-ph/9607032},
     year = 1997,
    month = feb,
   volume = 475,
    pages = {740-753},
      doi = {10.1086/303544},
   adsurl = {http://adsabs.harvard.edu/abs/1997ApJ...475..740N}
}

@INPROCEEDINGS{nix--94,
  author={Nix, D.A. and Weigend, A.S.},
  booktitle={Proceedings of 1994 IEEE International Conference on Neural Networks (ICNN'94)}, 
  title={Estimating the mean and variance of the target probability distribution}, 
  year={1994},
  volume={1},
  number={},
  pages={55-60 vol.1},
  doi={10.1109/ICNN.1994.374138}}

@ARTICLE{nomoto-w7,
   author = {{Nomoto}, K. and {Thielemann}, F.-K. and {Yokoi}, K.},
    title = "{Accreting white dwarf models of Type I supernovae. III - Carbon deflagration supernovae}",
  journal = {\apj},
     year = 1984,
    month = nov,
   volume = 286,
    pages = {644-658},
      doi = {10.1086/162639},
   adsurl = {http://adsabs.harvard.edu/abs/1984ApJ...286..644N}
}

@ARTICLE{nomoto--91,
   author = {{Nomoto}, K. and {Kondo}, Y.},
    title = "{Conditions for accretion-induced collapse of white dwarfs}",
  journal = {\apjl},
     year = 1991,
    month = jan,
   volume = 367,
    pages = {L19-L22},
      doi = {10.1086/185922},
   adsurl = {http://adsabs.harvard.edu/abs/1991ApJ...367L..19N}
}

@ARTICLE{nonaka--12,
       author = {{Nonaka}, A. and {Aspden}, A.~J. and {Zingale}, M. and {Almgren}, A.~S. and
         {Bell}, J.~B. and {Woosley}, S.~E.},
        title = "{High-resolution Simulations of Convection Preceding Ignition in Type Ia Supernovae Using Adaptive Mesh Refinement}",
      journal = {\apj},
         year = "2012",
        month = "Jan",
       volume = {745},
       number = {1},
          eid = {73},
        pages = {73},
          doi = {10.1088/0004-637X/745/1/73},
archivePrefix = {arXiv},
       eprint = {1111.3086},
 primaryClass = {astro-ph.HE},
       adsurl = {https://ui.adsabs.harvard.edu/abs/2012ApJ...745...73N}
}

@ARTICLE{11fe--nature,
   author = {{Nugent}, P.~E. and {Sullivan}, M. and {Cenko}, S.~B. and {Thomas}, R.~C. and 
	{Kasen}, D. and {Howell}, D.~A. and {Bersier}, D. and {Bloom}, J.~S. and 
	{Kulkarni}, S.~R. and {Kandrashoff}, M.~T. and {Filippenko}, A.~V. and 
	{Silverman}, J.~M. and {Marcy}, G.~W. and {Howard}, A.~W. and 
	{Isaacson}, H.~T. and {Maguire}, K. and {Suzuki}, N. and {Tarlton}, J.~E. and 
	{Pan}, Y.-C. and {Bildsten}, L. and {Fulton}, B.~J. and {Parrent}, J.~T. and 
	{Sand}, D. and {Podsiadlowski}, P. and {Bianco}, F.~B. and {Dilday}, B. and 
	{Graham}, M.~L. and {Lyman}, J. and {James}, P. and {Kasliwal}, M.~M. and 
	{Law}, N.~M. and {Quimby}, R.~M. and {Hook}, I.~M. and {Walker}, E.~S. and 
	{Mazzali}, P. and {Pian}, E. and {Ofek}, E.~O. and {Gal-Yam}, A. and 
	{Poznanski}, D.},
    title = "{Supernova SN 2011fe from an exploding carbon-oxygen white dwarf star}",
  journal = {\nat},
archivePrefix = "arXiv",
   eprint = {1110.6201},
     year = 2011,
    month = dec,
   volume = 480,
    pages = {344-347},
      doi = {10.1038/nature10644},
   adsurl = {http://adsabs.harvard.edu/abs/2011Natur.480..344N}
}

@ARTICLE{obrien--21,
       author = {{O'Brien}, John T. and {Kerzendorf}, Wolfgang E. and {Fullard}, Andrew and {Williamson}, Marc and {Pakmor}, R{\"u}diger and {Buchner}, Johannes and {Hachinger}, Stephan and {Vogl}, Christian and {Gillanders}, James H. and {Fl{\"o}rs}, Andreas and {van der Smagt}, Patrick},
        title = "{Probabilistic Reconstruction of Type Ia Supernova SN 2002bo}",
      journal = {\apjl},
         year = 2021,
        month = aug,
       volume = {916},
       number = {2},
          eid = {L14},
        pages = {L14},
          doi = {10.3847/2041-8213/ac1173},
archivePrefix = {arXiv},
       eprint = {2105.07910},
 primaryClass = {astro-ph.SR},
       adsurl = {https://ui.adsabs.harvard.edu/abs/2021ApJ...916L..14O}
}

@ARTICLE{obrien--24,
       author = {{O'Brien}, John T. and {Kerzendorf}, Wolfgang E. and {Fullard}, Andrew and {Pakmor}, R{\"u}diger and {Buchner}, Johannes and {Vogl}, Christian and {Chen}, Nutan and {van der Smagt}, Patrick and {Williamson}, Marc and {Singhal}, Jaladh},
        title = "{1991T-Like Type Ia Supernovae as an Extension of the Normal Population}",
      journal = {\apj},
         year = 2024,
        month = apr,
       volume = {964},
       number = {2},
          eid = {137},
        pages = {137},
          doi = {10.3847/1538-4357/ad2358},
archivePrefix = {arXiv},
       eprint = {2306.08137},
 primaryClass = {astro-ph.HE},
       adsurl = {https://ui.adsabs.harvard.edu/abs/2024ApJ...964..137O}
}

@ARTICLE{ogawa--23,
       author = {{Ogawa}, Mao and {Maeda}, Keiichi and {Kawabata}, Miho},
        title = "{Systematic Investigation of Very-early-phase Spectra of Type Ia Supernovae}",
      journal = {\apj},
         year = 2023,
        month = sep,
       volume = {955},
       number = {1},
          eid = {49},
        pages = {49},
          doi = {10.3847/1538-4357/acec74},
archivePrefix = {arXiv},
       eprint = {2308.03834},
 primaryClass = {astro-ph.HE},
       adsurl = {https://ui.adsabs.harvard.edu/abs/2023ApJ...955...49O}
}

@ARTICLE{padilla-gonzalez--23,
       author = {{Padilla Gonzalez}, E. and {Howell}, D. Andrew and {Burke}, J. and {Dong}, Yize and {Hiramatsu}, D. and {McCully}, C. and {Pellegrino}, C. and {Kerzendorf}, W. and {Modjaz}, M. and {Terreran}, G. and {Williamson}, M.},
        title = "{Peculiar Spectral Evolution of the Type I Supernova 2019eix: A Possible Double Detonation from a Helium Shell on a Sub-Chandrasekhar-mass White Dwarf}",
      journal = {\apj},
         year = 2023,
        month = aug,
       volume = {953},
       number = {1},
          eid = {25},
        pages = {25},
          doi = {10.3847/1538-4357/acdd6a},
archivePrefix = {arXiv},
       eprint = {2305.07708},
 primaryClass = {astro-ph.HE},
       adsurl = {https://ui.adsabs.harvard.edu/abs/2023ApJ...953...25P}
}

@ARTICLE{pakmor--10,
   author = {{Pakmor}, R. and {Kromer}, M. and {R{\"o}pke}, F.~K. and {Sim}, S.~A. and 
	{Ruiter}, A.~J. and {Hillebrandt}, W.},
    title = "{Sub-luminous type Ia supernovae from the mergers of equal-mass white dwarfs with mass \~{}0.9M$_{solar}$}",
  journal = {\nat},
archivePrefix = "arXiv",
   eprint = {0911.0926},
 primaryClass = "astro-ph.HE",
     year = 2010,
    month = jan,
   volume = 463,
    pages = {61-64},
      doi = {10.1038/nature08642},
   adsurl = {http://adsabs.harvard.edu/abs/2010Natur.463...61P}
}

@ARTICLE{pakmor--11,
       author = {{Pakmor}, R. and {Hachinger}, S. and {R{\"o}pke}, F.~K. and {Hillebrandt}, W.},
        title = "{Violent mergers of nearly equal-mass white dwarf as progenitors of subluminous Type Ia supernovae}",
      journal = {\aap},
         year = 2011,
        month = apr,
       volume = {528},
          eid = {A117},
        pages = {A117},
          doi = {10.1051/0004-6361/201015653},
archivePrefix = {arXiv},
       eprint = {1102.1354},
 primaryClass = {astro-ph.SR},
       adsurl = {https://ui.adsabs.harvard.edu/abs/2011A&A...528A.117P}
}

@ARTICLE{pakmor-2012,
   author = {{Pakmor}, R. and {Kromer}, M. and {Taubenberger}, S. and {Sim}, S.~A. and 
	{R{\"o}pke}, F.~K. and {Hillebrandt}, W.},
    title = "{Normal Type Ia Supernovae from Violent Mergers of White Dwarf Binaries}",
  journal = {\apjl},
archivePrefix = "arXiv",
   eprint = {1201.5123},
 primaryClass = "astro-ph.HE",
     year = 2012,
    month = mar,
   volume = 747,
      eid = {L10},
    pages = {L10},
      doi = {10.1088/2041-8205/747/1/L10},
   adsurl = {http://adsabs.harvard.edu/abs/2012ApJ...747L..10P}
}

@ARTICLE{pakmor--24,
       author = {{Pakmor}, R{\"u}diger and {Seitenzahl}, Ivo R. and {Ruiter}, Ashley J. and {Sim}, Stuart A. and {R{\"o}pke}, Friedrich K. and {Taubenberger}, Stefan and {Bieri}, Rebekka and {Blondin}, St{\'e}phane},
        title = "{Type Ia supernova explosion models are inherently multidimensional}",
      journal = {\aap},
         year = 2024,
        month = jun,
       volume = {686},
          eid = {A227},
        pages = {A227},
          doi = {10.1051/0004-6361/202449637},
archivePrefix = {arXiv},
       eprint = {2402.11010},
 primaryClass = {astro-ph.HE},
       adsurl = {https://ui.adsabs.harvard.edu/abs/2024A&A...686A.227P}
}

@ARTICLE{pytorch,
       author = {{Paszke}, Adam and {Gross}, Sam and {Massa}, Francisco and {Lerer}, Adam and {Bradbury}, James and {Chanan}, Gregory and {Killeen}, Trevor and {Lin}, Zeming and {Gimelshein}, Natalia and {Antiga}, Luca and {Desmaison}, Alban and {K{\"o}pf}, Andreas and {Yang}, Edward and {DeVito}, Zach and {Raison}, Martin and {Tejani}, Alykhan and {Chilamkurthy}, Sasank and {Steiner}, Benoit and {Fang}, Lu and {Bai}, Junjie and {Chintala}, Soumith},
        title = "{PyTorch: An Imperative Style, High-Performance Deep Learning Library}",
      journal = {arXiv e-prints},
         year = 2019,
        month = dec,
          eid = {arXiv:1912.01703},
        pages = {arXiv:1912.01703},
          doi = {10.48550/arXiv.1912.01703},
archivePrefix = {arXiv},
       eprint = {1912.01703},
 primaryClass = {cs.LG},
       adsurl = {https://ui.adsabs.harvard.edu/abs/2019arXiv191201703P}
}

@ARTICLE{patat--13,
       author = {{Patat}, F. and {Cordiner}, M.~A. and {Cox}, N.~L.~J. and {Anderson}, R.~I. and {Harutyunyan}, A. and {Kotak}, R. and {Palaversa}, L. and {Stanishev}, V. and {Tomasella}, L. and {Benetti}, S. and {Goobar}, A. and {Pastorello}, A. and {Sollerman}, J.},
        title = "{Multi-epoch high-resolution spectroscopy of SN 2011fe. Linking the progenitor to its environment}",
      journal = {\aap},
         year = 2013,
        month = jan,
       volume = {549},
          eid = {A62},
        pages = {A62},
          doi = {10.1051/0004-6361/201118556},
archivePrefix = {arXiv},
       eprint = {1112.0247},
 primaryClass = {astro-ph.SR},
       adsurl = {https://ui.adsabs.harvard.edu/abs/2013A&A...549A..62P}
}

@ARTICLE{pereira--13,
   author = {{Pereira}, R. and {Thomas}, R.~C. and {Aldering}, G. and {Antilogus}, P. and 
	{Baltay}, C. and {Benitez-Herrera}, S. and {Bongard}, S. and 
	{Buton}, C. and {Canto}, A. and {Cellier-Holzem}, F. and {Chen}, J. and 
	{Childress}, M. and {Chotard}, N. and {Copin}, Y. and {Fakhouri}, H.~K. and 
	{Fink}, M. and {Fouchez}, D. and {Gangler}, E. and {Guy}, J. and 
	{Hillebrandt}, W. and {Hsiao}, E.~Y. and {Kerschhaggl}, M. and 
	{Kowalski}, M. and {Kromer}, M. and {Nordin}, J. and {Nugent}, P. and 
	{Paech}, K. and {Pain}, R. and {P{\'e}contal}, E. and {Perlmutter}, S. and 
	{Rabinowitz}, D. and {Rigault}, M. and {Runge}, K. and {Saunders}, C. and 
	{Smadja}, G. and {Tao}, C. and {Taubenberger}, S. and {Tilquin}, A. and 
	{Wu}, C.},
    title = "{Spectrophotometric time series of SN 2011fe from the Nearby Supernova Factory}",
  journal = {\aap},
archivePrefix = "arXiv",
   eprint = {1302.1292},
     year = 2013,
    month = jun,
   volume = 554,
      eid = {A27},
    pages = {A27},
      doi = {10.1051/0004-6361/201221008},
   adsurl = {http://adsabs.harvard.edu/abs/2013A\%26A...554A..27P}
}

@ARTICLE{phillips--07,
   author = {{Phillips}, M.~M. and {Li}, W. and {Frieman}, J.~A. and {Blinnikov}, S.~I. and 
	{DePoy}, D. and {Prieto}, J.~L. and {Milne}, P. and {Contreras}, C. and 
	{Folatelli}, G. and {Morrell}, N. and {Hamuy}, M. and {Suntzeff}, N.~B. and 
	{Roth}, M. and {Gonz{\'a}lez}, S. and {Krzeminski}, W. and {Filippenko}, A.~V. and 
	{Freedman}, W.~L. and {Chornock}, R. and {Jha}, S. and {Madore}, B.~F. and 
	{Persson}, S.~E. and {Burns}, C.~R. and {Wyatt}, P. and {Murphy}, D. and 
	{Foley}, R.~J. and {Ganeshalingam}, M. and {Serduke}, F.~J.~D. and 
	{Krisciunas}, K. and {Bassett}, B. and {Becker}, A. and {Dilday}, B. and 
	{Eastman}, J. and {Garnavich}, P.~M. and {Holtzman}, J. and 
	{Kessler}, R. and {Lampeitl}, H. and {Marriner}, J. and {Frank}, S. and 
	{Marshall}, J.~L. and {Miknaitis}, G. and {Sako}, M. and {Schneider}, D.~P. and 
	{van der Heyden}, K. and {Yasuda}, N.},
    title = "{The Peculiar SN 2005hk: Do Some Type Ia Supernovae Explode as Deflagrations?1,}",
  journal = {\pasp},
   eprint = {astro-ph/0611295},
     year = 2007,
    month = apr,
   volume = 119,
    pages = {360-387},
      doi = {10.1086/518372},
   adsurl = {http://adsabs.harvard.edu/abs/2007PASP..119..360P}
}

@ARTICLE{plewa--04,
       author = {{Plewa}, T. and {Calder}, A.~C. and {Lamb}, D.~Q.},
        title = "{Type Ia Supernova Explosion: Gravitationally Confined Detonation}",
      journal = {\apj},
         year = "2004",
        month = "Sep",
       volume = {612},
       number = {1},
        pages = {L37-L40},
          doi = {10.1086/424036},
archivePrefix = {arXiv},
       eprint = {astro-ph/0405163},
 primaryClass = {astro-ph},
       adsurl = {https://ui.adsabs.harvard.edu/abs/2004ApJ...612L..37P}
}

@ARTICLE{plewa--07,
       author = {{Plewa}, Tomasz},
        title = "{Detonating Failed Deflagration Model of Thermonuclear Supernovae. I. Explosion Dynamics}",
      journal = {\apj},
         year = 2007,
        month = mar,
       volume = {657},
       number = {2},
        pages = {942-960},
          doi = {10.1086/511412},
archivePrefix = {arXiv},
       eprint = {astro-ph/0611776},
 primaryClass = {astro-ph},
       adsurl = {https://ui.adsabs.harvard.edu/abs/2007ApJ...657..942P}
}

@ARTICLE{polin--19,
       author = {{Polin}, Abigail and {Nugent}, Peter and {Kasen}, Daniel},
        title = "{Observational Predictions for Sub-Chandrasekhar Mass Explosions: Further Evidence for Multiple Progenitor Systems for Type Ia Supernovae}",
      journal = {\apj},
         year = "2019",
        month = "Mar",
       volume = {873},
       number = {1},
          eid = {84},
        pages = {84},
          doi = {10.3847/1538-4357/aafb6a},
archivePrefix = {arXiv},
       eprint = {1811.07127},
 primaryClass = {astro-ph.HE},
       adsurl = {https://ui.adsabs.harvard.edu/abs/2019ApJ...873...84P}
}

@ARTICLE{reinecke--02a,
   author = {{Reinecke}, M. and {Hillebrandt}, W. and {Niemeyer}, J.~C.},
    title = "{Refined numerical models for multidimensional type Ia supernova simulations}",
  journal = {\aap},
   eprint = {astro-ph/0111475},
     year = 2002,
    month = may,
   volume = 386,
    pages = {936-943},
      doi = {10.1051/0004-6361:20020323},
   adsurl = {http://adsabs.harvard.edu/abs/2002A\%26A...386..936R}
}

@ARTICLE{reinecke--02b,
   author = {{Reinecke}, M. and {Hillebrandt}, W. and {Niemeyer}, J.~C.},
    title = "{Three-dimensional simulations of type Ia supernovae}",
  journal = {\aap},
   eprint = {astro-ph/0206459},
     year = 2002,
    month = sep,
   volume = 391,
    pages = {1167-1172},
      doi = {10.1051/0004-6361:20020885},
   adsurl = {http://adsabs.harvard.edu/abs/2002A\%26A...391.1167R}
}

@ARTICLE{rigault--25,
       author = {{Rigault}, M. and {Smith}, M. and {Goobar}, A. and {Maguire}, K. and {Dimitriadis}, G. and {Johansson}, J. and {Nordin}, J. and {Burgaz}, U. and {Dhawan}, S. and {Sollerman}, J. and {Regnault}, N. and {Kowalski}, M. and {Nugent}, P. and {Andreoni}, I. and {Amenouche}, M. and {Aubert}, M. and {Barjou-Delayre}, C. and {Bautista}, J. and {Bellm}, E. and {Betoule}, M. and {Bloom}, J.~S. and {Carreres}, B. and {Chen}, T.~X. and {Copin}, Y. and {Deckers}, M. and {de Jaeger}, T. and {Feinstein}, F. and {Fouchez}, D. and {Fremling}, C. and {Galbany}, L. and {Ginolin}, M. and {Graham}, M. and {Groom}, S.~L. and {Harvey}, L. and {Kasliwal}, M.~M. and {Kenworthy}, W.~D. and {Kim}, Y.-L. and {Kuhn}, D. and {Kulkarni}, S.~R. and {Lacroix}, L. and {Laher}, R.~R. and {Masci}, F.~J. and {M{\"u}ller-Bravo}, T.~E. and {Miller}, A. and {Osman}, M. and {Perley}, D. and {Popovic}, B. and {Purdum}, J. and {Qin}, Y.-J. and {Racine}, B. and {Reusch}, S. and {Riddle}, R. and {Rosnet}, P. and {Rosselli}, D. and {Ruppin}, F. and {Senzel}, R. and {Rusholme}, B. and {Schweyer}, T. and {Terwel}, J.~H. and {Townsend}, A. and {Tzanidakis}, A. and {Wold}, A. and {Yan}, L.},
        title = "{ZTF SN Ia DR2: Overview}",
      journal = {\aap},
         year = 2025,
        month = feb,
       volume = {694},
          eid = {A1},
        pages = {A1},
          doi = {10.1051/0004-6361/202450388},
archivePrefix = {arXiv},
       eprint = {2409.04346},
 primaryClass = {astro-ph.CO},
       adsurl = {https://ui.adsabs.harvard.edu/abs/2025A&A...694A...1R}
}

@ARTICLE{roepke--hillebrandt--05,
       author = {{R{\"o}pke}, F.~K. and {Hillebrandt}, W.},
        title = "{Full-star type Ia supernova explosion models}",
      journal = {\aap},
         year = 2005,
        month = feb,
       volume = {431},
        pages = {635-645},
          doi = {10.1051/0004-6361:20041859},
archivePrefix = {arXiv},
       eprint = {astro-ph/0409286},
 primaryClass = {astro-ph},
       adsurl = {https://ui.adsabs.harvard.edu/abs/2005A&A...431..635R}
}

@ARTICLE{roepke--06,
   author = {{R{\"o}pke}, F.~K. and {Hillebrandt}, W. and {Niemeyer}, J.~C. and 
	{Woosley}, S.~E.},
    title = "{Multi-spot ignition in type Ia supernova models}",
  journal = {\aap},
   eprint = {astro-ph/0510474},
     year = 2006,
    month = mar,
   volume = 448,
    pages = {1-14},
      doi = {10.1051/0004-6361:20053926},
   adsurl = {http://adsabs.harvard.edu/abs/2006A\%26A...448....1R}
}

@ARTICLE{roepke--07,
   author = {{R{\"o}pke}, F.~K. and {Hillebrandt}, W. and {Schmidt}, W. and 
	{Niemeyer}, J.~C. and {Blinnikov}, S.~I. and {Mazzali}, P.~A.
	},
    title = "{A Three-Dimensional Deflagration Model for Type Ia Supernovae Compared with Observations}",
  journal = {\apj},
archivePrefix = "arXiv",
   eprint = {0707.1024},
     year = 2007,
    month = oct,
   volume = 668,
    pages = {1132-1139},
      doi = {10.1086/521347},
   adsurl = {http://adsabs.harvard.edu/abs/2007ApJ...668.1132R}
}

@ARTICLE{roepke--18,
       author = {{R{\"o}pke}, Friedrich K. and {Sim}, Stuart A.},
        title = "{Models for Type Ia Supernovae and Related Astrophysical Transients}",
      journal = {\ssr},
         year = 2018,
        month = jun,
       volume = {214},
       number = {4},
          eid = {72},
        pages = {72},
          doi = {10.1007/s11214-018-0503-8},
archivePrefix = {arXiv},
       eprint = {1805.07268},
 primaryClass = {astro-ph.SR},
       adsurl = {https://ui.adsabs.harvard.edu/abs/2018SSRv..214...72R}
}

@ARTICLE{ruiter--25,
       author = {{Ruiter}, Ashley Jade and {Seitenzahl}, Ivo Rolf},
        title = "{Type Ia supernova progenitors: a contemporary view of a long-standing puzzle}",
      journal = {\aapr},
         year = 2025,
        month = dec,
       volume = {33},
       number = {1},
          eid = {1},
        pages = {1},
          doi = {10.1007/s00159-024-00158-9},
archivePrefix = {arXiv},
       eprint = {2412.01766},
 primaryClass = {astro-ph.SR},
       adsurl = {https://ui.adsabs.harvard.edu/abs/2025A&ARv..33....1R}
}

@ARTICLE{05hk--400days,
   author = {{Sahu}, D.~K. and {Tanaka}, M. and {Anupama}, G.~C. and {Kawabata}, K.~S. and 
	{Maeda}, K. and {Tominaga}, N. and {Nomoto}, K. and {Mazzali}, P.~A. and 
	{Prabhu}, T.~P.},
    title = "{The Evolution of the Peculiar Type Ia Supernova SN 2005hk over 400 Days}",
  journal = {\apj},
archivePrefix = "arXiv",
   eprint = {0710.3636},
     year = 2008,
    month = jun,
   volume = 680,
    pages = {580-592},
      doi = {10.1086/587772},
   adsurl = {http://adsabs.harvard.edu/abs/2008ApJ...680..580S}
}

@ARTICLE{sasdelli--14,
   author = {{Sasdelli}, M. and {Mazzali}, P.~A. and {Pian}, E. and {Nomoto}, K. and 
	{Hachinger}, S. and {Cappellaro}, E. and {Benetti}, S.},
    title = "{Abundance stratification in Type Ia supernovae - IV. The luminous, peculiar SN 1991T}",
  journal = {\mnras},
archivePrefix = "arXiv",
   eprint = {1409.0116},
 primaryClass = "astro-ph.SR",
     year = 2014,
    month = nov,
   volume = 445,
    pages = {711-725},
      doi = {10.1093/mnras/stu1777},
   adsurl = {http://adsabs.harvard.edu/abs/2014MNRAS.445..711S}
}

@ARTICLE{sasdelli--17,
       author = {{Sasdelli}, Michele and {Hillebrandt}, W. and {Kromer}, M. and {Ishida}, E.~E.~O. and {R{\"o}pke}, F.~K. and {Sim}, S.~A. and {Pakmor}, R. and {Seitenzahl}, I.~R. and {Fink}, M.},
        title = "{A metric space for Type Ia supernova spectra: a new method to assess explosion scenarios}",
      journal = {\mnras},
         year = 2017,
        month = apr,
       volume = {466},
       number = {4},
        pages = {3784-3809},
          doi = {10.1093/mnras/stw3323},
archivePrefix = {arXiv},
       eprint = {1612.07104},
 primaryClass = {astro-ph.HE},
       adsurl = {https://ui.adsabs.harvard.edu/abs/2017MNRAS.466.3784S}
}

@ARTICLE{seitenzahl--13,
   author = {{Seitenzahl}, I.~R. and {Ciaraldi-Schoolmann}, F. and {R{\"o}pke}, F.~K. and 
	{Fink}, M. and {Hillebrandt}, W. and {Kromer}, M. and {Pakmor}, R. and 
	{Ruiter}, A.~J. and {Sim}, S.~A. and {Taubenberger}, S.},
    title = "{Three-dimensional delayed-detonation models with nucleosynthesis for Type Ia supernovae}",
  journal = {\mnras},
archivePrefix = "arXiv",
   eprint = {1211.3015},
 primaryClass = "astro-ph.SR",
     year = 2013,
    month = feb,
   volume = 429,
    pages = {1156-1172},
      doi = {10.1093/mnras/sts402},
   adsurl = {http://adsabs.harvard.edu/abs/2013MNRAS.429.1156S}
}

@ARTICLE{seitenzahl--16,
       author = {{Seitenzahl}, Ivo R. and {Kromer}, Markus and {Ohlmann}, Sebastian T. and
         {Ciaraldi-Schoolmann}, Franco and {Marquardt}, Kai and {Fink}, Michael and
         {Hillebrandt}, Wolfgang and {Pakmor}, R{\"u}diger and
         {R{\"o}pke}, Friedrich K. and {Ruiter}, Ashley J.},
        title = "{Three-dimensional simulations of gravitationally confined detonations compared to observations of SN 1991T}",
      journal = {\aap},
         year = "2016",
        month = "Jul",
       volume = {592},
          eid = {A57},
        pages = {A57},
          doi = {10.1051/0004-6361/201527251},
archivePrefix = {arXiv},
       eprint = {1606.00089},
 primaryClass = {astro-ph.SR},
       adsurl = {https://ui.adsabs.harvard.edu/abs/2016A&A...592A..57S}
}

@INBOOK{seitenzahl--17,
       author = {{Seitenzahl}, Ivo Rolf and {Townsley}, Dean M.},
        title = "{Nucleosynthesis in Thermonuclear Supernovae}",
    booktitle = {Handbook of Supernovae},
         year = 2017,
       editor = {{Alsabti}, Athem W. and {Murdin}, Paul},
        pages = {1955},
          doi = {10.1007/978-3-319-21846-5_87},
       adsurl = {https://ui.adsabs.harvard.edu/abs/2017hsn..book.1955S}
}

@ARTICLE{shappee--11,
       author = {{Shappee}, Benjamin J. and {Stanek}, K.~Z.},
        title = "{A New Cepheid Distance to the Giant Spiral M101 Based on Image Subtraction of Hubble Space Telescope/Advanced Camera for Surveys Observations}",
      journal = {\apj},
         year = 2011,
        month = jun,
       volume = {733},
       number = {2},
          eid = {124},
        pages = {124},
          doi = {10.1088/0004-637X/733/2/124},
archivePrefix = {arXiv},
       eprint = {1012.3747},
 primaryClass = {astro-ph.CO},
       adsurl = {https://ui.adsabs.harvard.edu/abs/2011ApJ...733..124S}
}

@ARTICLE{shappee--17,
       author = {{Shappee}, B.~J. and {Stanek}, K.~Z. and {Kochanek}, C.~S. and {Garnavich}, P.~M.},
        title = "{Whimper of a Bang: Documenting the Final Days of the Nearby Type Ia Supernova 2011fe}",
      journal = {\apj},
         year = 2017,
        month = may,
       volume = {841},
       number = {1},
          eid = {48},
        pages = {48},
          doi = {10.3847/1538-4357/aa6eab},
archivePrefix = {arXiv},
       eprint = {1608.01155},
 primaryClass = {astro-ph.HE},
       adsurl = {https://ui.adsabs.harvard.edu/abs/2017ApJ...841...48S}
}

@ARTICLE{shen--14,
       author = {{Shen}, Ken J. and {Bildsten}, Lars},
        title = "{The Ignition of Carbon Detonations via Converging Shock Waves in White Dwarfs}",
      journal = {\apj},
         year = 2014,
        month = apr,
       volume = {785},
       number = {1},
          eid = {61},
        pages = {61},
          doi = {10.1088/0004-637X/785/1/61},
archivePrefix = {arXiv},
       eprint = {1305.6925},
 primaryClass = {astro-ph.HE},
       adsurl = {https://ui.adsabs.harvard.edu/abs/2014ApJ...785...61S}
}

@ARTICLE{shen--14b,
       author = {{Shen}, Ken J. and {Moore}, Kevin},
        title = "{The Initiation and Propagation of Helium Detonations in White Dwarf Envelopes}",
      journal = {\apj},
         year = 2014,
        month = dec,
       volume = {797},
       number = {1},
          eid = {46},
        pages = {46},
          doi = {10.1088/0004-637X/797/1/46},
archivePrefix = {arXiv},
       eprint = {1409.3568},
 primaryClass = {astro-ph.HE},
       adsurl = {https://ui.adsabs.harvard.edu/abs/2014ApJ...797...46S}
}

@ARTICLE{shen--18,
       author = {{Shen}, Ken J. and {Kasen}, Daniel and {Miles}, Broxton J. and
         {Townsley}, Dean M.},
        title = "{Sub-Chandrasekhar-mass White Dwarf Detonations Revisited}",
      journal = {The Astrophysical Journal},
         year = "2018",
        month = "Feb",
       volume = {854},
       number = {1},
          eid = {52},
        pages = {52},
          doi = {10.3847/1538-4357/aaa8de},
archivePrefix = {arXiv},
       eprint = {1706.01898},
 primaryClass = {astro-ph.HE},
       adsurl = {https://ui.adsabs.harvard.edu/abs/2018ApJ...854...52S}
}

@ARTICLE{shen--21,
       author = {{Shen}, Ken J. and {Boos}, Samuel J. and {Townsley}, Dean M. and {Kasen}, Daniel},
        title = "{Multidimensional Radiative Transfer Calculations of Double Detonations of Sub-Chandrasekhar-mass White Dwarfs}",
      journal = {\apj},
         year = 2021,
        month = nov,
       volume = {922},
       number = {1},
          eid = {68},
        pages = {68},
          doi = {10.3847/1538-4357/ac2304},
archivePrefix = {arXiv},
       eprint = {2108.12435},
 primaryClass = {astro-ph.SR},
       adsurl = {https://ui.adsabs.harvard.edu/abs/2021ApJ...922...68S}
}

@ARTICLE{silverman--ia,
   author = {{Silverman}, J.~M. and {Foley}, R.~J. and {Filippenko}, A.~V. and 
	{Ganeshalingam}, M. and {Barth}, A.~J. and {Chornock}, R. and 
	{Griffith}, C.~V. and {Kong}, J.~J. and {Lee}, N. and {Leonard}, D.~C. and 
	{Matheson}, T. and {Miller}, E.~G. and {Steele}, T.~N. and {Barris}, B.~J. and 
	{Bloom}, J.~S. and {Cobb}, B.~E. and {Coil}, A.~L. and {Desroches}, L.-B. and 
	{Gates}, E.~L. and {Ho}, L.~C. and {Jha}, S.~W. and {Kandrashoff}, M.~T. and 
	{Li}, W. and {Mandel}, K.~S. and {Modjaz}, M. and {Moore}, M.~R. and 
	{Mostardi}, R.~E. and {Papenkova}, M.~S. and {Park}, S. and 
	{Perley}, D.~A. and {Poznanski}, D. and {Reuter}, C.~A. and 
	{Scala}, J. and {Serduke}, F.~J.~D. and {Shields}, J.~C. and 
	{Swift}, B.~J. and {Tonry}, J.~L. and {Van Dyk}, S.~D. and {Wang}, X. and 
	{Wong}, D.~S.},
    title = "{Berkeley Supernova Ia Program - I. Observations, data reduction and spectroscopic sample of 582 low-redshift Type Ia supernovae}",
  journal = {\mnras},
archivePrefix = "arXiv",
   eprint = {1202.2128},
     year = 2012,
    month = sep,
   volume = 425,
    pages = {1789-1818},
      doi = {10.1111/j.1365-2966.2012.21270.x},
   adsurl = {http://adsabs.harvard.edu/abs/2012MNRAS.425.1789S}
}

@ARTICLE{silverman--12b,
       author = {{Silverman}, Jeffrey M. and {Kong}, Jason J. and {Filippenko}, Alexei V.},
        title = "{Berkeley Supernova Ia Program - II. Initial analysis of spectra obtained near maximum brightness}",
      journal = {\mnras},
         year = 2012,
        month = sep,
       volume = {425},
       number = {3},
        pages = {1819-1888},
          doi = {10.1111/j.1365-2966.2012.21269.x},
archivePrefix = {arXiv},
       eprint = {1202.2129},
 primaryClass = {astro-ph.CO},
       adsurl = {https://ui.adsabs.harvard.edu/abs/2012MNRAS.425.1819S}
}

@ARTICLE{sim--13,
   author = {{Sim}, S.~A. and {Seitenzahl}, I.~R. and {Kromer}, M. and {Ciaraldi-Schoolmann}, F. and 
	{R{\"o}pke}, F.~K. and {Fink}, M. and {Hillebrandt}, W. and 
	{Pakmor}, R. and {Ruiter}, A.~J. and {Taubenberger}, S.},
    title = "{Synthetic light curves and spectra for three-dimensional delayed-detonation models of Type Ia supernovae}",
  journal = {\mnras},
archivePrefix = "arXiv",
   eprint = {1308.4833},
 primaryClass = "astro-ph.HE",
     year = 2013,
    month = nov,
   volume = 436,
    pages = {333-347},
      doi = {10.1093/mnras/stt1574},
   adsurl = {http://adsabs.harvard.edu/abs/2013MNRAS.436..333S}
}

@ARTICLE{sluijterman--23,
       author = {{Sluijterman}, Laurens and {Cator}, Eric and {Heskes}, Tom},
        title = "{Optimal Training of Mean Variance Estimation Neural Networks}",
      journal = {arXiv e-prints},
         year = 2023,
        month = feb,
          eid = {arXiv:2302.08875},
        pages = {arXiv:2302.08875},
          doi = {10.48550/arXiv.2302.08875},
archivePrefix = {arXiv},
       eprint = {2302.08875},
 primaryClass = {stat.ML},
       adsurl = {https://ui.adsabs.harvard.edu/abs/2023arXiv230208875S}
}

@ARTICLE{springel--10,
       author = {{Springel}, Volker},
        title = "{Smoothed Particle Hydrodynamics in Astrophysics}",
      journal = {\araa},
         year = 2010,
        month = sep,
       volume = {48},
        pages = {391-430},
          doi = {10.1146/annurev-astro-081309-130914},
archivePrefix = {arXiv},
       eprint = {1109.2219},
 primaryClass = {astro-ph.CO},
       adsurl = {https://ui.adsabs.harvard.edu/abs/2010ARA&A..48..391S}
}

@ARTICLE{srivastav--22,
       author = {{Srivastav}, Shubham and {Smartt}, S.~J. and {Huber}, M.~E. and {Chambers}, K.~C. and {Angus}, C.~R. and {Chen}, T.-W. and {Callan}, F.~P. and {Gillanders}, J.~H. and {McBrien}, O.~R. and {Sim}, S.~A. and {Fulton}, M. and {Hjorth}, J. and {Smith}, K.~W. and {Young}, D.~R. and {Auchettl}, K. and {Anderson}, J.~P. and {Pignata}, G. and {de Boer}, T.~J.~L. and {Lin}, C.-C. and {Magnier}, E.~A.},
        title = "{SN 2020kyg and the rates of faint Iax supernovae from ATLAS}",
      journal = {\mnras},
         year = 2022,
        month = apr,
       volume = {511},
       number = {2},
        pages = {2708-2731},
          doi = {10.1093/mnras/stac177},
archivePrefix = {arXiv},
       eprint = {2111.09491},
 primaryClass = {astro-ph.HE},
       adsurl = {https://ui.adsabs.harvard.edu/abs/2022MNRAS.511.2708S}
}

@ARTICLE{stanishev--18,
       author = {{Stanishev}, V. and {Goobar}, A. and {Amanullah}, R. and {Bassett}, B. and {Fantaye}, Y.~T. and {Garnavich}, P. and {Hlozek}, R. and {Nordin}, J. and {Okouma}, P.~M. and {{\"O}stman}, L. and {Sako}, M. and {Scalzo}, R. and {Smith}, M.},
        title = "{Type Ia supernova Hubble diagram with near-infrared and optical observations}",
      journal = {\aap},
         year = 2018,
        month = jul,
       volume = {615},
          eid = {A45},
        pages = {A45},
          doi = {10.1051/0004-6361/201732357},
       adsurl = {https://ui.adsabs.harvard.edu/abs/2018A&A...615A..45S}
}

@ARTICLE{stehle--05,
   author = {{Stehle}, M. and {Mazzali}, P.~A. and {Benetti}, S. and {Hillebrandt}, W.
	},
    title = "{Abundance stratification in Type Ia supernovae - I. The case of SN 2002bo}",
  journal = {\mnras},
   eprint = {astro-ph/0409342},
     year = 2005,
    month = jul,
   volume = 360,
    pages = {1231-1243},
      doi = {10.1111/j.1365-2966.2005.09116.x},
   adsurl = {http://adsabs.harvard.edu/abs/2005MNRAS.360.1231S}
}

@ARTICLE{tides,
       author = {{Swann}, E. and {Sullivan}, M. and {Carrick}, J. and {Hoenig}, S. and {Hook}, I. and {Kotak}, R. and {Maguire}, K. and {McMahon}, R. and {Nichol}, R. and {Smartt}, S.},
        title = "{4MOST Consortium Survey 10: The Time-Domain Extragalactic Survey (TiDES)}",
      journal = {The Messenger},
         year = 2019,
        month = mar,
       volume = {175},
        pages = {58-61},
          doi = {10.18727/0722-6691/5129},
archivePrefix = {arXiv},
       eprint = {1903.02476},
 primaryClass = {astro-ph.IM},
       adsurl = {https://ui.adsabs.harvard.edu/abs/2019Msngr.175...58S}
}

@ARTICLE{tanaka--11,
       author = {{Tanaka}, Masaomi and {Mazzali}, Paolo A. and {Stanishev}, Vallery and
         {Maurer}, Immanuel and {Kerzendorf}, Wolfgang E. and {Nomoto}, Ken'ichi},
        title = "{Abundance stratification in Type Ia supernovae - III. The normal SN 2003du}",
      journal = {Monthly Notices of the Royal Astronomical Society},
         year = "2011",
        month = "Jan",
       volume = {410},
       number = {3},
        pages = {1725-1738},
          doi = {10.1111/j.1365-2966.2010.17556.x},
archivePrefix = {arXiv},
       eprint = {1008.3140},
 primaryClass = {astro-ph.SR},
       adsurl = {https://ui.adsabs.harvard.edu/abs/2011MNRAS.410.1725T}
}

@ARTICLE{taubenberger--13,
       author = {{Taubenberger}, S. and {Kromer}, M. and {Pakmor}, R. and {Pignata}, G. and {Maeda}, K. and {Hachinger}, S. and {Leibundgut}, B. and {Hillebrandt}, W.},
        title = "{[O I] {\ensuremath{\lambda}}{\ensuremath{\lambda}}6300, 6364 in the Nebular Spectrum of a Subluminous Type Ia Supernova}",
      journal = {\apjl},
         year = 2013,
        month = oct,
       volume = {775},
       number = {2},
          eid = {L43},
        pages = {L43},
          doi = {10.1088/2041-8205/775/2/L43},
archivePrefix = {arXiv},
       eprint = {1308.3145},
 primaryClass = {astro-ph.SR},
       adsurl = {https://ui.adsabs.harvard.edu/abs/2013ApJ...775L..43T}
}

@ARTICLE{synapps,
       author = {{Thomas}, R.~C. and {Nugent}, P.~E. and {Meza}, J.~C.},
        title = "{SYNAPPS: Data-Driven Analysis for Supernova Spectroscopy}",
      journal = {\pasp},
         year = 2011,
        month = feb,
       volume = {123},
       number = {900},
        pages = {237},
          doi = {10.1086/658673},
       adsurl = {https://ui.adsabs.harvard.edu/abs/2011PASP..123..237T}
}

@ARTICLE{ting--19,
       author = {{Ting}, Yuan-Sen and {Conroy}, Charlie and {Rix}, Hans-Walter and {Cargile}, Phillip},
        title = "{The Payne: Self-consistent ab initio Fitting of Stellar Spectra}",
      journal = {\apj},
         year = 2019,
        month = jul,
       volume = {879},
       number = {2},
          eid = {69},
        pages = {69},
          doi = {10.3847/1538-4357/ab2331},
archivePrefix = {arXiv},
       eprint = {1804.01530},
 primaryClass = {astro-ph.SR},
       adsurl = {https://ui.adsabs.harvard.edu/abs/2019ApJ...879...69T}
}

@ARTICLE{townsley--19,
       author = {{Townsley}, Dean M. and {Miles}, Broxton J. and {Shen}, Ken J. and
         {Kasen}, Daniel},
        title = "{Double Detonations with Thin, Modestly Enriched Helium Layers can Make Normal Type Ia Supernovae}",
      journal = {The Astrophysical Journal},
         year = "2019",
        month = "Jun",
       volume = {878},
       number = {2},
          eid = {L38},
        pages = {L38},
          doi = {10.3847/2041-8213/ab27cd},
archivePrefix = {arXiv},
       eprint = {1903.10960},
 primaryClass = {astro-ph.SR},
       adsurl = {https://ui.adsabs.harvard.edu/abs/2019ApJ...878L..38T}
}

@ARTICLE{vanrossum--2012,
   author = {{van Rossum}, D.~R.},
    title = "{Radiation Energy Balance Method for Calculating the Time Evolution of Type Ia Supernovae during the Post-explosion Phase}",
  journal = {\apj},
archivePrefix = "arXiv",
   eprint = {1206.5463},
 primaryClass = "astro-ph.SR",
     year = 2012,
    month = sep,
   volume = 756,
      eid = {31},
    pages = {31},
      doi = {10.1088/0004-637X/756/1/31},
   adsurl = {http://adsabs.harvard.edu/abs/2012ApJ...756...31V}
}

@ARTICLE{vogl--19,
       author = {{Vogl}, C. and {Sim}, S.~A. and {Noebauer}, U.~M. and {Kerzendorf}, W.~E. and {Hillebrandt}, W.},
        title = "{Spectral modeling of type II supernovae. I. Dilution factors}",
      journal = {\aap},
         year = 2019,
        month = jan,
       volume = {621},
          eid = {A29},
        pages = {A29},
          doi = {10.1051/0004-6361/201833701},
archivePrefix = {arXiv},
       eprint = {1811.02543},
 primaryClass = {astro-ph.HE},
       adsurl = {https://ui.adsabs.harvard.edu/abs/2019A&A...621A..29V}
}

@ARTICLE{webbink--84,
   author = {{Webbink}, R.~F.},
    title = "{Double white dwarfs as progenitors of R Coronae Borealis stars and Type I supernovae}",
  journal = {\apj},
     year = 1984,
    month = feb,
   volume = 277,
    pages = {355-360},
      doi = {10.1086/161701},
   adsurl = {http://adsabs.harvard.edu/abs/1984ApJ...277..355W}
}

@ARTICLE{woosley--94,
   author = {{Woosley}, S.~E. and {Weaver}, T.~A.},
    title = "{Sub-Chandrasekhar mass models for Type IA supernovae}",
  journal = {\apj},
     year = 1994,
    month = mar,
   volume = 423,
    pages = {371-379},
      doi = {10.1086/173813},
   adsurl = {http://adsabs.harvard.edu/abs/1994ApJ...423..371W}
}

@ARTICLE{woosley--07,
       author = {{Woosley}, S.~E.},
        title = "{Type Ia Supernovae: Burning and Detonation in the Distributed Regime}",
      journal = {\apj},
         year = "2007",
        month = "Oct",
       volume = {668},
       number = {2},
        pages = {1109-1117},
          doi = {10.1086/520835},
archivePrefix = {arXiv},
       eprint = {0709.4237},
 primaryClass = {astro-ph},
       adsurl = {https://ui.adsabs.harvard.edu/abs/2007ApJ...668.1109W}
}

@ARTICLE{woosley--09,
       author = {{Woosley}, S.~E. and {Kerstein}, A.~R. and {Sankaran}, V. and
         {Aspden}, A.~J. and {R{\"o}pke}, F.~K.},
        title = "{Type Ia Supernovae: Calculations of Turbulent Flames Using the Linear Eddy Model}",
      journal = {\apj},
         year = "2009",
        month = "Oct",
       volume = {704},
       number = {1},
        pages = {255-273},
          doi = {10.1088/0004-637X/704/1/255},
archivePrefix = {arXiv},
       eprint = {0811.3610},
 primaryClass = {astro-ph},
       adsurl = {https://ui.adsabs.harvard.edu/abs/2009ApJ...704..255W}
}

@ARTICLE{yadavalli--24,
       author = {{Yadavalli}, S. Karthik and {Villar}, V. Ashley and {Izzo}, Luca and {Zenati}, Yossef and {Foley}, Ryan J. and {Wheeler}, J. Craig and {Angus}, Charlotte R. and {B{\'a}nhidi}, Dominik and {Auchettl}, Katie and {B{\'\i}r{\'o}}, Barna Imre and {B{\'o}di}, Attila and {Bodola}, Zs{\'o}fia and {de Boer}, Thomas and {Chambers}, Kenneth C. and {Chornock}, Ryan and {Coulter}, David A. and {Cs{\'a}nyi}, Istv{\'a}n and {Cseh}, Borb{\'a}la and {Dandu}, Srujan and {Davis}, Kyle W. and {Dickinson}, Connor Braden and {Farias}, Diego and {Farah}, Joseph and {Gall}, Christa and {Gao}, Hua and {Howell}, D. Andrew and {Jacobson-Galan}, Wynn V. and {Khetan}, Nandita and {Kilpatrick}, Charles D. and {K{\"o}nyves-T{\'o}th}, R{\'e}ka and {Kriskovics}, Levente and {LeBaron}, Natalie and {Loertscher}, Kayla and {Le Saux}, X.~K. and {Margutti}, Raffaella and {Magnier}, Eugene A. and {McCully}, Curtis and {McGill}, Peter and {Miao}, Hao-Yu and {Newsome}, Megan and {Padilla Gonzalez}, Estefania and {P{\'a}l}, Andr{\'a}s and {P{\'a}l}, Bor{\'o}ka H. and {Pan}, Yen-Chen and {Politsch}, Collin A. and {Ransome}, Conor L. and {Ramirez-Ruiz}, Enrico and {Rest}, Armin and {Rest}, Sofia and {Robinson}, Olivia and {Sears}, Huei and {Scheer}, Jackson and {S{\'o}dor}, {\'A}d{\'a}m and {Swift}, Jonathan and {Sz{\'e}kely}, P{\'e}ter and {Szak{\'a}ts}, R{\'o}bert and {Szalai}, Tam{\'a}s and {Taggart}, Kirsty and {Terreran}, Giacomo and {Venkatraman}, Padma and {Vink{\'o}}, J{\'o}zsef and {Yang}, Grace and {Zhou}, Henry},
        title = "{SN 2022oqm: A Bright and Multipeaked Calcium-rich Transient}",
      journal = {\apj},
         year = 2024,
        month = sep,
       volume = {972},
       number = {2},
          eid = {194},
        pages = {194},
          doi = {10.3847/1538-4357/ad5a7c},
archivePrefix = {arXiv},
       eprint = {2308.12991},
 primaryClass = {astro-ph.HE},
       adsurl = {https://ui.adsabs.harvard.edu/abs/2024ApJ...972..194Y}
}

@ARTICLE{wiserep,
   author = {{Yaron}, O. and {Gal-Yam}, A.},
    title = "{WISeREP - An Interactive Supernova Data Repository}",
  journal = {\pasp},
archivePrefix = "arXiv",
   eprint = {1204.1891},
 primaryClass = "astro-ph.IM",
     year = 2012,
    month = jul,
   volume = 124,
    pages = {668-681},
      doi = {10.1086/666656},
   adsurl = {http://adsabs.harvard.edu/abs/2012PASP..124..668Y}
}

%\begin{thebibliography}{99}
%\end{thebibliography}

%%%%%%%%%%%%%%%%%%%%%%%%%%%%%%%%%%%%%%%%%%%%%%%%%%

%%%%%%%%%%%%%%%%% APPENDICES %%%%%%%%%%%%%%%%%%%%%

\appendix

\section{Sampling new input parameters from HESMA models}
\label{sect:appdx:scatter}
Here we present additional discussion and figures showing how new input parameters were sampled from existing explosion scenarios. Accounting for all systematics relevant to explosion modelling is non-trivial. Some existing suites explore the impact of parameters such as metallicity or ignition conditions, but this is incomplete and non-exhaustive. Where appropriate, we generally base our assumed scatter during sampling on results from the existing models, but allow for somewhat larger variations in an effort to address these systematics. Table~\ref{tab:scatter} lists the assumed scatter boundaries when sampling input parameters.

\renewcommand{\arraystretch}{1.4} % 1.5 times the default height
\begin{table}
\centering
\caption{Boundaries for scatter on parameters during sampling}\tabularnewline
\label{tab:scatter}\tabularnewline
\begin{tabular}{llllll}\hline
\hline
Parameter    & DEF         & DDT    & DOD    &  GCD           & VM        \tabularnewline
\hline
$M_{\textrm{ej}}$           & $^{+100}_{-20}$   & $\cdots$  & $\cdots$ & $\cdots$ & $\cdots$  \tabularnewline
$f_C$                       & $\cdots$          & $\cdots$  & $\cdots$ & $\pm20$ & $\cdots$  \tabularnewline
$f_C^b$                     & $\pm20$           & $\pm20^{\textrm{*}}$   & $\pm20^{\textrm{†}}$ & $\pm20$ & $\pm20$  \tabularnewline
$f_C^{\textrm{IGE}}$        & $\pm20$           & $\pm20$   & $\pm20$ & $\pm20$ & $\pm50$  \tabularnewline
$f_C^{\textrm{Ni}}$         & $\pm20$           & $\pm20$   & $\pm20^{\textrm{‡}}$ & $\pm20$ & $\cdots$  \tabularnewline
$f_S^b$                     & $\cdots$          & $\cdots$  & $^{+20}_{-80}$ & $^{+20}_{-80}$ & $\cdots$  \tabularnewline
$f_S^{\textrm{IGE}}$        & $\cdots$          & $\cdots$  & $\cdots$ & $\pm20$ & $\cdots$  \tabularnewline
$f_S^{\textrm{Ni}}$         & $\cdots$          & $\cdots$  & $\pm90$ & $\pm20$ &  $\cdots$ \tabularnewline
$f_S^{\textrm{C/O}}$        & $\cdots$          & $\cdots$  & $\pm80$ & $\pm20$ &  $\cdots$ \tabularnewline
$KE$                        & $\pm30$           & $\pm30$   & $\pm30$ & $\pm30$ &  $\pm30$ \tabularnewline
\hline
\hline
\multicolumn{6}{p{\dimexpr\linewidth-12\tabcolsep}}{%
\textit{Note:} Parameters without values are either sampled independently or are not relevant to that model. See the relevant sections for further details. Values are given as percentages.%
} \tabularnewline
\multicolumn{6}{p{\dimexpr\linewidth-12\tabcolsep}}{*: Up to $f_C^b$ = 0.96, the maximum predicted by the underlying models.%
}
\tabularnewline
\multicolumn{6}{p{\dimexpr\linewidth-12\tabcolsep}}{†: Up to $f_C^b$ = 0.99, the maximum predicted by the underlying models.%
}
\tabularnewline
\multicolumn{6}{p{\dimexpr\linewidth-12\tabcolsep}}{‡: Up to $f_C^{\textrm{Ni}}$ = 0.90, the maximum predicted by the underlying models.%
}

\end{tabular}
\end{table}
\renewcommand{\arraystretch}{1.0} % 1.5 times the default height

%%%%%%%%%%%%%%%%%%%%%%%%%%%%%%%%%%%%%%%%%%%%%%%%%%%%%%%%%%%%%%%%%%%%%%%%%%%%%%%%%%%%%%%%%%%%%%%%%%%%%%%%%%%%%%
%%%%%%%%%%%%%%%%%%%%%%%%%%%%%%%%%%%%%%%%%%%%%%%%%%%%%%%%%%%%%%%%%%%%%%%%%%%%%%%%%%%%%%%%%%%%%%%%%%%%%%%%%%%%%%

\subsection{Pure deflagrations (DEF)}
\label{sect:appdx:def}

\begin{figure}
\centering
\includegraphics[width=\columnwidth]{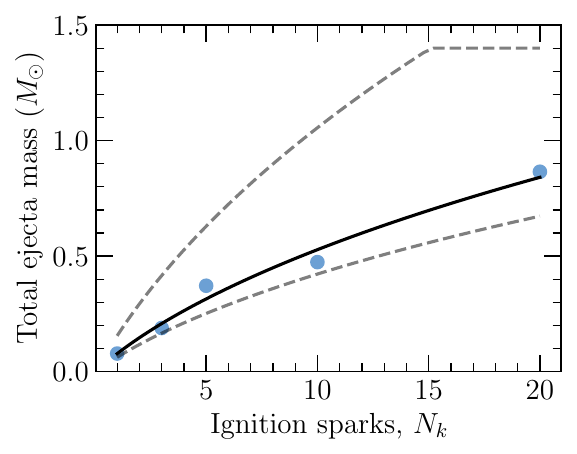}
\caption{Correlation between ignition sparks, $N_k$, and total ejecta mass used to sample training data models shown as a black solid line. Circles show predictions from the \citealt{fink-2014} pure deflagration models. The upper and lower limits of the training data samples are shown as dashed lines.}
\label{fig:def_param1}
\centering
\end{figure}

Figure~\ref{fig:def_param1} shows the relationship between the number of ignition sparks, $N_k$, and the total ejecta mass for the \cite{fink-2014} models. As discussed in Sect.~\ref{sect:def}, we limit our sampling of $N_k$ to 1 -- 20. The number of sparks is an artificial parameter used to control the strength of the explosion therefore it is unsurprising that a larger number of sparks corresponds to an increased ejecta mass. Ejecta masses measured among SNe~Iax range from $\sim$0.1 -- 1.4~$M_{\odot}$ \citep{obs--05hk--08a, srivastav--22}, while those predicted by the \cite{fink-2014} models for $1 \leq N_k \leq 20$ range from $\sim$0.1 -- 0.9~$M_{\odot}$ and are typically systematically lower than those observed in SNe~Iax of similar luminosities. When sampling the ejecta mass based on $N_{k,x}$, we allow for a scatter of $+100$~per~cent, up to $1.4~M_{\odot}$, to allow for models with similar ejecta masses to those observed in SNe~Iax. Whether such an increased ejecta mass for a given luminosity could be achieved within the pure deflagration scenario remains to be seen.

\begin{figure}
\centering
\includegraphics[width=\columnwidth]{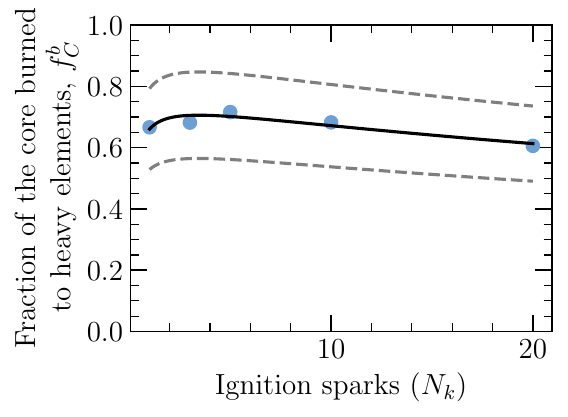}
\caption{As in Fig.~\ref{fig:def_param1} for ignition sparks, $N_k$, and the fraction of the core burned to heavy elements, $f^b_C$.}
\label{fig:def_param3}
\centering
\end{figure}

In Fig.~\ref{fig:def_param3} we show the relation between the fraction of the core burned to heavy elements and the number of ignition sparks in the \cite{fink-2014} models. In general, the models show only minor variations in how much material is burned. For larger values of $N_k$ ($\gtrsim$200), the \cite{fink-2014} models show systematically lower $f^b_C$. \cite{fink-2014} also investigate the impact of the white dwarf central density on the resulting explosion for the case of $N_k = 100$ and find that this results in a $\sim$15~per~cent difference in the fraction of burned material relative to their fiducial value. We therefore allow for a conservative scatter of 20~per~cent when sampling. Likewise, Fig.~\ref{fig:def_param5} shows the fraction of IGE burned to $^{56}$Ni in the core as a function of $N_k$. The \cite{fink-2014} models again show only minor variations and a $\sim$15~per~cent difference in the $^{56}$Ni to IGE fraction for models with different central densities, therefore we allow for 20~per~cent scatter when sampling.

\begin{figure}
\centering
\includegraphics[width=\columnwidth]{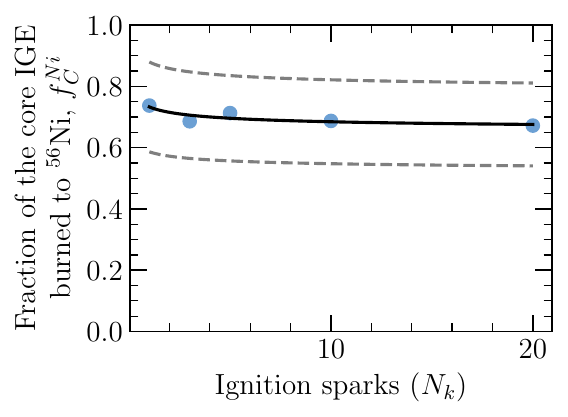}
\caption{As in Fig.~\ref{fig:def_param1} for ignition sparks, $N_k$, and the fraction of the core IGE burned to $^{56}$Ni, $f^{Ni}_C$.}
\label{fig:def_param5}
\centering
\end{figure}

Figure~\ref{fig:def_param10} shows the relation between kinetic energy and total ejecta mass. Models with $N_k$ $\gtrsim$ 150 all result in 1.4~$M_{\odot}$ of ejecta, but predict variations in kinetic energy of up to approximately 20~per~cent. When sampling kinetic energy we therefore take a conservative scatter of 30~per~cent.

\begin{figure}
\centering
\includegraphics[width=\columnwidth]{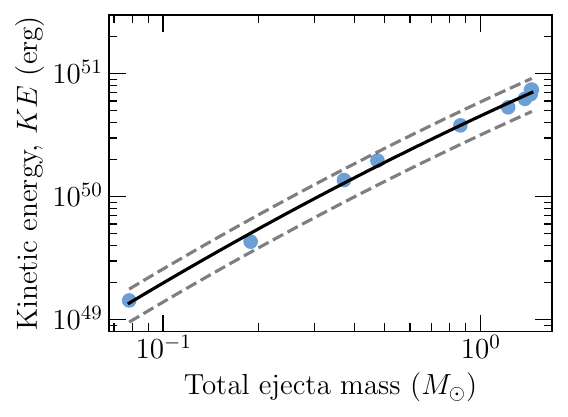}
\caption{As in Fig.~\ref{fig:def_param1} for total ejecta mass and kinetic energy.}
\label{fig:def_param10}
\centering
\end{figure}

%%%%%%%%%%%%%%%%%%%%%%%%%%%%%%%%%%%%%%%%%%%%%%%%%%%%%%%%%%%%%%%%%%%%%%%%%%%%%%%%%%%%%%%%%%%%%%%%%%%%%%%%%%%%%%
%%%%%%%%%%%%%%%%%%%%%%%%%%%%%%%%%%%%%%%%%%%%%%%%%%%%%%%%%%%%%%%%%%%%%%%%%%%%%%%%%%%%%%%%%%%%%%%%%%%%%%%%%%%%%%
\subsection{Delayed detonations (DDT)}
\label{sect:appdx:ddt}

\begin{figure}
\centering
\includegraphics[width=\columnwidth]{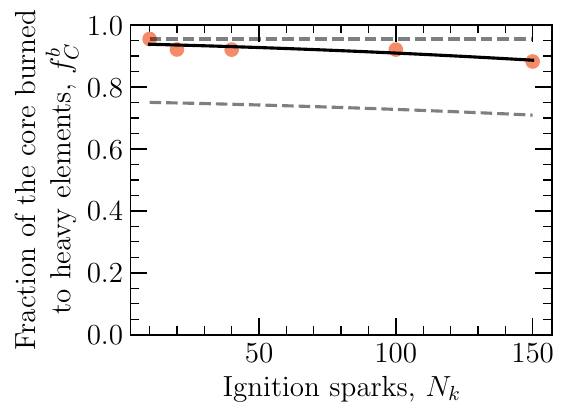}
\caption{Correlation between ignition sparks, $N_k$, and the fraction of the core burned to heavy elements, $f^b_C$, used to sample training data models shown as a black line. Circles show predictions from the \citealt{seitenzahl--13} delayed detonation models. The upper and lower limits of the training data samples are shown as dashed lines.}
\label{fig:ddt_param3}
\centering
\end{figure}

The relationship between the fraction of the core burned to heavy elements and the number of ignition sparks for the \cite{seitenzahl--13} delayed detonation models is shown in Fig.~\ref{fig:ddt_param3}. For $N_k$ ranging from 10 -- 150, the \cite{seitenzahl--13} models show only minor variations. \cite{seitenzahl--13} test the impact of variations in the white dwarf central density and metallicity for N100 specifically, finding the resultant fraction of the core that is burned varies by up to approximately 5~per~cent. \cite{seitenzahl--13} also generate variations with more compact distributions of ignition kernels (N300C and N1600C). In the case of the N1600C model, the fraction of burned material varies by $\sim$10~per~cent. It it is unclear what effect compounded changes in initial conditions would have, in particular for models with $10 \leq N_k \leq 150$, therefore in the absence of such models we again take a conservative approach and allow for 20~per~cent scatter when sampling. To allow for at least some unburned material to remain in the core, we set an upper limit equal to that of the $N_k = 10$ model, 0.96. The fraction of IGE burned to $^{56}$Ni shows similar variations and therefore here we also assume 20~per~cent scatter (Fig.~\ref{fig:ddt_param5}).

\begin{figure}
\centering
\includegraphics[width=\columnwidth]{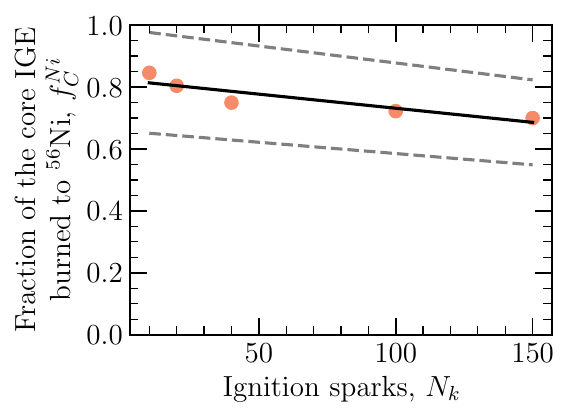}
\caption{As in Fig.~\ref{fig:ddt_param3} for ignition sparks, $N_k$, and the fraction of the core IGE burned to $^{56}$Ni, $f^{Ni}_C$.}
\label{fig:ddt_param5}
\centering
\end{figure}

\begin{figure}
\centering
\includegraphics[width=\columnwidth]{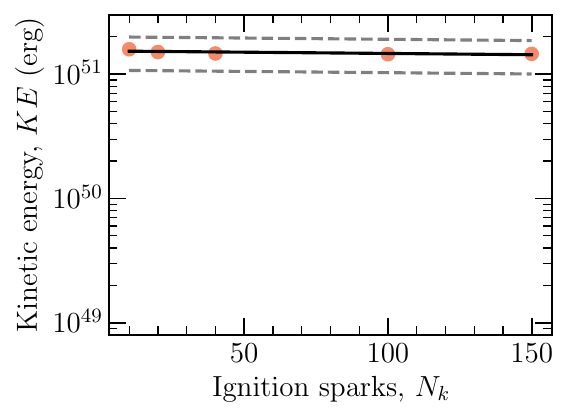}
\caption{As in Fig.~\ref{fig:ddt_param3} for ignition sparks, $N_k$, and kinetic energy.}
\label{fig:ddt_param10}
\centering
\end{figure}

As discussed in Sect.~\ref{sect:dedet}, all delayed detonation models presented by \cite{seitenzahl--13} produce $\sim$1.4~$M_{\odot}$ of ejecta. Unlike our pure deflagration models, we therefore sample the kinetic energy based on the number of ignition sparks. The \cite{seitenzahl--13} models show only minor variations in kinetic energy as a function of $N_k$, while changes in the central density or compactness of the ignition kernels can result in variations of $\sim$10~per~cent (Fig.~\ref{fig:ddt_param10}). As with the pure deflagration models, we take a conservative approach and allow up to 30~per~cent scatter when sampling.

%%%%%%%%%%%%%%%%%%%%%%%%%%%%%%%%%%%%%%%%%%%%%%%%%%%%%%%%%%%%%%%%%%%%%%%%%%%%%%%%%%%%%%%%%%%%%%%%%%%%%%%%%%%%%%
%%%%%%%%%%%%%%%%%%%%%%%%%%%%%%%%%%%%%%%%%%%%%%%%%%%%%%%%%%%%%%%%%%%%%%%%%%%%%%%%%%%%%%%%%%%%%%%%%%%%%%%%%%%%%%
\subsection{Double detonations (DOD)}
\label{sect:appdx:dod}

\begin{figure}
\centering
\includegraphics[width=\columnwidth]{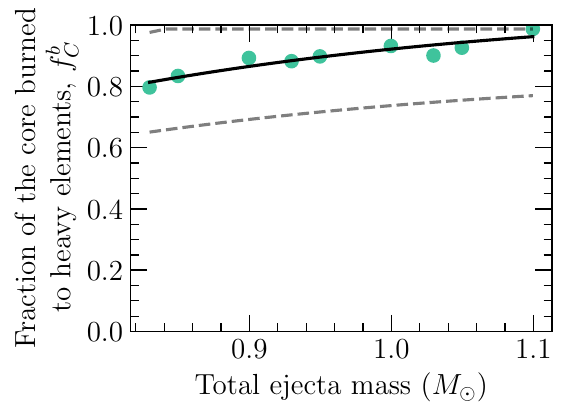}
\caption{Correlation between total ejecta mass and the fraction of the core burned to heavy elements, $f^b_C$, used to sample training data models shown as a black line. Circles show predictions from the \citealt{gronow--21} double detonation models. The upper and lower limits are shown as dashed lines.}
\label{fig:dod_param3}
\centering
\end{figure}

Figure~\ref{fig:dod_param3} shows the fraction of the core burned to heavy elements as a function of the total ejecta mass (core mass + shell mass) for the \cite{gronow--21} models. The models show a strong correlation, with larger ejecta masses corresponding to larger fractions of the core burned to heavy elements. \cite{gronow--21} test the impact of mixing between the shell and core and find that removing mixing can result in differences of up to a few percent in the final abundances of the core after explosion. In addition, \cite{gronow--21b} test the impact of metallicity and find that, depending on the total ejecta mass, the final abundances in the core can vary by $\sim$10 -- 15~per~cent. We therefore again take a conservative approach and adopt scatter of up to 20~per~cent. The fraction of IGE burned to $^{56}$Ni in the core shows a similar, although less pronounced variation, and we again adopt conservative scatter of up to 20~per~cent (Fig.~\ref{fig:dod_param5}). To avoid scenarios where all of the core is burned to heavy elements or all of the IGE are burned to $^{56}$Ni, we set upper limits equal to the maxima of the \cite{gronow--21} models.

\begin{figure}
\centering
\includegraphics[width=\columnwidth]{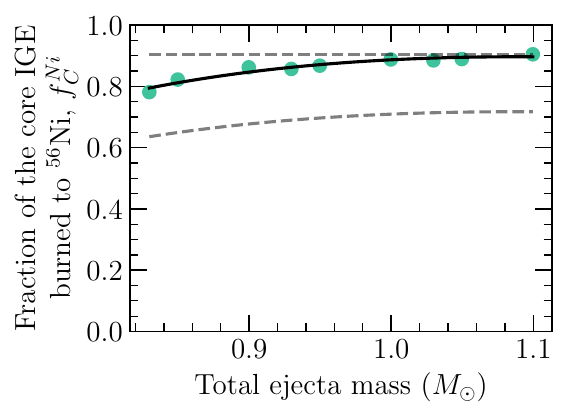}
\caption{As in Fig.~\ref{fig:dod_param3} for total ejecta mass and the fraction of the core IGE burned to $^{56}$Ni, $f^{Ni}_C$.}
\label{fig:dod_param5}
\centering
\end{figure}

\begin{figure}
\centering
\includegraphics[width=\columnwidth]{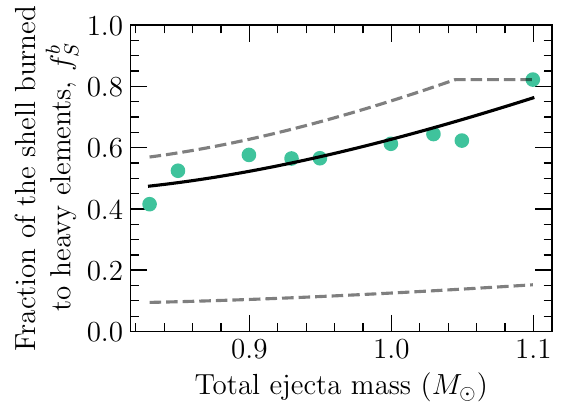}
\caption{As in Fig.~\ref{fig:dod_param3} for total ejecta mass and the fraction of the shell burned to heavy elements, $f^{b}_S$.}
\label{fig:dod_param6}
\centering
\end{figure}

In Fig.~\ref{fig:dod_param6} we show the fraction of the shell burned to heavy elements as a function of the total ejecta mass. The \cite{gronow--21} models generally show a correlation in which larger total ejecta masses result in a higher fraction of the helium shell burning to heavy elements, but there is considerably more scatter than for properties of the core. In addition, the composition of the helium shell is highly asymmetric following explosion and the side opposite the point of ignition shows considerably less burned material. Finally, as discussed in Sect.~\ref{sect:dodet}, \cite{shen--14} and \cite{townsley--19} demonstrate that pollution in the helium shell before explosion can result in models producing little or no burned material in the shell. To account for these various effects, we allow for scatter of up to $-80$~per~cent or $+20$~per~cent. The choice of a $-80$~per~cent lower limit is arbitrary. Future models should explore the variation expected from different levels of pollution in multi-dimensional models. None of the \cite{gronow--21} models predict the entire shell is burned therefore we set an upper limit equal to the $1.1~M_{\odot}$ model, 0.82.

\begin{figure}
\centering
\includegraphics[width=\columnwidth]{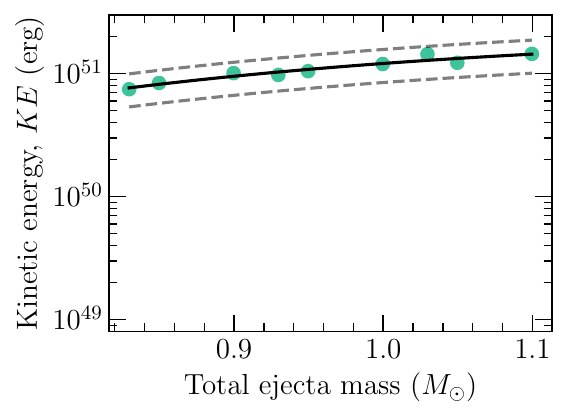}
\caption{As in Fig.~\ref{fig:dod_param3} for total ejecta mass and kinetic energy.}
\label{fig:dod_param10}
\centering
\end{figure}

As with the \cite{fink-2014} pure deflagration models, the \cite{gronow--21} double detonation models show a strong correlation between kinetic energy and total ejecta mass (Fig.~\ref{fig:dod_param10}). The full abundance profiles for the metallicity-dependant double detonation models presented by \cite{gronow--21b} are not available on HESMA therefore we are unable to directly calculate the impact of metallicity on kinetic energy. Mixing between the core and shell however can result in a $\sim$20~per~cent change in the kinetic energy. In keeping with our DEF and DDT models, we therefore again adopt scatter of 30~per~cent when sampling the kinetic energy.

%%%%%%%%%%%%%%%%%%%%%%%%%%%%%%%%%%%%%%%%%%%%%%%%%%%%%%%%%%%%%%%%%%%%%%%%%%%%%%%%%%%%%%%%%%%%%%%%%%%%%%%%%%%%%%
%%%%%%%%%%%%%%%%%%%%%%%%%%%%%%%%%%%%%%%%%%%%%%%%%%%%%%%%%%%%%%%%%%%%%%%%%%%%%%%%%%%%%%%%%%%%%%%%%%%%%%%%%%%%%%
\subsection{Gravitationally confined detonations (GCD)}
\label{sect:appdx:gcd}

\cite{lach--22b} present the gravitationally confined detonation models that form the basis of our training data. Here, the initial deflagration is ignited at a variety of white dwarf central densities and offset radii. These models follow a similar set-up as the pure deflagration models presented by \cite{lach--22}, but do not explore the impact of metallicity or rotation as in \cite{lach--22}. Therefore the expected level of variation for our gravitationally confined models is less well-constrained and we use the \cite{lach--22} models, in addition to our other training data, to inform suitable ranges for scatter.

\begin{figure}
\centering
\includegraphics[width=\columnwidth]{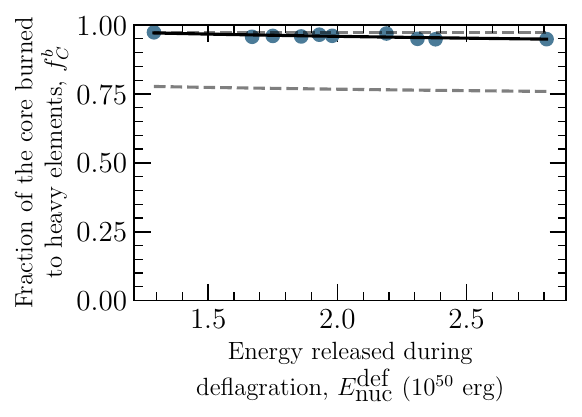}
\caption{Correlation between nuclear energy released during the deflagration phase and the fraction of the core burned to heavy elements, $f^b_C$, used to sample training data models shown as a black line. Circles show predictions from the \citealt{lach--22b} gravitationally detonation models. The upper and lower limits are shown as dashed lines.}
\label{fig:gcd_param3}
\centering
\end{figure}

Figure~\ref{fig:gcd_param3} shows the correlation between the fraction of the core burned to heavy elements and the nuclear energy released by the deflagration. Generally, the models show only minor variations. Among the \cite{lach--22} models, metallicity produces up to $\sim$5~per~cent variation in the amount of burned material. Differences in rotation however produce a larger change of $\sim$15~per~cent. In the \cite{lach--22} models, the progenitor white dwarf is not fully disrupted and only a relatively small amount of ejecta is produced therefore it is unclear if the fully disrupted gravitationally confined detonation models would show similar levels of variation. Nevertheless, based on the \cite{lach--22} models and in keeping with our other training datasets, we allow for scatter of up to 20~per~cent during sampling. Likewise, the \cite{lach--22} models show variations of up to $\sim$10~per~cent in the fraction of IGE burned to $^{56}$Ni depending on the initial carbon fraction of the white dwarf, hence we again allow for scatter of up to 20~per~cent (Fig.~\ref{fig:gcd_param5}).

\begin{figure}
\centering
\includegraphics[width=\columnwidth]{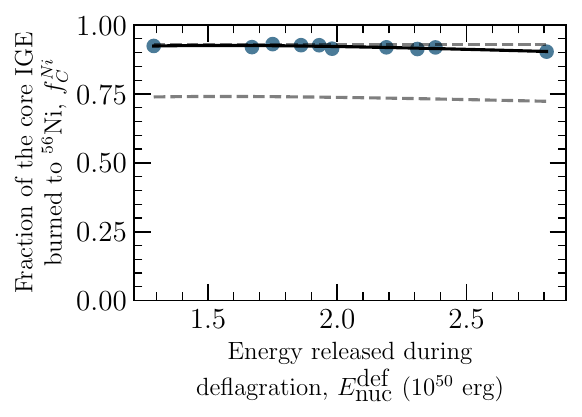}
\caption{As in Fig.~\ref{fig:gcd_param3} for total ejecta mass and the fraction of the core IGE burned to $^{56}$Ni, $f^{Ni}_C$.}
\label{fig:gcd_param5}
\centering
\end{figure}

Following the initial deflagration, in the \cite{lach--22b} gravitationally confined detonation models the deflagration ash rises towards the surface of the white dwarf and wraps around the core before subsequently triggering a detonation. The result is an asymmetric ejecta with varying amounts of burned material in the shell along different viewing angles. The \cite{lach--22b} models show a weak correlation between the amount of material in the shell burned to heavy elements and the amount of nuclear energy released during the deflagration. The \cite{lach--22} models however do not predict a shell and therefore cannot be used to inform our gravitationally confined detonation models in this case. To account for asymmetries in the ejecta and produce models viewed along different lines of sight, we follow the same approach as for our double detonation models and allow for scatter of $-80$ -- $+20$~per~cent. Again we note that the lower limit of $-80$~per~cent is arbitrary and not well-motivated physically, but is designed to follow a similar approach as our DOD models.

\begin{figure}
\centering
\includegraphics[width=\columnwidth]{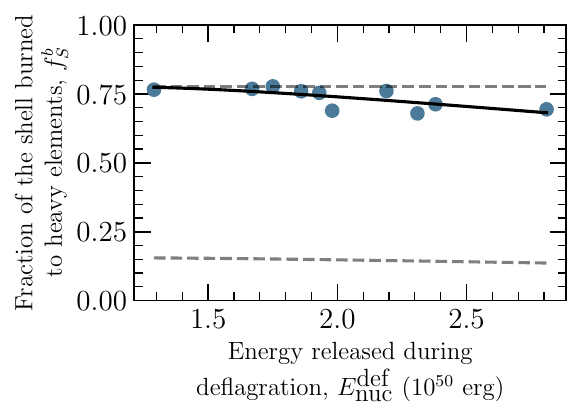}
\caption{As in Fig.~\ref{fig:gcd_param3} for total ejecta mass and the fraction of the shell burned to heavy elements, $f^{b}_S$.}
\label{fig:gcd_param6}
\centering
\end{figure}

\cite{lach--22b} models with larger amounts of energy released during the initial deflagration phase generally show smaller kinetic energies (Fig.~\ref{fig:gcd_param10}). All of these models produce 1.4~$M_{\odot}$ of ejecta, while the \cite{lach--22} pure deflagration models predict varying amounts of ejecta. The \cite{lach--22} models show a strong correlation between kinetic energy and ejecta mass, but models with similar ejecta masses can vary in kinetic energy by $\sim$15~per~cent. In keeping with our other training data, we therefore adopt a conservative scatter of 30~per~cent.

\begin{figure}
\centering
\includegraphics[width=\columnwidth]{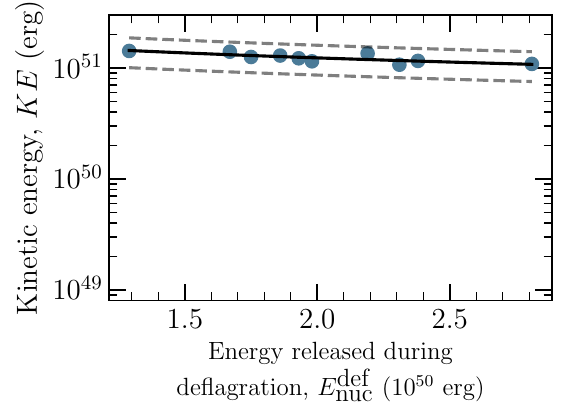}
\caption{As in Fig.~\ref{fig:gcd_param3} for total ejecta mass and kinetic energy.}
\label{fig:gcd_param10}
\centering
\end{figure}

%%%%%%%%%%%%%%%%%%%%%%%%%%%%%%%%%%%%%%%%%%%%%%%%%%%%%%%%%%%%%%%%%%%%%%%%%%%%%%%%%%%%%%%%%%%%%%%%%%%%%%%%%%%%%%
%%%%%%%%%%%%%%%%%%%%%%%%%%%%%%%%%%%%%%%%%%%%%%%%%%%%%%%%%%%%%%%%%%%%%%%%%%%%%%%%%%%%%%%%%%%%%%%%%%%%%%%%%%%%%%
\subsection{Violent mergers (VM)}
\label{sect:appdx:vm}

As discussed in Sect.~\ref{sect:vm}, our violent merger models are based on only three existing explosion models available from HESMA and are therefore less constrained and require more assumptions. As with our gravitationally confined detonation models, our assumed scatter is designed to broadly cover the predictions from the existing models and is also informed from our other training datasets. Figure~\ref{fig:vm_param3} shows that models with higher total masses generally result in more of the ejecta being burned to heavy elements. Again however this is based on only three models. In keeping with our other training datasets, we allow for 20~per~cent scatter, which covers the predictions from models. Figure~\ref{fig:vm_param10} likewise shows that the three HESMA merger models show a correlation between kinetic energy and total ejecta mass. Following our other training datasets, we again assume scatter of up to 30~per~cent when sampling new parameters.

\begin{figure}
\centering
\includegraphics[width=\columnwidth]{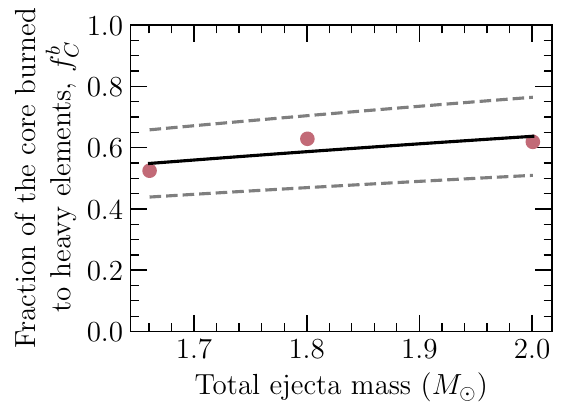}
\caption{}
\label{fig:vm_param3}
\centering
\end{figure}

\begin{figure}
\centering
\includegraphics[width=\columnwidth]{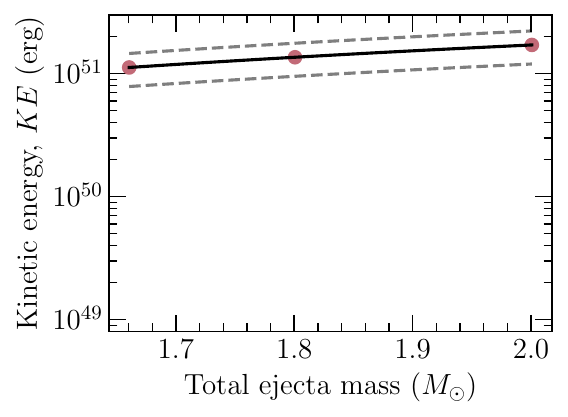}
\caption{As in Fig.~\ref{fig:vm_param3} for total ejecta mass and kinetic energy.}
\label{fig:vm_param10}
\centering
\end{figure}

%%%%%%%%%%%%%%%%%%%%%%%%%%%%%%%%%%%%%%%%%%%%%%%%%%

\subsection{Rise time}
\label{sect:appdx:rise}

\begin{figure}
\centering
\includegraphics[width=\columnwidth]{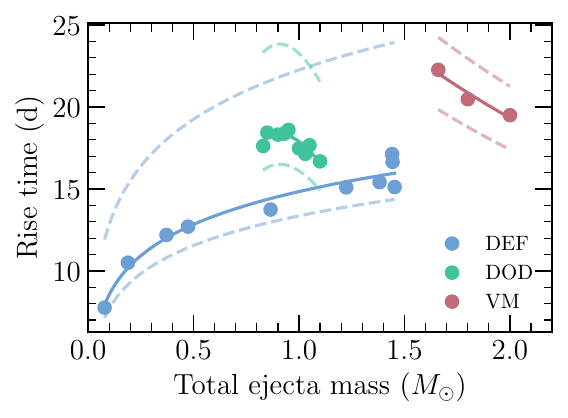}
\caption{Rise times of pure deflagration, double detonation, and violent merger models as a function of ejecta mass from which our training samples are generated. Solid lines show the correlation observed among the models while dashed lines show the upper and lower limits of the sampled values in the training data.}
\label{fig:rise}
\centering
\end{figure}

Rise times of the same SNe~Ia models calculated with different radiative transfer codes can show differences of up to a few days \citep{blondin--22}. In addition, the models used generally show shorter rise times than most SNe~Ia (e.g. \citealt{miller--20a}). To account for this systematic and ensure that our models are able to match the observed luminosities of SNe~Ia, we allow for changes in the rise time, $t_r$, relative to the underlying model and assume additional scatter (typically with a higher upper boundary). For the pure deflagration, double detonation, and violent merger models, $t_r$ is proportional to $M_{\textrm{ej}}$ (Fig.~\ref{fig:rise}). We calculate a new $t_{r,x}$ for each model based on the sampled $M_{\textrm{ej},x}$. As discussed in Sect.~\ref{sect:def}, the \cite{fink-2014} models generally show lower ejecta masses and faster rise times than observed among some SNe Iax. We therefore allow for scatter of up $-10$ -- $+50$~per~cent when sampling rise times to produce values comparable to slower-evolving SNe~Iax \citep{magee--25}. For the double detonations, we allow for scatter of $-10$ -- $+30$~per~cent. The merger models show comparable rise times to SNe~Ia, therefore we use smaller scatter of $\pm$10~per~cent. The delayed detonation models do not show any correlation between rise time and explosion strength, therefore we uniformly sample $t_{r,x}$ between 16 -- 23\,d, which broadly covers the range of rise times observed for normal SNe~Ia. In the gravitationally confined detonation scenario, $t_r$ is proportional to $E_{\textrm{nuc}}^{\textrm{def}}$, therefore we select a new $t_{r,x}$ based on the sampled $E_{\textrm{nuc},x}^{\textrm{def}}$ allowing for scatter of $\pm$10~per~cent, as in the merger models. All interpolated light curves are then stretched to match $t_{r,x}$ and the luminosity at the required phase, $L_x$, is calculated. The parameterisation of rise times for our training samples is somewhat arbitrary and may lead to biases in our fits. Nevertheless, in the absence of full simulations testing the impact of different assumptions and systematics, our pragmatic approach at least has clearly defined priors.

% Don't change these lines
\bsp	% typesetting comment
\label{lastpage}
\end{document}